\documentclass[12pt]{iopart}
\UseRawInputEncoding
\usepackage[utf8]{inputenc} 
\usepackage[T1]{fontenc}
\usepackage{graphicx}
\graphicspath{{./}{./figures/}}

\expandafter\let\csname equation*\endcsname\relax
\expandafter\let\csname endequation*\endcsname\relax
\usepackage{amsmath,amssymb}
\DeclareMathOperator{\rank}{rank}
\newcommand{\ds}{\displaystyle}
\newcommand{\threej}[6]{%
	\left(
		\begin{array}{@{}ccc@{}}
			\scriptstyle #1 & \scriptstyle #2 & \scriptstyle #3 \\
			\scriptstyle #4 & \scriptstyle #5 & \scriptstyle #6
		\end{array}
\right)}

\usepackage{url}
\usepackage[hidelinks]{hyperref}

\begin{document}

\title[Gravitational-Wave Sky Mapping with Pulsar Timing Arrays]{Gravitational-Wave Sky Mapping with Pulsar Timing Arrays:\\
	The Full Earth-Pulsar Response and Resolution Operational Regimes
}

\author[orcid=0009-0002-5322-3964, gname=Stepan, sname=Andrianov]{S. A. Andrianov}
\address{Lebedev Physical Institute, Astro Space Center, Pushchino Radio Astronomy Observatory, 142290, Moscow, Russia}
\ead{andrianovs@prao.ru}  
\author[orcid=0000-0002-4866-1532, gname=Sergei, sname=Kopeikin]{S. M. Kopeikin} 
\address{Department of Physics \& Astronomy, University of Missouri, 322 Physics Bldg., Columbia, Missouri 65211, USA}
\ead{kopeikins@missouri.edu}

\begin{abstract}
	
	Pulsar timing arrays (PTAs) are usually analyzed in the Earth-term
	approximation, although the exact response of an Earth--pulsar baseline
	contains additional angular structure through the pulsar term. We develop
	a tensor-spherical-harmonic framework for PTA sky mapping of a
	deterministic, band-limited gravitational-wave strain field, retaining the
	full Earth--pulsar response. The single-baseline response is reduced to a
	compact analytical closed form involving two spherical Bessel functions. Its
	asymptotic analysis reveals four regimes: Earth-term dominance, transition,
	pulsar-term dominance, and an effective diffraction cutoff at
	$l_{\rm cut}\simeq\omega L$. This scale characterizes the intrinsic
	exponential suppression of an individual baseline, while the practical
	angular resolution of a PTA is primarily governed by the finite number,
	sky distribution and  timing noise of the pulsars.

	Formulating PTA sky mapping as a linear inverse problem; we derive the associated Fisher information matrix and construct an singular-value-decomposition-based reconstruction framework that explicitly separates observable and poorly constrained sky modes. The analysis clarifies the relationship between intrinsic detector resolution and array-sampling limitations and provides a quantitative assessment of the circumstances under which coherent pulsar-term information becomes relevant for full-sky reconstruction. Idealized scaling arguments suggest that exploiting the transition/pulsar-term-dominated regime in global sky mapping imposes a severe $\sim l^4$ penalty on the array size required to maintain uniform reconstruction fidelity at high multipoles. The formalism extends naturally to anisotropic stochastic backgrounds and provides a unified framework for deterministic field reconstruction and angular-power anisotropy inference.

\end{abstract}

\section{INTRODUCTION}
\label{sec:introduction}

The existence of gravitational waves (GWs) is one of the fundamental predictions of Einstein's general theory of relativity \cite{einstein1918}. In the linearized weak-field approximation, gravitational waves are transverse tensor perturbations of the spacetime metric that propagate at the speed of light and transport energy, linear momentum, and angular momentum away from their sources \cite{MTW1973,Maggiore2008,Abbott2017GW170817,Monitor2017}. The existence of gravitational radiation is supported indirectly by the orbital decay of binary pulsars \cite{taylor1982} and directly by observations of compact-binary mergers, beginning with the first detection of GW150914 by the Advanced LIGO observatories \cite{ligo2016} and continuing with observations by the global LIGO--Virgo--KAGRA detector network \cite{abbott2021gwtc3}. 

The gravitational-wave spectrum spans many decades in frequency, from ultra-low frequencies $f\sim10^{-18}\,\mathrm{Hz}$, corresponding to primordial cosmological perturbations and inflationary relic backgrounds probed through cosmic microwave background polarization measurements \cite{KamionkowskiKosowsky1999,CapriniFigueroa2018}, up to frequencies of several kilohertz generated by rapidly varying astrophysical phenomena such as neutron-star mergers and core-collapse supernovae \cite{Maggiore2008,Andresen2022}. This enormous frequency range makes gravitational-wave astronomy a unique probe of both strong-field gravity and physical processes occurring in the early Universe \cite{Maggiore2008,CapriniFigueroa2018}.

The first experimental efforts to detect gravitational waves (GWs) utilized resonant-mass ``Weber bars'' in the 1960s, but subsequent independent experiments failed to reproduce these claimed detections \cite{weber1960,weber1969,garwin1974,schutz2009}. Robust indirect confirmation instead emerged from precision timing of the binary pulsar PSR B1913+16 \cite{hulse1975}, whose orbital decay matched the energy loss predicted by the general relativistic quadrupole formula \cite{peters1963,peters1964,landaulifshitz1975,taylor1982,weisberg2016}. Subsequent discoveries of relativistic binaries---including PSR B1534+12, PSR J1141$-$6545, and the double pulsar PSRJ 0737$-$3039A/B enabled stringent strong-field tests, solidifying the empirical foundations of GW physics \cite{wolszczan1991,kaspi2000,burgay2003,lyne2004,stairs2003,kramer2006,wex2014,kramer2021} and paving the way for direct detection by Advanced LIGO in 2015 \cite{ligo2016}.

Following these ground-based successes, GW astronomy has expanded across the frequency spectrum, with pulsar timing providing a unique window into the nanohertz regime. The concept of using pulsars as Galactic-scale GW detectors emerged in the late 1970s \cite{sazhin1978,detweiler1979}, demonstrating that passing GWs induce fractional frequency shifts in pulse times of arrival (TOAs). Early searches for an isotropic stochastic GW background (SGWB) monitored individual millisecond pulsars (MSPs) for red-noise signatures \cite{detweiler1979,kaspi1994,jenet2006}, while binary pulsars offered a probe of even lower frequencies via secular orbital perturbations \cite{kopeikin1997,kopeikin1999}. Although these efforts initially yielded only upper limits \cite{kaspi1994,jenet2006,shannon2015}, they established the methodology.

However, single-pulsar timing residuals are dominated by non-GW noise, such as spin irregularities, interstellar medium effects, ephemeris errors, and instrumental noise. To overcome this, \cite{hellings1983} proposed cross-correlating pairs of pulsars, showing that an isotropic SGWB induces a characteristic quadrupolar angular correlation (the \emph{Hellings--Downs curve}) dependent only on the pair's angular separation. \cite{foster1990} expanded this into the \emph{pulsar timing array} (PTA) concept, demonstrating that a network of sky-distributed MSPs can distinguish spatial noise signatures: monopolar for clock errors, dipolar for Solar-System ephemeris errors, and the Hellings--Downs quadrupolar correlation for GWs \cite{allen2023PhRvDa,allen2023PhRvDb,allen2024CQGra,allen2025PhRvL}.

Operating in the nanohertz band, PTAs have recently reported strong evidence for an SGWB via a common-spectrum process exhibiting the expected spatial correlations \cite{agazie2023,eptacollaboration2023,reardon2023,xu2023,miles2025}. While the leading interpretation is a cosmic population of supermassive black-hole binaries \cite{jaffe2003,sesana2008,sesana2010,sesana2013}, cosmological sources like relic GWs, cosmic strings, and phase transitions remain viable \cite{grishchuk1976,vilenkin1981,damour2005,caprini2018}. Disentangling these origins necessitates moving beyond isotropic detection toward measuring the angular structure of the GW sky via anisotropy studies and sky mapping \cite{mingarelli2013,grunthal2024,curylo2026}.

In the standard PTA response formalism, the GW-induced timing residual is decomposed into an Earth term and a pulsar term \cite{hellings1983,anholm2009}. Physically, the timing residual measures the difference between the GW metric perturbation evaluated at the Earth when the pulse is received and at the pulsar when the pulse was emitted. These two contributions are known respectively as the Earth term and the pulsar term. The Earth term is common to all pulsars and therefore generates the spatial correlations exploited by PTA analyses, whereas the pulsar term depends on the individual pulsar distance, $L$, the GW source direction and frequency, $\omega$. Because pulsar distances are generally known with insufficient precision to determine the pulsar-term phase accurately, and because $\omega L\gg1$ for most PTA pulsars, the pulsar term is commonly treated as incoherent between different pulsars and effectively absorbed into the noise budget of stochastic-background analyses. Consequently, current searches for stochastic backgrounds and GW sky maps rely primarily on correlations generated by the Earth term. However, the pulsar term contains additional phase information and sensitivity to higher-order angular structure of the gravitational-wave field \cite{lee2011,ellis2012, mingarelli2014, tsai2026}. Neglecting it therefore discards potentially valuable information and ultimately limits the angular resolution achievable by a PTA.

With next-generation facilities such as the Square Kilometre Array (SKA) \cite{janssen2015, shannon2025} and the Large European Array for Pulsars (LEAP) \cite{kramer2022}, together with continuing advances in very-long-baseline interferometry (VLBI) astrometry \cite{deller2011}, pulsar distances are expected to become sufficiently precise to permit coherent use of the pulsar term in PTA analyses. Under these conditions, the theoretical description of PTA observations must be extended to incorporate the full Earth-plus-pulsar response, thereby enabling a realistic assessment of the ultimate imaging and source-localization capabilities of a Galactic-scale gravitational-wave telescope.

A series of recent works have addressed related questions regarding the full Earth-pulsar response. References \cite{ng2022,liu2022, bernardo2023} studied the multipole response of PTAs to anisotropic SGWBs in both general relativity and alternative gravity theories, incorporating pulsar-term information via the Sachs--Wolfe integral \cite{sachs1967} to develop the line-of-sight (LOS) integral formalism. Domenech \cite{domenech2024} explored an alternative approach using analytical approximations of these integrals.  The total-angular-momentum (TAM) formalism introduced in \cite{qin2019} is successfully used in this series.

Regarding achievable PTA resolution and map-making, \cite{jow2026, tsai2026} investigated the resolution limits in the context of anisotropic SGWBs via Monte-Carlo simulations, while \cite{grunthal2024, grunthal2026} optimized mapping techniques by exploiting the high-multipole leakage effect in the resulting statistical sky-power distributions. 

Furthermore, \cite{agarwal2026a,semenzato2026} studied possible biases caused by high-multipole contamination in regularized solutions for the PTA spatial frequency response, utilizing Fisher information matrix and SVD approaches. 

Finally, \cite{curylo2026} developed a robust practical application framework for the linear GW strain sky-mapping problem.

The primary objective of this work is to establish a comprehensive theoretical framework for PTA sky mapping that retains the full Earth-plus-pulsar response. We reformulate the PTA response using a tensor spherical harmonic decomposition of the GW field, following \cite{gair2014}. This approach yields a compact, closed-form analytical response function for an elementary Earth-pulsar baseline, expressed entirely in terms of two spherical Bessel functions. A rigorous asymptotic analysis of this exact response reveals four distinct operational regimes: Earth-term dominance, transition, pulsar-term dominance, and an effective diffraction cutoff at $l_{\rm cut}\simeq\omega L$. This scale characterizes the intrinsic exponential suppression of an individual baseline, clarifying which angular scales of the GW sky can be probed in principle.

To quantify the recoverable information, we formulate PTA sky mapping as a linear inverse problem. Using the Fisher Information Matrix (FIM) and Singular Value Decomposition (SVD), we explicitly separate observable and poorly constrained sky modes. We demonstrate that the practical angular resolution of a PTA is strongly dominated by geometric conditioning (the number and distribution of pulsars) rather than just the physical diffraction cutoff. Specifically, we derive an $\sim l^4$ scaling law for the array size required to maintain uniform reconstruction fidelity at high multipoles. Under the assumptions of the idealized model developed in this paper, the scaling analysis suggests that depending on the GW frequency, the fiducial PTA population required to access the pulsar-term-sensitive regime for generic full-sky reconstruction spans $\sim 10^5$ to $\sim 10^{10}$. Because these requirements generally exceed the foreseeable observable millisecond-pulsar population, the analysis supports the continued use of the Earth-term approximation for full-sky GW field mapping.

We further extend this approach to searches for anisotropic stochastic gravitational-wave backgrounds, connecting our spherical-harmonic treatment to previous PTA sky-mapping and anisotropy formalisms
\cite{mingarelli2013,cornish2014,gair2014,gair2015}. Finally, as a direct consequence of this exact unified framework, we demonstrate that the TAM and LOS-integral formalism are mathematically equivalent to our tensor-harmonic approach. This equivalence eliminates the need for the numerical integration previously required in LOS-integral-based formalisms.

The resulting framework provides a mathematically rigorous foundation for future PTA sky-mapping efforts and for exploiting the enhanced capabilities expected from next-generation facilities such as the SKA and advanced VLBI astrometry.

The remainder of this paper develops this unified framework for understanding the directional response, imaging capabilities, and geometrical limitations of pulsar timing arrays. 

In Section~\ref{sec:formalism}, we formulate PTA sky mapping as a linear inverse problem setting up the required formalism used throughout this paper. Following \cite{gair2014} the gravitational-wave field is expanded in tensor spherical harmonics, and closed-form response functions are derived that retain both the Earth and pulsar terms. This representation leads naturally to a detector response matrix connecting the GW sky to PTA timing residuals, from which we construct the Fisher Information Matrix and introduce an SVD-based regularization scheme for stable map reconstruction.

Section~\ref{sec:three-operation-regimes} examines the directional response of an elementary pulsar--Earth baseline. A novel asymptotic analysis of the derived compact multipole response function reveals four distinct sensitivity regimes of an elementary PTA and establishes a diffraction-like angular-resolution limit controlled by the GW frequency and pulsar distance. These results clarify which angular scales of the GW sky can be probed in principle and which are effectively inaccessible (more precisely exponentially suppressed).

Building upon the single-baseline analysis, Section~\ref{sec:conditionning} investigates the collective response of a complete PTA via the means of FIM and SVD-modes. We study the conditioning of the inverse problem, derive geometric criteria for optimal pulsar placement, quantify spurious polarization (Einsteinian $+$ and $\times$) mixing and apparent leakage effects, and assess the consequences of non-uniform sky coverage. The analysis yields both intrinsic and practical limits on PTA imaging performance and identifies the conditions under which coherent use of the pulsar term can improve sky reconstruction.

Section~\ref{sec:numerical-validation} presents numerical tests of the theoretical predictions. Using simulated PTAs, we examine the singular-value spectrum of the response matrix, evaluate the performance of the SVD reconstruction procedure, and demonstrate consistency with the predicted transitions between sensitivity regimes, rank deficiencies, and sky-recovery limits.

Since a principal scientific objective of PTA experiments is the detection and characterization of a stochastic gravitational-wave background, Section~\ref{sec:anisotropy} extends the formalism to anisotropic background analyses. Following \cite{gair2014, lentati2013} we derive the associated mode-coupling relations, establish the connection with generalized overlap-reduction functions, and formulate a likelihood framework compatible with existing PTA pipelines while retaining the geometric interpretation developed throughout the paper.

Finally, Section~\ref{sec:conclusion} summarizes the principal results and discusses their implications for future PTA observations, gravitational-wave imaging, and next-generation radio facilities. Technical details of the tensor-harmonic formalism, detector response calculations, asymptotic derivations, and the relation to the TAM and LOS-integral formalisms are collected in \ref{app:basis-defs}--\ref{app:TAM-LOS-adaptation}.

\section{Spherical-Harmonic Decomposition of PTA Sky Maps} 
\label{sec:formalism}

The objective of the considered PTA sky mapping problem is to reconstruct the angular distribution of generic gravitational-wave (GW) strain on the celestial sphere from a finite set of pulsar timing measurements. As discussed in the Introduction, this problem is intrinsically an inverse problem: the PTA does not observe the GW field directly, but rather measures timing residuals produced by the integrated response of individual pulsar--Earth baselines to incident GWs. Recovering the underlying sky distribution therefore requires a representation of the GW field that naturally separates its angular degrees of freedom (multipole modes) and allows the detector response to be expressed in a tractable linear form.

A convenient and physically transparent framework is provided by the decomposition of the GW metric perturbation into gradient (electric/E-mode) and curl (magnetic/B-mode) tensor spherical harmonics \cite{kamionkowski1997, allen1999, gair2014, gair2015}. This representation is analogous to the spherical-harmonic analysis widely used in cosmic microwave background polarization studies and provides a natural basis for characterizing anisotropic GW signals on the sky. Within this formalism, the PTA timing-residual response can be written as a linear mapping between the spherical-harmonic coefficients of the GW field and the measured pulsar timing residuals.

In this section we begin with the standard transverse--traceless plane-wave description of the GW field and derive the timing-residual response of an individual pulsar. We then project the GW field onto tensor spherical harmonics and obtain closed-form analytical expressions for the elementary detector response functions while explicitly retaining both the Earth and pulsar terms, assuming the distances to the pulsars to be precisely determined. Finally, we formulate the PTA sky-mapping problem as a Gaussian linear inverse problem, introduce the corresponding FIM, and describe an SVD-based regularization scheme for treating the inevitable rank deficiencies and ill-conditioning of the reconstruction problem \cite{gair2014, cornish2015, mingarelli2013}.

\subsection{The Signal Plane Wave Expansion}
The primary observables in a PTA are the timing residuals (TRs), defined as the differences between the observed and model-predicted pulse times of arrival (TOAs). A passing GW perturbs the spacetime metric along the radio pulse trajectory connecting the pulsar and the Earth, producing fluctuations in the measured arrival times \cite{estabrook1975, sazhin1978, detweiler1979, anholm2009}. In the weak-field approximation, the spacetime metric is written as
\begin{equation}\label{eq-1}
	g_{\mu\nu}=\eta_{\mu\nu}+h_{\mu\nu},
\end{equation}
where $\eta_{\mu\nu}$ is the Minkowski metric and $h_{\mu\nu}$ is a small perturbation satisfying $|h_{\mu\nu}|\ll1$. Throughout this paper, Greek indices $\mu,\nu=0,1,2,3$ denote spacetime coordinates, $x^\mu=(t,\vec x)$, while Latin indices $a,b=1,2,3$ denote spatial coordinates. We also adopt the geometric units where the fundamental speed $c = 1$. Here we consider the flat, stationary Universe approximation that is sufficient for intra-Galactic pulsar observations.

In the transverse--traceless (TT) gauge, the temporal components of the metric perturbation vanish, $h_{0\mu}=0$, and the physical degrees of freedom of the gravitational-wave (GW) field are encoded in the spatial metric perturbation $h_{ab}$. The latter can be expanded as a superposition of monochromatic plane waves \cite{allen1999, anholm2009},
\begin{equation}
	h_{ab}(t,\vec x)=
	\sum_A
	\ds\int_{-\infty}^{\infty}d\omega
	\oint_{\mathbb{S}^2}d\hat\Omega\,
	h_A(\omega,\hat\Omega)\,
	e^A_{ab}(\hat\Omega)\,
	e^{i\omega(t-\hat\Omega\cdot\vec x)},
	\label{eq:h-fourier}
\end{equation}
where $i=\sqrt{-1}$ denotes the imaginary unit, $\omega=2\pi f$ is the angular frequency, $\hat\Omega$ is the unit vector pointing toward the GW propagation direction, and $d\hat\Omega=\sin\theta\,d\theta\,d\phi$ is the solid-angle element on the unit sphere $\mathbb{S}^2$. The quantities $h_A(\omega,\hat\Omega)$ are the Fourier amplitudes of the GW field, while $e^A_{ab}(\hat\Omega)$ are the GW polarization tensors defined in \ref{app:basis-defs}. In general relativity, the polarization index $A\in\{+,\times\}$ labels the two tensor polarization states.

The GW-induced timing residual is given by \cite{estabrook1975, sazhin1978, detweiler1979}
\begin{equation}
	\delta t(t)=
	\ds\int_0^t dt'\,
	\ds\frac{\hat p^a\hat p^b}
	{2(1+\hat\Omega\!\cdot\!\hat p)}
	\,\Delta h_{ab}(t',\hat\Omega),
	\label{eq:tr-def}
\end{equation}
where $\hat p$ is the unit vector from the Earth to the pulsar and
\begin{equation}
	\Delta h_{ab}(t,\hat\Omega)
	=
	h_{ab}(t_{\rm e},\vec x_{\rm e})
	-
	h_{ab}(t_{\rm p},\vec x_{\rm p})
\end{equation}
is the difference between the metric perturbation evaluated at the reception event on Earth and at the pulse-emission event at the pulsar.

Choosing the coordinate system of \ref{app:basis-defs} with the Earth located at the spatial origin,
\begin{equation}
	\begin{aligned}
		(t_{\rm p},\vec x_{\rm p}) &= (t-L,L\hat p),\\
		(t_{\rm e},\vec x_{\rm e}) &= (t,\vec 0),
		\label{eq:time-coordinates-def}
	\end{aligned}
\end{equation}
where $L$ is the pulsar distance. Substituting Eq.~\eqref{eq:time-coordinates-def} into Eq.~\eqref{eq:h-fourier} and then into Eq.~\eqref{eq:tr-def} yields
\begin{equation}
	\delta t(t)
	=
	\sum_A
	\ds\int_{-\infty}^{\infty}d\omega
	\oint_{\mathbb{S}^2}d\hat\Omega\,
	h_A(\omega,\hat\Omega)
	\ds\frac{e^{i\omega t}}{i\omega}
	R^A(\omega,\hat\Omega),
	\label{eq:deltat-fourier}
\end{equation}
where
\begin{equation}
	R^A(\omega,\hat\Omega)
	=
	\left[
		1-e^{-i\omega L(1+\hat\Omega\cdot\hat p)}
	\right]
	F^A(\hat\Omega)
	\label{eq:R-def}
\end{equation}
is the pulsar timing response function. The first term represents the Earth term, while the exponential factor
accounts for the pulsar term. The geometric antenna pattern is
\begin{equation}
	F^A(\hat\Omega)
	=
	\ds\frac{1}{2}
	\ds\frac{\hat p^a\hat p^b}
	{1+\hat\Omega\cdot\hat p}
	e^A_{ab}(\hat\Omega),
	\label{eq:F-def}
\end{equation}
which describes the angular sensitivity of the Earth--pulsar baseline
to a GW of polarization $A$ arriving from the direction $\hat\Omega$.
Throughout this paper, repeated spatial indices
($a,b,c,\ldots$) are assumed to be summed over according to the
Einstein summation convention unless stated otherwise.

\subsection{The Signal Spherical Expansion}

The primary objectives, plane-wave amplitudes $h_A(\omega,\hat{\Omega})$ ($A=+,\times$), describe the GW field as a function of propagation direction on the celestial sphere. For studies of anisotropy and sky reconstruction, however, it is often advantageous to replace this continuous directional description by a discrete multipolar representation. This is achieved by expanding the transverse-traceless metric perturbation in a complete basis of tensor spherical harmonics \cite{kamionkowski1997, thorne1980, gair2014}. The explicit construction of the tensor-harmonic basis, its orthogonality properties, and its relation to the polarization tensors are summarized in \ref{app:basis-defs}; see in particular Eqs.~\eqref{eq:Ytens-def}--\eqref{eq:GA-def}.

In the linear polarization basis, the spatial metric perturbation is written as
\begin{equation}
	h_{ab}(\omega,\hat{\Omega})
	=
	h_+(\omega,\hat{\Omega})\,e^+_{ab}(\hat{\Omega})
	+
	h_\times(\omega,\hat{\Omega})\,e^\times_{ab}(\hat{\Omega}),
	\label{eq:h-polarization-decomposition}
\end{equation}
where $e^+_{ab}$ and $e^\times_{ab}$ are the transverse-traceless polarization tensors \eqref{eq:lm-def}--\eqref{eq:eA-def}.

Alternatively, the same metric perturbation can be expanded in the tensor-harmonic basis,
\begin{equation}
	h_{ab}(\omega,\hat{\Omega})
	=
	\sum_{lm}
	\Big[
		a^G_{lm}(\omega)\,
		Y^G_{(lm)ab}(\hat{\Omega})
		+
		a^C_{lm}(\omega)\,
		Y^C_{(lm)ab}(\hat{\Omega})
	\Big],
	\label{eq:h-spherical-decomposition}
\end{equation}
where $Y^G_{(lm)ab}$ and $Y^C_{(lm)ab}$ are the gradient (electric-type) and curl (magnetic-type) tensor spherical harmonics, respectively. The indices $l$ and $m$ are the usual angular multipole numbers, $a,b$ denote tensor components on three-dimensional space, and
\[
	\sum_{lm}
	\equiv
	\sum_{l=2}^{\infty}\sum_{m=-l}^{l}.
\]
The summation begins at $l=2$ because the tensor spherical harmonics are generated from scalar spherical harmonics by the differential operators defined in Eq.~\eqref{eq:Ytens-def}. These operators annihilate the monopole ($l=0$) and dipole ($l=1$) modes, so nontrivial transverse-traceless tensor harmonics exist only for $l\ge2$. Physically, Einstainian gravitational waves represent a symmetric trace-free tensor field on the sphere possesses neither monopolar nor dipolar degrees of freedom.

The coefficients $a^G_{lm}$ and $a^C_{lm}$ characterize the gradient and curl components of the GW sky. They provide a discrete description of the angular structure of the gravitational-wave field and play a role analogous to the $E$- and $B$-mode coefficients used in cosmic-microwave-background polarization analyses. The correspondence between the plane-wave amplitudes and the tensor-harmonic coefficients is \cite{gair2014}\newline
\begin{subequations}\label{eq:a-h-transform}
	\begin{align}
		h_+(\omega,\hat \Omega)
&=
\sum_{lm}
\frac{N_l}{2}
\Big[
	a^G_{lm}(\omega)\,
	W_{lm}(\hat{\Omega})
	-
	a^C_{lm}(\omega)\,
	X_{lm}(\hat{\Omega})
\Big],
\label{eq:a-h-transform-a}
\\[1ex]
h_\times(\omega,\hat \Omega)
&=
\sum_{lm}
\frac{N_l}{2}
\Big[
	a^G_{lm}(\omega)\,
	X_{lm}(\hat{\Omega})
	+
	a^C_{lm}(\omega)\,
	W_{lm}(\hat{\Omega})
\Big].
\label{eq:a-h-transform-b}
	\end{align}
\end{subequations}
\newline\noindent with inverse relations
\newline
\begin{subequations}\label{eq:h-a-transform}
	\begin{align}
		a^G_{lm}(\omega)
	&=
	N_l
	\oint_{\mathbb{S}^2} d\hat\Omega
	\Big[
		h_+(\omega,\hat\Omega)\,
		W^*_{lm}(\hat\Omega)
		+
		h_\times(\omega,\hat\Omega)\,
		X^*_{lm}(\hat\Omega)
	\Big],
	\\\\
	a^C_{lm}(\omega)
	&=
	N_l
	\oint_{\mathbb{S}^2} d\hat\Omega
	\Big[
		h_\times(\omega,\hat\Omega)\,
		W^*_{lm}(\hat\Omega)
		-
		h_+(\omega,\hat\Omega)\,
		X^*_{lm}(\hat\Omega)
	\Big].
	\end{align}
\end{subequations}
\newline\noindent
Here $N_l$, $W_{lm}$, and $X_{lm}$ are defined in \ref{app:basis-defs}, and the asterisk denotes a complex conjugation. Equations~(\ref{eq:a-h-transform}) and~(\ref{eq:h-a-transform}) establish the correspondence between the conventional plane-wave description and its tensor-harmonic decomposition. In the remainder of this paper, the coefficients $a^G_{lm}$ and $a^C_{lm}$ serve as the fundamental variables of the sky-mapping problem. The PTA response derived in the following subsection acts directly on these multipole coefficients, allowing the timing residuals to be expressed as a linear mapping between the observed data and the angular structure of the GW field.

\subsection{The Elementary Detector Response Functions}

The timing response of an individual Earth--pulsar baseline can be expressed naturally in the same tensor-harmonic basis. Substituting the decomposition \eqref{eq:h-spherical-decomposition} into Eq.~\eqref{eq:deltat-fourier} gives\newline\noindent 
\begin{equation}
	\delta t(t)
	=
	\ds\int_{-\infty}^{\infty} d\omega\,
	\ds\frac{e^{i\omega t}}{i\omega}
	\sum_{lm}
	\Big[
		a^G_{lm}(\omega)\,
		R^G_{lm}(\omega)
		+
		a^C_{lm}(\omega)\,
		R^C_{lm}(\omega)
	\Big],
	\label{eq:deltat-sperical}
\end{equation} 
where the multipolar detector response functions are
\newline\noindent
\begin{subequations}\label{eq:RA-RP-transform}
	\begin{align}
		R^G_{lm}(\omega)
	&=
	\ds\ds\frac{N_l}{2}
	\oint_{\mathbb{S}^2} d\hat{\Omega}\,
	\Big[
		R^+(\omega,\hat{\Omega})
		W_{lm}(\hat{\Omega})
		+
		R^\times(\omega,\hat{\Omega})
		X_{lm}(\hat{\Omega})
	\Big],
	\\\nonumber\\
	R^C_{lm}(\omega)
	&=
	\ds\ds\frac{N_l}{2}
	\oint_{\mathbb{S}^2} d\hat{\Omega}\,
	\Big[
		R^\times(\omega,\hat{\Omega})
		W_{lm}(\hat{\Omega})
		-
		R^+(\omega,\hat{\Omega})
		X_{lm}(\hat{\Omega})
	\Big].
	\end{align}
\end{subequations}
\newline\noindent These relations may be inverted to recover the plane-wave response functions,\newline\noindent
\begin{subequations}
	\begin{align}
		R^+(\omega,\hat \Omega)
	&=
	\sum_{lm}N_l
	\Big[
		R^G_{lm}(\omega)
		W^*_{lm}(\hat\Omega)
		-
		R^C_{lm}(\omega)
		X^*_{lm}(\hat\Omega)
	\Big],
	\\\nonumber\\
	R^\times(\omega,\hat \Omega)
	&=
	\sum_{lm}N_l
	\Big[
		R^G_{lm}(\omega)
		X^*_{lm}(\hat\Omega)
		+
		R^C_{lm}(\omega)
		W^*_{lm}(\hat\Omega)
	\Big].
	\label{eq:RP-RA-transform}
	\end{align}
\end{subequations}\newline\noindent
A major advantage of the tensor-harmonic formulation is that the angular integrals in Eq.~\eqref{eq:RA-RP-transform} can be evaluated analytically. Following \cite{gair2014,gair2015}, one finds
\newline\noindent
\begin{align}
	R^G_{lm}(\omega)
	&=
	2\pi\,
	Y_{lm}(\hat p)\,
	(-i)^l\,
	e^{-iy}\,
	f_l(y),
	\label{eq:RG-analyt}
	\\\nonumber\\
	R^C_{lm}(\omega)
	&=
	0,
	\label{eq:RC-analyt}
\end{align}\newline\noindent
where $y=\omega L$ and $f_l(y)$ in its compact form is a linear combination of two spherical Bessel functions defined in Eq.~\eqref{eq:fl-def}. The derivation is given in \ref{app:response}.

From Eqs.~\eqref{eq:RG-analyt}--\eqref{eq:RC-analyt}, it is evident that the curl-mode response vanishes identically ($R^C(\omega) = 0$). Consequently, although both the plus and cross polarization maps can be formally reconstructed via Eq.~\eqref{eq:a-h-transform}, they are synthesized entirely from the gradient-mode coefficients $a^G_{lm}(\omega)$, with the unobservable curl modes $a^C_{lm}(\omega)$ necessarily set to zero.

\subsection{The Likelihood Function}

Having derived the response of a single Earth--pulsar baseline, we now extend the formalism to an array of pulsars. The PTA data set consists of timing residuals measured for multiple pulsars at discrete observing epochs. For pulsar $i$ observed at epoch $t_j$, the residual is modeled as
\begin{equation}
	s_i(t_j)=\delta t_i(t_j)+n_i(t_j),
\end{equation}
where $\delta t_i(t_j)$ is the GW-induced signal and $n_i(t_j)$ denotes the noise contribution. The latter includes radiometer noise, pulse-phase jitter, chromatic and achromatic red-noise processes, and other instrumental or astrophysical systematics \cite{ellis2014,agazie2023a,eptacollaboration2023a,miles2025}.

Throughout this work we assume the standard Gaussian-noise model commonly adopted in PTA analyses \cite{vanhaasteren2009,ellis2013apj,agazie2023a}. The angular brackets $\langle \cdots \rangle$ denote an ensemble average over realizations of the noise process. We assume that the noise has zero mean,
\begin{equation}\label{noise1}
	\langle n_i(t_j)\rangle =0,
\end{equation}
and covariance
\begin{equation}\label{noise2}
	\langle n_i(t_j)n_{i'}(t_{j'})\rangle
	=C_{iji'j'},
\end{equation}
where $n_i(t_j)$ is the noise contribution to the timing residual of pulsar $i$ measured at observing epoch $t_j$, and $C_{iji'j'}$ is the corresponding noise covariance matrix in component form.
When convenient, the pulsar and time indices are combined into a single
compound index so that $C_{iji'j'}$ is represented by the matrix
$\mathbf C$. In what follows we use boldface symbols to denote matrices
and vectors, while retaining explicit indices whenever they help clarify
the angular dependence of the problem.

Collecting all timing residuals into a single data vector,
\begin{equation}
	\mathbf{s}
	=
	\bigl[
		s_1(t_1),\ldots,s_1(t_{N_1}),
		s_2(t_1),\ldots,
		s_{N_{\rm pul}}(t_{N_{\rm TOA}})
	\bigr]^T,
\end{equation}
the PTA measurement model may be written compactly as
\begin{equation}
	\mathbf{s}
	=
	\mathbf{R}\mathbf{a}
	+
	\mathbf{n},
	\label{eq:linear-model}
\end{equation}
where $\mathbf{a}$ contains the tensor-harmonic sky coefficients and
$\mathbf{R}$ is the corresponding response (statistically speaking, design) matrix. 

Under the Gaussian approximation, the likelihood for the data is
\begin{equation}
	p(\mathbf{s}\mid\mathbf{a})
	=
	\ds\frac{1}
	{\sqrt{\det(2\pi\mathbf{C})}}
	\exp
	\!\left[
		-\ds\frac12
		(\mathbf{s}-\mathbf{R}\mathbf{a})^\dagger
		\mathbf{C}^{-1}
		(\mathbf{s}-\mathbf{R}\mathbf{a})
	\right],
	\label{eq:likelihood}
\end{equation}
where the superscript $\dagger$ denotes Hermitian conjugation and
$\mathbf C=\langle\mathbf n\mathbf n^\dagger\rangle$ is the noise
covariance matrix \eqref{noise2}. For the tensor-harmonic representation derived above, the model prediction entering Eq.~\eqref{eq:linear-model} is\newline\noindent
\begin{equation}
	(\mathbf{R}\mathbf{a})_j
	=
	\ds\int_{-\infty}^{\infty}
	d\omega\,
	\ds\frac{e^{i\omega t_j}}{i\omega}
	\sum_{lm}
	a^G_{lm}(\omega)
	R^G_{lm}(\omega),
	\label{eq:Ra-product}
\end{equation}
where the index $j$ labels individual PTA measurements. In principle the coefficients $a^G_{lm}$ are frequency dependent. Several parameterizations are possible \cite{gair2014, talbot2021}. In this work we analyze one Fourier frequency $\omega$ at a time. For a fixed frequency bin, the Fourier amplitudes of the timing residuals from all pulsars are assembled into the data vector $\mathbf{s}$, and the frequency dependence of $\mathbf{R}$ and $\mathbf{a}$ is suppressed for notational simplicity. The likelihood \eqref{eq:likelihood} then retains the same form.

Maximizing Eq.~\eqref{eq:likelihood} yields the generalized least-squares estimator
\begin{equation}
	\hat{\mathbf a}
	=
	\left(
		\mathbf R^\dagger
		\mathbf C^{-1}
		\mathbf R
	\right)^{-1}
	\mathbf R^\dagger
	\mathbf C^{-1}
	\mathbf s,
	\label{eq:likelihood-solution}
\end{equation}
which is the so-called clean map \cite{curylo2026}. The matrix
\begin{equation}
	\mathbf F
	=
	\mathbf R^\dagger
	\mathbf C^{-1}
	\mathbf R
	\label{eq:FIM-def}
\end{equation}
plays a central role in the inverse problem and will appear repeatedly throughout the remainder of the paper. It is the Fisher Information Matrix , which characterizes how strongly the measured timing residuals constrain the GW strain field sky coefficients. The FIM determines both the statistical uncertainties of the reconstructed sky map and the set of sky modes that can be measured by the pulsar array. Its rank gives the number of independently accessible modes, while its condition number quantifies the degree of ill-conditioning of the reconstruction.

To simplify the numerical analysis we factorize the noise covariance matrix via Cholesky decomposition \cite{golub2013}:
\begin{equation}
	\mathbf C
	=
	\sigma^2
	\mathbf L\mathbf L^\dagger,
	\label{eq:Cholesky-def}
\end{equation}
where $\mathbf L$ is 
a lower triangular matrix that capture the structure of the unit-variance covariance matrix, and $\sigma^2$ acts as a global variance scale. Separating the scale $\sigma^2$ allows it to be subsequently estimated from the data residuals. Defining the whitened quantities
\begin{equation}
	\tilde{\mathbf R}
	=
	\mathbf L^{-1}\mathbf R,
	\qquad
	\tilde{\mathbf s}
	=
	\mathbf L^{-1}\mathbf s,
	\qquad
	\tilde{\mathbf n}
	=
	\mathbf L^{-1}\mathbf n,
	\label{eq:likelihood-cholesky}
\end{equation}
one obtains
\begin{equation}
	\langle
	\tilde{\mathbf n}
	\tilde{\mathbf n}^\dagger
	\rangle
	=
	\sigma^2\mathbf I.
\end{equation}
Then the whitened Fisher matrix $\mathbf{F}$ is then
\begin{equation}
	\mathbf F
	= \sigma^2
	\tilde{\mathbf R}^\dagger
	\tilde{\mathbf R}.
\end{equation}

The maximum-likelihood solution is obtained using a singular-value
decomposition (SVD) of the whitened response matrix
\cite{hansen1998},
\begin{equation}
	\tilde{\mathbf R}
	=
	\mathbf U
	\mathbf\Sigma
	\mathbf V^\dagger ,
	\label{eq:svd-def}
\end{equation}
where $\mathbf U$ and $\mathbf V$ are unitary matrices and
$\mathbf\Sigma$ is a diagonal matrix whose elements are the singular
values. The SVD decomposes the response operator into orthogonal
data-space modes (the columns of $\mathbf U$) and orthogonal
parameter-space modes (the columns of $\mathbf V$), while the singular
values quantify the sensitivity of the PTA to each mode. Modes
associated with small singular values are only weakly constrained by
the data and, if inverted directly, would strongly amplify measurement
noise.

Because the response matrix is generally rectangular and may be
rank-deficient, a conventional matrix inverse does not exist. The
inverse problem is therefore solved using the Moore--Penrose
pseudoinverse \cite{moore1920,penrose1955}, denoted by the superscript
``$+$''. The pseudoinverse generalizes matrix inversion to non-square
and singular matrices and yields the minimum-norm least-squares
solution. Equivalently, among all parameter vectors that fit the data
equally well in the least-squares sense, it selects the one having the
smallest Euclidean norm.

The Moore--Penrose pseudoinverse is constructed from the SVD according
to
\begin{equation}
	\tilde{\mathbf R}^{+}
	=
	\mathbf V
	\mathbf\Sigma^{+}
	\mathbf U^\dagger ,
\end{equation}
where $\mathbf\Sigma^{+}$ is obtained by replacing each retained
singular value $\sigma_i$ with its reciprocal $\sigma_i^{-1}$ and
setting singular values below a prescribed threshold to zero. This
regularization removes poorly constrained modes to which the detector
has little or no sensitivity.

The corresponding maximum-likelihood estimator is
\begin{equation}
	\hat{\mathbf a}
	=
	\tilde{\mathbf R}^{+}
	\tilde{\mathbf s}
	=
	\mathbf V
	\mathbf\Sigma^{+}
	\mathbf U^\dagger
	\tilde{\mathbf s},
	\label{eq:likelihood-svd}
\end{equation}
which reconstructs the sky multipoles within the subspace spanned by
the retained singular modes. The discarded modes form an effective null
space of the detector response and therefore cannot be reconstructed
reliably from the observations.

The covariance matrix of the reconstructed multipoles is
\begin{equation}
	{\rm Cov}(\hat{\mathbf a})
	=
	\hat{\sigma}^2
	\mathbf V
	(\mathbf\Sigma^{+})^2
	\mathbf V^\dagger ,
	\label{eq:likelihood-svd-cov}
\end{equation}
where $\hat{\sigma}^2$ is the residual variance estimate,
\begin{equation}
	\hat{\sigma}^2
	=
	\ds\frac{
		\|
		\tilde{\mathbf s}
		-
		\tilde{\mathbf R}\hat{\mathbf a}
		\|^2
	}
	{
		N_{\rm data}
		-
		N_{\rm non\mbox{-}null}
	},
	\label{eq:likelihood-svd-noise}
\end{equation}
with $\|\mathbf a\|=(\mathbf a\!\cdot\!\mathbf a^*)^{1/2}$ denoting the
Euclidean norm, $N_{\rm data}$ the number of PTA measurements, and
$N_{\rm non\mbox{-}null}$ the number of retained singular modes. The
quantity $\hat{\sigma}^2$ therefore measures the residual variance
remaining after the data have been projected onto the observable
subspace of the detector response. 
This residual variance $\hat{\sigma}^2$ serves as the empirical maximum-likelihood estimator for the unknown global scaling parameter $\sigma^2$ introduced in the noise covariance decomposition, Eq.~\eqref{eq:Cholesky-def}. 

The SVD formulation is particularly advantageous for PTA sky reconstruction because it remains stable in the presence of rank deficiency and strong parameter degeneracies. Unlike Eq.~\eqref{eq:likelihood-solution}, it avoids explicitly forming the inverse Fisher matrix and therefore mitigates the quadratic amplification of condition numbers associated with the normal equations.

In summary, the tensor-harmonic formalism recasts PTA sky mapping as the linear inverse problem
\(
\mathbf s=\mathbf R\mathbf a+\mathbf n
\),
in which the data are related directly to the multipolar structure of the GW sky. The analytic response functions derived in the previous section incorporate the full Earth and pulsar terms and reveal that an isolated pulsar--Earth baseline is insensitive to curl modes. The Fisher matrix characterizes the recoverability of the sky modes, while the SVD-based pseudoinverse provides a robust estimator when the response matrix is singular or poorly conditioned. In the following section we investigate the asymptotic properties of the response function $f_l(y)$ and identify the regimes that govern PTA sky reconstruction.

\section{Angular-Sensitivity Regimes of an  Earth--pulsar Response}
\label{sec:three-operation-regimes}

\begin{figure}
	\centering
	\includegraphics[width=.9\textwidth]{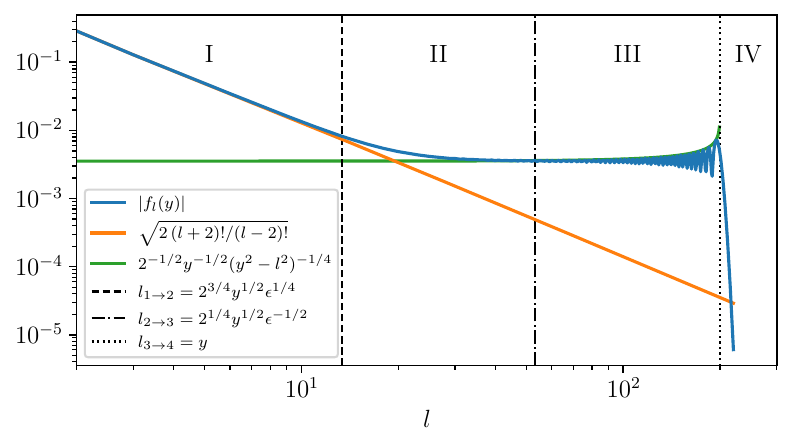}
	\caption{\scriptsize 
		Absolute value of the sensitivity function $f_l(y)$, with
		$y=\omega L=200$, shown as a function of multipole number $l$. A relatively low value of $y$ is taken for illustration; the tolerance parameter for the regime boundaries is $\epsilon = 10\%$.
		Four asymptotic regimes can be identified:
		(I) an Earth-term-dominated;
		(II) a transition; 
		(III) a pulsar-term-dominated; and
		(IV) an effective diffraction cutoff at $l\gtrsim \omega L$.
		The cutoff follows from the turning-point (Airy) asymptotics of the underlying spherical Bessel functions;
	}
	\label{fig:fl}
\end{figure}

The spherical-harmonic formulation developed in Section~\ref{sec:formalism} demonstrates that the angular sensitivity of an elementary pulsar--Earth detector is completely determined by the multipole response function $f_l(y)$ appearing in the analytical solution \eqref{eq:RG-analyt}--\eqref{eq:RC-analyt}, where the dimensionless parameter
\[
	y\equiv\omega L
\]
is the product of the GW angular frequency $\omega$ and the pulsar distance $L$. The function $f_l(y)$ therefore encapsulates the connection between GW wavelength, detector baseline length, and sky-map angular scale on the response of the PTA. 

As shown in Eq.~\eqref{eq:fl-def} and \ref{app:asymptotics}, $f_l(y)$ is expressed as a linear combination of spherical Bessel functions of the first kind and exhibits qualitatively different asymptotic behavior in different regions of the $(l,y)$ parameter space. Consequently, the response of a pulsar timing detector to a multipolar GW sky is not uniform across angular scales. Instead, the detector passes through a sequence of distinct sensitivity regimes as the multipole order $l$ is varied relative to the characteristic scale $y$.

These regimes admit a clear physical interpretation:
\begin{enumerate}
	\item \textbf{Earth-term-dominated regime}, in which the detector response is governed primarily by the metric perturbation at the Solar-System barycenter;

	\item \textbf{Transition regime}, in which the contribution originating at the pulsar becomes significant and both the Earth and pulsar terms exhibit similar amplitudes;

	\item \textbf{Pulsar-term dominated regime}, in which the Earth term decays significantly and the pulsar term dominates the multipole response;

	\item \textbf{Effective diffraction cutoff regime}, in which  geometric averaging strongly (more precisely, exponentially) suppresses the detector response to increasingly fine angular structure.
\end{enumerate}

The existence of these regimes determines which angular scales of the GW sky can in principle be reconstructed by a PTA and which modes become effectively inaccessible. In this section we describe the asymptotic behavior of $f_l(y)$ in each regime, establish the corresponding scaling laws for the detector response, and discuss their physical implications for PTA sky mapping and multipole recovery. The presented regimes are summarized in Fig.~\ref{fig:fl}.

All the asymptotic behavior refered in this section is related to the function $f_l(y)$ itself. For the discussion of the spherical Bessel function assympotics see \ref{app:asymptotics}.

\subsection{The Earth-Term Dominated Regime}
\label{sec:earth-term-regime}

The Earth-term-dominated regime corresponds to the operating conditions of most contemporary PTA experiments and is therefore the regime most commonly assumed in studies of isotropic/anisotropic gravitational-wave backgrounds and PTA sky mapping \cite{anholm2009, mingarelli2013, taylor2021}. Within the tensor-spherical-harmonic formalism, the asymptotic detector response in this limit was derived by \cite{gair2014} and is given by Eq.~\eqref{eq:RG-large-y}.

This regime is defined by two conditions. First, the multipole number $l$ remains finite, corresponding to a sky map of finite angular resolution. Second, the dimensionless parameter
\begin{equation}
	y=\omega L \gg 1,
\end{equation}
which is satisfied for typical PTA observations. For example, a GW with frequency $f=10~\mathrm{nHz}$ observed using a pulsar at a distance of $L=1~\mathrm{kpc}$ yields ${y\approx6\times10^3}$. Physically, this means that many GW wavelengths fit within the Earth--pulsar baseline.

In this limit the detector response becomes independent of $y$. The rapidly oscillating pulsar term averages out, suppressing its contribution to the sky response, while the Earth term remains unaffected. Consequently, the detector response is governed almost entirely by the Earth term, which is common to all pulsars in the array. 

The corresponding multipole sensitivity follows the asymptotic scaling, \eqref{eq:RG-large-y}--\eqref{eq:N_l-asympt},
\begin{equation}
	f_l(y)\propto l^{-2},
\end{equation}
demonstrating a rapid loss of sensitivity with increasing angular complexity of the GW sky.  As a result, low-order multipoles dominate the measured signal, whereas progressively finer angular structure becomes increasingly difficult to reconstruct.

\subsection{The Transition Regime}
\label{sec:transition-regime}

The large-$y$ asymptotics of $f_l(y)$, Eq.~\eqref{eq:fl-large-y},
show that the Earth-term approximation cannot remain valid at
arbitrarily high multipoles. In the Earth-term-dominated regime, the
sensitivity decreases approximately as $l^{-2}$, whereas the
pulsar-term contribution approaches a finite oscillatory amplitude.
Consequently, a range of multipoles must exist in which the two
contributions become comparable. Within this interval neither term can
be neglected, and the PTA response evolves continuously from
Earth-term dominance to pulsar-term dominance (Fig.~\ref{fig:fl}).

\ref{app:asymptotics} derives quantitative estimates of this
crossover by examining the domains of validity of the Earth-term and
pulsar-term asymptotic expansions. The resulting characteristic
multipoles, denoted by $l_{1\to2}$ and $l_{2\to3}$, should not be
interpreted as sharp physical boundaries. Rather, they represent
crossover scales at which an asymptotic approximation ceases to be
self-consistent at a prescribed accuracy level. The multipole
$l_{1\to2}$ marks the onset of the transition regime, where corrections
to the Earth-term approximation become significant, while
$l_{2\to3}$ marks the onset of the pulsar-term-dominated regime.

The numerical values of these boundaries depend on the tolerance
parameter $\epsilon$ introduced in \ref{app:asymptotics},
which specifies the acceptable fractional error of the asymptotic
expansions. Since Eqs.~\eqref{eq:l-12-boundary}
and~\eqref{eq:l-23-boundary} are themselves obtained from asymptotic
approximations, the boundaries are intended only as order-of-magnitude
estimates.
We adopt the order of magnitude comparison value	$\epsilon=10\%$.

For a typical Galactic pulsar and a gravitational-wave source in the
PTA band, Eq.~\eqref{eq:l-12-boundary} yields
\begin{equation}
	l_{1\to2}
	\simeq
	\left(\omega L\right)^{1/2}
	\approx
	80
	\left(
		\ds\frac{\omega}{2\pi\cdot10~\mathrm{nHz}}
	\right)^{1/2}
	\left(
		\ds\frac{L}{1~\mathrm{kpc}}
	\right)^{1/2}.
	\label{eq:fl-1-2-boundary-number}
\end{equation}
Below this scale the detector response is predominantly determined by
the Earth term. As $l$ approaches $l_{1\to2}$, however, pulsar-term
corrections become increasingly important and the Earth-term
approximation gradually loses predictive accuracy.

The upper end of the transition region is determined by the condition
that the pulsar-term asymptotic expansion becomes self-consistent.
Using the same tolerance level, $\epsilon=10\%$,
Eq.~\eqref{eq:l-23-boundary} gives
\begin{equation}
	l_{2\to3}
	\simeq 4
	\left(\omega L\right)^{1/2}
	\approx
	320
	\left(
		\ds\frac{\omega}{2\pi\cdot10~\mathrm{nHz}}
	\right)^{1/2}
	\left(
		\ds\frac{L}{1~\mathrm{kpc}}
	\right)^{1/2}.
	\label{eq:fl-2-3-boundary-number}
\end{equation}

For multipoles
\[
	l_{1\to2}\lesssim l \lesssim l_{2\to3},
\]
the Earth-term and pulsar-term contributions are comparable in
magnitude, and neither asymptotic approximation is individually
adequate. Beyond the upper boundary,
\[
	l\gtrsim l_{2\to3},
\]
the detector enters the pulsar-term-dominated regime, where the
Earth-term contribution becomes subdominant and the sensitivity is
described by the large-$l$ asymptotics derived in
\ref{app:pulsar-regime}.

The existence of the finite interval
$[\,l_{1\to2},\,l_{2\to3}\,]$ demonstrates that the commonly used
Earth-term approximation occupies only a restricted region of the
$(l,y)$ parameter space. As the multipole number increases, the PTA
response must pass through a transition regime before reaching the
pulsar-term-dominated sensitivity regime.

\subsection{The Pulsar-Term Dominated Regime}
\label{sec:pulsar-term-regime}

As the multipole number increases beyond the transition region discussed above, the detector enters a regime in which the pulsar term provides the dominant contribution to the response function. This transition occurs when the multipole number $l$ becomes comparable to the dimensionless parameter
\(
y=\omega L,
\)
that is,
\(
l \sim y.
\)
Throughout this subsection we additionally assume
$
l \gg 1,
$
which allows the asymptotic expansions derived in \ref{app:asymptotics} to be applied.

In this regime the sensitivity function is described by the pulsar-term asymptotic solution \eqref{eq:fl-ly-asympt}, which is a harmonic approximation with steadily increasing envelope.  The behavior of the detector is controlled primarily by the ratio
\begin{equation}
	\xi_l=\ds\frac{l+1/2}{y}.
\end{equation}
Immediately above the transition regime boundary, where $\xi_l\ll1$, the oscillation amplitude remains small and the sensitivity approaches
\begin{equation}
	f_l(y)\sim 
	2^{-1/2}y^{-1}.
\end{equation}
To leading order, the response becomes nearly independent of the multipole number. This is a remarkable departure from the Earth-term-dominated regime, where the sensitivity decreases as $l^{-2}$. Consequently, after the transition at $l_{2\to3}$, the detector retains approximately uniform sensitivity to a broad range of multipoles extending up to the next transition boundary, denoted by $l_{3\to4}$. In this sense, the pulsar term allows access to angular scales that would be severely suppressed in the conventional Earth-term approximation.

As $\xi_l$ approaches unity, the character of the response changes. Both the oscillation amplitude and the oscillation period in multipole space increase, and the pulsar-term contribution acquires the asymptotic scaling \eqref{eq:fl-assimp-jy-assymp}
\begin{equation}
	f_l(y) \sim(1-\xi_l^2)^{-1/4}.
\end{equation}
The resulting enhancement originates from the stationary-phase structure of the pulsar-term integral and may be interpreted as \emph{a geometric resonance between the angular scale of the GW distribution and the characteristic phase variation across the Earth--pulsar baseline}. The response therefore reaches its largest pulsar-term amplitude in the vicinity of the pulsar-term to sensitivity-cutoff transition.

It is important to emphasize, however, that this enhancement does not restore the sensitivity level achieved in the Earth-term-dominated regime. Even at its maximum, the pulsar-term response remains substantially smaller than the peak Earth-term response. The maximum Earth-term sensitivity occurs at the lowest allowed multipole, $l_{\rm e}=2$, whereas the maximum pulsar-term sensitivity is attained near the Airy turning point, $l_{\rm p}\simeq l_{\rm cut}\simeq \omega L$. Using Eqs.~\eqref{eq:fl-large-y} and \eqref{eq:fl-secondary-peak}, the corresponding maxima are
\begin{equation}
	\begin{aligned}
		f_{l_\mathrm{e}}(y)
	&\approx  3 \times 10^{-1}, \\
	f_{l_\mathrm{p}}(y)
	&\approx
	4\times10^{-4}
	\left(
		\ds\frac{\omega}
		{2\pi\cdot10~\mathrm{nHz}}
	\right)^{-5/6}
	\left(
		\ds\frac{L}
		{1~\mathrm{kpc}}
	\right)^{-5/6}.
	\end{aligned}
\end{equation}

For typical PTA parameters, the maximum pulsar-term response is therefore several orders of magnitude smaller than the maximum Earth-term response. The principal significance of the pulsar-term-dominated regime is not an increase in absolute sensitivity, but rather the extension of measurable multipole content beyond the range accessible in the Earth-term approximation. This extension persists until the detector reaches the final regime, where the response becomes exponentially suppressed and sky reconstruction effectively ceases.

\subsection{Effective diffraction cutoff regime}
\label{sec:cutoff-regime}

The pulsar-term-dominated regime terminates at the third characteristic transition, denoted by $l_{3\to4}$. In the vicinity of this transition the sensitivity is described by an Airy-function approximation \eqref{eq:fl-Ai}, which provides a uniform matching between the two asymptotic regimes. To leading order, the transition occurs at
\begin{equation}
	l_{3\to 4} = \omega L
	\approx
	6.5\times10^3
	\left(
		\ds\frac{\omega}{2\pi\cdot10~\mathrm{nHz}}
	\right)
	\left(
		\ds\frac{L}{1~\mathrm{kpc}}
	\right).
	\label{eq:fl-3-4-boundary-number}
\end{equation}
This marks the end of the pulsar-term-dominated regime. The geometric resonance discussed in Section~\ref{sec:pulsar-term-regime} reaches its maximum effectiveness near this boundary and can no longer compensate for the increasingly rapid phase cancellation produced by finer angular structure in the GW sky.

According to Eq.~\eqref{eq:fl-jl-approx}, the response function directly inherits the spherical Bessel function of the first kind characteristic behavior in the vicinity of the turning point. 
For multipoles below the transition, $l<l_{3\to4}$, the detector response oscillates as a function of $l$ with a slowly varying amplitude. As the transition is approached, the oscillations are progressively pumped. Beyond the boundary, $l>l_{3\to4}$, the sensitivity function enters an exponential decay regime. The response to higher multipoles then decreases so rapidly that these modes become exponentially suppressed. Consequently, $l_{3\to4}$ represents a  diffraction  scale for  the angular structure that can be probed by an elementary Earth--pulsar detector.

The transition multipole $l_{3\to4}$ therefore determines the intrinsic diffraction scale of the detector. Using the standard correspondence between multipole number and angular scale, we obtain
\begin{equation}
	\theta =
	\ds\frac{180^\circ}{l_{3\to 4}}
	\approx
	1.7~\mathrm{arcmin}
	\left(
		\ds\frac{\omega}{2\pi\cdot10~\mathrm{nHz}}
	\right)^{-1}
	\left(
		\ds\frac{L}{1~\mathrm{kpc}}
	\right)^{-1}.
	\label{eq:resolution}
\end{equation}
This result admits a simple physical interpretation. Since
\begin{equation}
	l_{3\to4}
	\simeq \omega L
	=
	\ds\frac{2\pi L}{\lambda},
\end{equation}
where $\lambda$ is the GW wavelength, Eq.~\eqref{eq:resolution} can be written  as
\begin{equation}
	\theta \simeq \ds\frac{\lambda}{2 L}.
	\label{eq:difraction-limit}
\end{equation}
Comparing this relation to optical/radio interferometers' characteristic resolution,
the Earth--pulsar distance acts as an effective baseline that determines the finest angular structure resolvable in the GW sky.

A signal-power-dependent estimation of the PTA resolution limit, $1/\mathrm{SNR}\cdot \lambda /L$, was provided  in \cite{tsai2026}. We note that this resolution, beyond the effective difraction cutoff, \eqref{eq:difraction-limit} is achievable via the means of compressed sensing algorithms \cite{foucart2013}. Assuming \textit{a-priory} defined model of the signal, e.g., a set of point sources, substantially bright sources might be localized more precisely than the detector's point spread function characteristic width. These techniques are an information trade-off: adding priors limits the range of possible signal sources while leading to more precise position determination.

In summary, the asymptotic analysis of the elementary detector response function $f_l(y)$ reveals four distinct angular-sensitivity regimes. At low multipoles, the response is dominated by the  Earth term and decreases approximately as a power law. With an increasing multipole number, the Earth-term sensitivity drops substantially and becomes comparable to the pulsar term, marking the transition region. At intermediate multipoles, the pulsar term becomes important and produces an oscillatory response that extends sensitivity to substantially finer angular scales. Finally, near $l_{3\to4}\simeq\omega L$, the detector encounters a  diffraction sensitivity cutoff governed by Airy-function asymptotics, beyond which the response decays exponentially.

When generalized to a PTA, these single-baseline properties directly determine the structure and conditioning of the response matrix $\mathbf{R}$ appearing in the likelihood function \eqref{eq:likelihood}. The exponential cutoff imposes a physical upper limit on the recoverable multipoles, while the gradual suppression of the response over a broad range of $l$ generates increasingly small singular values and a corresponding degradation of the FIM. The resulting inverse problem closely resembles sparse interferometric imaging \cite{hogbom1974}: only a finite range of angular modes can be reconstructed reliably, whereas modes beyond the intrinsic resolution limit remain  inaccessible.

\section{Array Conditioning and Sky-Mapping Performance}
\label{sec:conditionning}

Section~\ref{sec:formalism} established that PTA sky mapping can be formulated--see Eq.~\eqref{eq:linear-model}--as a linear inverse problem,
\(
\mathbf{s}=\mathbf{R}\mathbf{a}+\mathbf{n},
\)
where the response matrix $\mathbf{R}$ encodes the timing residuals response of the PTA to the spherical-harmonic modes of the GW sky. Section~\ref{sec:three-operation-regimes} then identified the asymptotic angular-sensitivity regimes of an individual pulsar--Earth baseline and demonstrated that the elementary response becomes strongly suppressed beyond a characteristic multipole cutoff. We now investigate how these single-baseline properties determine the conditioning of the PTA inverse problem and, consequently, the fidelity of the reconstructed sky maps. 

We assume that pulsars' intrinsic and measurement noise (correlation matrix), their ephemerides, including distances, are precisely known. Hence, we consider explicitly geometrical properties of PTA, i.e. the inverse-problem-driven biases.

The stability of the maximum-likelihood solution \eqref{eq:likelihood-solution} is governed by the FIM
\begin{equation}
	\mathbf{F}=
	\mathbf{R}^{\dagger}
	\mathbf{C}^{-1}
	\mathbf{R},
\end{equation}
defined in Eq.~\eqref{eq:FIM-def}. Since the FIM contains all information about parameter identifiability and statistical uncertainties, its rank and condition number directly determine the number of GW modes that can be reconstructed reliably from a given PTA data set.

The linear structure of the likelihood allows the properties of the FIM to be inferred directly from the response matrix. Using the whitened representation introduced in Section~\ref{sec:formalism},
\begin{equation}
	\rank \mathbf{F}
	=
	\rank
	\left(
		\tilde{\mathbf{R}}^\dagger
		\tilde{\mathbf{R}}
	\right)
	=
	\rank \tilde{\mathbf{R}}
	=
	\rank \mathbf{R},
	\label{eq:FIM-rank}
\end{equation}
where $\tilde{\mathbf{R}}$ is defined by Eq.~\eqref{eq:likelihood-cholesky}. Equation~\eqref{eq:FIM-rank} shows that all rank deficiencies of the likelihood problem originate from the geometry and sampling properties of the PTA encoded in $\mathbf{R}$. Consequently, the conditioning of the full array can be understood by studying how the responses of individual pulsars combine within the response matrix.

For each Fourier frequency of the PTA data, the response matrix has dimensions
\(
N_{\mathrm{pulsars}}
\times
N_{\mathrm{modes}},
\)
where $N_{\mathrm{modes}}$ is the number of spherical-harmonic coefficients included in the sky model. Therefore,
\begin{equation}
	\rank \mathbf{F}
	\le
	\min
	\{
		N_\mathrm{pulsars},
		N_\mathrm{modes}
	\}.
	\label{eq:FIM-max-rank}
\end{equation}
Equation~\eqref{eq:FIM-max-rank} provides a purely algebraic upper bound on the recoverable sky complexity: the number of independently measurable modes can never exceed the number of pulsars in the array. Equivalently, the maximum  algebraic sky-map resolution is reached when
\begin{equation}
	N_{\mathrm{modes}}
	=
	N_{\mathrm{pulsars}}.
	\label{eq:N-modes-max}
\end{equation}
This result was initially obtained independently by \cite{cornish2014} and \cite{gair2014}.

It is important to distinguish this  algebraic limitation from the physical diffraction scale derived in Section~\ref{sec:cutoff-regime}. The latter is imposed by the elementary detector response and determines the highest multipole that can be measured   a single baseline. The former arises from the finite number of pulsars available to sample those modes. Together, these two effects define the ultimate sky-mapping capability of a PTA.

\subsection{The Geometric Optimality Conditions}

Equation~\eqref{eq:FIM-max-rank} provides an upper bound on the rank of the FIM. Achieving this bound, however, requires the rows and columns of the response matrix $\mathbf{R}$ to remain as linearly independent as possible. Geometrically, this implies that different pulsars should probe complementary combinations of GW-sky modes, while different spherical-harmonic modes should produce distinguishable signatures across the PTA. The degree of linear dependence can be quantified through the angles between vectors in the response matrix.

Let $\mathbf{r}_A$ denote the $A$-th row of $\mathbf{R}$, corresponding to the response of pulsar $A$ to the set of GW-sky modes included in the reconstruction. The cosine of the angle between two response vectors $\mathbf{r}_A$ and $\mathbf{r}_B$ is
\begin{equation}
	r = \ds\frac{\mathbf{r}_A \cdot \mathbf{r}_B^{*}}{\lVert \mathbf{r}_A \rVert  \,\lVert \mathbf{r}_B \rVert },
	\label{eq:dot-rows}
\end{equation}
where the symbol ``$\cdot$'' denotes the  inner product,
and $\lVert \mathbf{a}\rVert=(\mathbf{a}\cdot\mathbf{a}^*)^{1/2}$ is the corresponding Euclidean norm. The case $|r|=1$ in Eq.~\eqref{eq:dot-rows} corresponds to complete linear dependence and $|r|=0$ corresponds to orthogonal responses.

Using the analytical detector response \eqref{eq:RG-analyt} together with the spherical-harmonic addition theorem,
\begin{equation}
	P_l(\hat p_A \cdot \hat p_B)
	=
	\ds\frac{4\pi}{2l+1}\sum_m
	Y_{lm}(\hat p_A)
	Y^*_{lm}(\hat p_B),
\end{equation}
the correlation between two pulsar response vectors may be written as
\begin{equation}
	|r| \propto
	\left|
	\sum_{l=2}^{l_\mathrm{max}}
	(2l+1)f_l(y_A) f_l^*(y_B)
	P_{l} (\hat{p}_A \cdot \hat{p}_B)
	\right|.
	\label{eq:dot-rows-explicit}
\end{equation}
Here $l_{\max}$ denotes the highest multipole retained in the sky model. It should be distinguished from the  detector's diffraction cutoff $l_{\rm cut}\simeq\omega L$ discussed in Section~\ref{sec:cutoff-regime}. While $l_{\rm cut}$ sets the finest angular scale to which an individual Earth--pulsar baseline is intrinsically sensitive, $l_{\max}$ is a reconstruction parameter determining the number of modeled sky modes,
\begin{equation}\label{eq:modmax}
	N_{\rm modes}
	=
	\sum_{l=2}^{l_{\max}}(2l+1)
	=
	(l_{\max}+1)^2-4
	\simeq l_{\max}^2 .
\end{equation}
The practically achievable value of $l_{\max}$ is therefore governed not only by the fundamental detector response but also by the rank, Eq. \eqref{eq:FIM-rank}, and conditioning, Eq. \eqref{eq:FIM-max-rank}, of the PTA Fisher Information Matrix ${\bf F}$.

Equation~\eqref{eq:dot-rows-explicit} provides a quantitative measure of redundancy between pulsars. Large values of $|r|$ indicate that the corresponding pulsars respond to nearly identical combinations of GW modes and therefore contribute little independent information to the array. Conversely, small values of $|r|$ correspond to nearly orthogonal response vectors and maximize the improvement in the rank and conditioning of the Fisher matrix.

This immediately leads to a geometric optimality criterion for PTA design. The scientific impact of adding a new pulsar is maximized when its response vector is as orthogonal as possible to the responses of all pulsars already present in the array. In practice, this suggests that pulsar surveys should preferentially target sky regions where the correlation function \eqref{eq:dot-rows-explicit} is minimized. Pulsars located near the minima of this function provide the largest increase in the information content of the array.

The finite dimensionality of the PTA response space imposes, however, a  limitation on this strategy. Since Eq.~\eqref{eq:dot-rows-explicit} contains only multipoles up to $l_{\max}$, the corresponding correlation function possesses only a finite angular complexity on the celestial sphere. Its nodal curves identify directions for which the response of a newly added pulsar is approximately orthogonal to the responses of the existing PTA. As the array grows, these near-optimal directions become progressively exhausted, and newly added pulsars become increasingly correlated with those already present. Consequently, enlarging a PTA eventually yields diminishing returns: although additional pulsars continue to improve sensitivity, the amount of independent geometric information contributed by each new pulsar decreases.

An equivalent perspective is obtained by examining the columns of the response matrix. Whereas row orthogonality quantifies the diversity of pulsar responses, column orthogonality measures the distinguishability of different GW-sky modes. Let $\mathbf{c}_{lm}$ denote the column of $\tilde{\mathbf{R}}$ corresponding to mode $(lm)$. The cosine of the angle between two column vectors is defined through their normalized Hermitian inner product,
\begin{equation}
	c =
	\ds\frac{
		\mathbf{c}_{lm}\cdot \mathbf{c}_{l'm'}^*
	}{
		\lVert \mathbf{c}_{lm} \rVert
		\,
		\lVert \mathbf{c}_{l'm'} \rVert
	},
	\label{eq:dot-columns}
\end{equation}
As in Eq.~\eqref{eq:dot-rows}, $|c|=1$ corresponds to two linearly dependent columns, while $|c|=0$ corresponds to orthogonal modes.
Substituting the analytical response \eqref{eq:RG-analyt} into Eq.~\eqref{eq:dot-columns} yields
\begin{equation}
	|c|
	\propto
	\left|
	\sum_{A=1}^{N_{\mathrm{pulsars}}}
	Y_{lm}(\hat p_A)
	Y_{l'm'}^{*}(\hat p_A)
	f_l(y_A)
	f_{l'}^{*}(y_A)
	\right|.
	\label{eq:dot-columns-explicit}
\end{equation}
Two limiting cases are particularly instructive.

The first corresponds to a \emph{filled aperture}. Suppose that the number of pulsars becomes very large,
\[
	N_{\mathrm{pulsars}}\rightarrow\infty,
	\qquad l~\text{is finite},
\]
and that the response amplitudes $f_l(y)$ vary only weakly across the array, as occurs in the Earth-term-dominated regime. The discrete sum in Eq.~\eqref{eq:dot-columns-explicit} then approaches an integral over the celestial sphere,
\begin{equation}
	|c|
	\to
	\mathrm{const}
	\left|
	\oint_{\mathbb{S}^{2}}
	d\hat p_A\,
	Y_{lm}(\hat p_A)
	Y_{l'm'}^{*}(\hat p_A)
	\right|
	=
	\mathrm{const}\,
	\delta_{ll'}\delta_{mm'}.
	\label{eq:dot-columnes-filled-aperture}
\end{equation}
The limit
$
|c|\propto\delta_{ll'}\delta_{mm'}
$
demonstrates that distinct spherical-harmonic modes become mutually orthogonal and therefore statistically independent. In this idealized limit, every additional mode contributes an independent degree of freedom to the reconstruction problem. The PTA then behaves analogously to an interferometric array with complete $uv$-plane coverage. Just as complete $uv$ coverage regularizes image reconstruction in radio interferometry, uniform sampling of the celestial sphere regularizes GW sky mapping by suppressing mode degeneracies.

The second limiting case is the large-$l$ regime:
\[
	l \to \infty, \qquad
	N_{\rm pulsars}~\text{is finite}.
\]
In this case all pulsars enter the  effective diffraction-cutoff regime discussed in Section~\ref{sec:cutoff-regime}. The exponential suppression of the response causes the sum in Eq.~\eqref{eq:dot-columns-explicit} to become dominated by the least-suppressed pulsar, which is generally the most distant pulsar in the array. Again denoting this pulsar by $A$, one obtains
\begin{equation}
	|c|
	\to
	\ds\frac{
		\left|
		Y_{lm}(\hat p_A)
		Y_{l'm'}^{*}(\hat p_A)
		\right|
	}{
		|Y_{lm}(\hat p_A)|
		\,|Y_{l'm'}(\hat p_A)|
	}
	=1.
	\label{eq:dot-columns-exp}
\end{equation}
Equation~\eqref{eq:dot-columns-exp} implies that all sufficiently high-order modes become nearly parallel in the response matrix. Consequently, adding additional spherical-harmonic modes no longer increases the rank of the FIM, and the corresponding GW-sky modes become observationally indistinguishable. In linear-algebra terms, these modes collapse onto an effectively one-dimensional subspace of the response matrix. This behavior is the matrix manifestation of the sensitivity-cutoff regime discussed in Section~\ref{sec:cutoff-regime}.

The two limiting cases reveal the  factors controlling PTA sky-mapping performance. Equation~\eqref{eq:dot-columns-exp} demonstrates that the ultimate physical angular resolution of a PTA is determined by its most distant pulsars, since they possess the largest cutoff multipole, $l_{\rm cut}\simeq\omega L$, and therefore remain sensitive to the finest angular structure of the GW sky. In contrast, Eq.~\eqref{eq:FIM-max-rank} shows that the practically achievable angular resolution is constrained by the finite rank of the FIM and, hence, by the number of pulsars available to sample the accessible mode space.

The distinction between these limits is important. The sensitivity cutoff defines the maximum multipole that can be measured in principle by an arbitrarily large PTA, whereas the rank bound determines how many of those physically accessible modes can actually be reconstructed from a finite set of pulsars. Thus, the sky-mapping capability of a PTA is governed jointly by its longest Earth--pulsar baselines and by the number and sky distribution of pulsars comprising the array.

Therefore, PTA imaging is constrained by two complementary limitations. The first is a diffraction-scale sensitivity limit imposed by the detector response itself and determined by the longest Earth--pulsar baselines. The second is a sampling limit imposed by the finite dimensionality of the PTA and determined by the number and sky distribution of pulsars in the array. Improving the angular resolution of PTA sky maps therefore requires progress along both fronts simultaneously: discovering more distant pulsars extends the range of physically accessible multipoles, while discovering additional pulsars improves the sampling of those modes and enhances the conditioning of the inverse problem.

These considerations may be summarized by two practical design principles for PTA sky mapping:

\begin{itemize}
	\item \textit{The highest physically accessible multipoles are determined by the most
			distant pulsars in the array, since these possess the largest cutoff
		multipoles, $l_{\rm cut}\simeq \omega L$.}

	\item \textit{The maximum angular resolution achievable in practice is determined by the number and sky distribution of pulsars available for observation.}
\end{itemize}

The first statement reflects the physical sensitivity cutoff discussed in Section~\ref{sec:cutoff-regime}, whereas the second follows from the rank bound \eqref{eq:FIM-max-rank} and the conditioning properties of the response matrix. Together they define the  diffraction and geometric limits of PTA-based gravitational-wave sky reconstruction.

\subsection{PTA Size Required to Reach the Transition Regime}

The transition discussed in Section~\ref{sec:three-operation-regimes} is a property of the asymptotic response of an individual Earth--pulsar baseline. The question addressed in the present subsection is different. Even if the detector response formally enters the transition regime at sufficiently high multipoles, a finite PTA may not contain enough independent measurements to reconstruct the corresponding GW-sky modes. We therefore ask: how many pulsars are required for a PTA to reach the multipoles at which the Earth-term approximation begins to break down, reaching the boundary $l_{1\to2}$?

The answer follows from the distinction between the diffraction sensitivity limit and the sampling limit established in the previous subsection. The response analysis of Section~\ref{sec:three-operation-regimes} determines which multipoles are accessible by a single baseline, whereas the rank of the FIM determines how many of those multipoles can be reconstructed by a finite PTA.

For multipoles below the transition multipole number $l_{1\to2}$, the detector response is accurately described by the Earth-term asymptotic solution, Eqs.~\eqref{eq:fl-large-y}, \eqref{eq:N_l-asympt} for which
\begin{equation}
	f_l(y)\propto l^{-2}.
	\label{eq:fl-lminus2}
\end{equation}
Moreover, the number of spherical-harmonic modes retained up to multipole $l_{\max}$ is
\begin{equation}
	N_{\rm modes}
	=
	\sum_{l=2}^{l_{\max}}(2l+1)
	=
	(l_{\max}+1)^2-4
	\simeq
	l_{\max}^2 .
	\label{eq:Nmodes-lmax}
\end{equation}
Consequently, improving the angular resolution of a PTA sky map requires not only sensitivity to progressively weaker modes, but also the reconstruction of a rapidly growing number of independent degrees of freedom.

To estimate the required PTA size, we examine how the modal reconstruction variance scales with multipole number at fixed array size. The estimate follows from the SVD of the response matrix. Let $\sigma_l$ denote a representative singular value of the whitened response matrix $\tilde{\mathbf{R}}$ for modes of multipole number $l$.

For the order-of-magnitude estimation we assume that the PTA pulsars
\begin{enumerate}
	\item are uniformly distributed on the celestial sphere
		\[
			\sigma_{lm} \simeq \sigma_{lm'} \simeq \sigma_l;
		\]  

	\item share the same characteristic distance scale $L$:
		\[
			L_A \simeq L_B,
		\]
		where $A$ and $B$ denote different pulsars;

	\item contribute statistically independent information to the response matrix,  i.e., contribute orthogonal spherical modes:
		\[
			|c| = 0,
		\] 
		where $c$ is defined in Eq.~\eqref{eq:dot-columns}.
\end{enumerate}

The response matrices' singular-value spectrum is then expected to follow the same asymptotic scaling as the detector response:
\begin{equation}
	\sigma_l
	\propto
	(N_{\rm pulsars})^{1/2}|f_{l}(y)|
	\propto
	(N_{\rm pulsars})^{1/2} l^{-2}.
	\label{eq:sigma-scaling}
\end{equation}

The variance of the maximum-likelihood estimator for a mode of multipole $l$ scales inversely with the square of the singular value, Eq.~\eqref{eq:likelihood-svd-cov}:
\begin{equation}
	\text{Var}(\hat{a}_{lm}) \propto \sigma_l^{-2} \propto \frac{l^4}{N_{\rm pulsars}}.
	\label{eq:var-scaling}
\end{equation}
To maintain a fixed reconstruction fidelity---defined as a uniform upper bound on the modal variance, $\text{Var}(\hat{a}_{lm}) \le \epsilon$, for all resolved multipoles up to $l_{\max}$---the total array size must satisfy the sensitivity scaling:
\begin{equation}
	N_{\rm pulsars} \propto l_{\max}^4.
	\label{eq:l4-scaling}
\end{equation}
This must be satisfied concurrently with the separate algebraic rank condition, Eq.~\eqref{eq:FIM-max-rank},
\[
	N_{\rm pulsars} \ge N_{\rm modes} \simeq l_{\max}^2,
\] 
which ensures the response matrix is not rank-deficient. Because $l^4 \gg l^2$ for large $l$, the sensitivity requirement dominates the array sizing.

Equation~\eqref{eq:l4-scaling} therefore provides an idealized parametric scaling estimate of the pulsar population required to probe progressively higher multipoles within the Earth-term-dominated regime and ultimately reach the transition multipole $l_{1\to2}$. This estimate assumes that every pulsar contributes statistically independent information. In reality, Eq.~\eqref{eq:dot-rows-explicit} shows that the response vectors of different pulsars become increasingly correlated as the PTA grows. Consequently, the actual pulsar population required to reach a given multipole will generally exceed the idealized scaling predicted by Eq.~\eqref{eq:l4-scaling}.

To convert the proportionality in Eq.~\eqref{eq:l4-scaling} into a numerical estimate, we must fix its overall normalization.The natural anchor is the lowest multipole accessible to tensor perturbations, $l_{\min}=2$. At this multipole the rank bound, Eq.~\eqref{eq:FIM-max-rank}, requires only the five quadrupole modes to be spanned, so the minimal self-consistent array that can represent the quadrupole subspace contains $N_{\text{ref}} = 5$ pulsars. We adopt this purely as a mathematical reference point to normalize the variance scaling. This yields the heuristic reference curve
\begin{equation}
	N_{\rm pulsars}(l_{\max}) \simeq
	N_{\text{ref}} 
	\left( \frac{l_{\max}}{l_{\min}} \right)^4 = 
	5\times \left( \frac{l_{\max}}{2} \right)^4.
	\label{eq:Np-heuristic}
\end{equation}
Evaluating this at the reference transition multipole $l_{1\to2} \simeq 80$, Eq.~\eqref{eq:fl-1-2-boundary-number} gives the fiducial single-number estimate
\begin{equation}
	N_{\text{trans}} \sim 5 \times \left( \frac{80}{2} \right)^4 \sim 10^7.
	\label{eq:transition-number}
\end{equation}

Equation~\eqref{eq:transition-number} is an evaluation at a representative PTA frequency and pulsar distance, not a rigid threshold. Because \(N_{\mathrm{pulsars}}\propto l_{\mathrm{max}}^{4}\) and \(l_{1\to2}\simeq(\omega L)^{1/2}\), the required array size at the transition multipole scales as \(N_{\mathrm{trans}}\propto(\omega L)^{2}\). This is a steep frequency dependence, and a single number therefore conceals an order-of-magnitude span. To exhibit that span we evaluate Eq.~\eqref{eq:Np-heuristic} at the two ends of the PTA band: \(f=1/(30\,\text{yr})\), set by the observing baseline, and \(f=1/(1\,\text{month})\), set by the Nyquist frequency of a typical cadence. For a fiducial \(L=1\)~kpc, Eq.~\eqref{eq:fl-1-2-boundary-number} gives \(l_{1\to2}\approx 26\) and \(l_{1\to2}\approx 500\), respectively, and Eq.~\eqref{eq:Np-heuristic} gives
\begin{subequations}
	\begin{align}
		N_\mathrm{trans} &\sim 
		5 \times \left(
			\ds\frac{26}{2}
		\right)^4
		\sim
		10^5, & f = 1/(30~\mathrm{years}),
		\\
		N_\mathrm{trans} &\sim 
		5 \times \left(
			\ds\frac{500}{2}
		\right)^4 
		\sim 2\cdot10^{10}, & f = 1/(1~\mathrm{month}).
	\end{align}
\end{subequations}
The single fiducial value of Eq.~\eqref{eq:transition-number} is thus a mid-band representative of a frequency- and characteristic-distance-dependent range, \(N_{\mathrm{trans}}\sim 10^{5} \div 10^{10}\), whose endpoints are set by the observing baseline and cadence rather than by the GW physics.

We arrive at a practical consequence: for a fixed, realistic number of pulsars, lowering the observational frequency shifts the pulsar-term sensitivity plateau $l_{1\to2}$ to lower multipoles, bringing it out of the effective null space of the response operator. Thus, the pulsar term can provide additional directional information at lower frequencies, even though the diffraction scale $\theta \sim \lambda/2L$ is correspondingly coarsened.

Current astrophysical models place the total Galactic millisecond-pulsar population at approximately ${3\times10^4}$ potentially observable objects \cite{lorimer2013}. For the upper part of the PTA band the required
array size exceeds this population by several orders of magnitude,
and generic full-sky strain reconstruction remains well inside the
Earth-term regime. Only at the lowest accessible frequencies, where
\(N_{\rm trans}\sim 10^{5}\) approaches the Galactic census, does
the requirement come within reach of an array containing a
substantial fraction of the Galactic millisecond-pulsar population.
For current arrays of order \(10^{2}\) pulsars the Earth-term
approximation therefore remains appropriate for generic full-sky
strain mapping across the entire PTA band.

A related but distinct question has been addressed in \cite{tsai2026} for coherent localization with well-characterized pulsars and in \cite{ellis2013,taylor2026} for single point-source searches. These studies show that a small number of pulsars, provided their distances are measured well enough to keep the pulsar terms coherent, is sufficient to form sharp sensitivity peaks and to localize a targeted source near the diffraction limit. This result does not conflict with the present estimate. Targeted localization is a low-dimensional inference problem: the source model fixes the functional form of the sky signal, and only a few source parameters are estimated. Generic full-sky mapping, by contrast, must determine \(O(l_{\max}^{2})\) independent strain coefficients, with no prior on their functional form. Our calculation is complementary in that it quantifies the array size required for the latter problem: under the idealized fixed-variance criterion, generic full-sky reconstruction of a band-limited strain field requires
\(N_{\rm pulsars}\sim 10^{5}\div10^{10}\) pulsars across the PTA band, several orders of magnitude above what targeted searches need.

\subsection{The Uneven Sky Sensitivity Distribution Consequence}

The estimates derived in the previous subsection assume that pulsars are distributed uniformly over the celestial sphere. Real PTAs deviate significantly from this idealization because pulsars are concentrated in specific regions of the sky and are observed with different sensitivities. Such nonuniformities introduce additional conditioning effects that modify the PTA response and ultimately affect the quality of the reconstructed GW sky maps.

\begin{figure}
	\centering
	\includegraphics[page=1, width=1\textwidth]{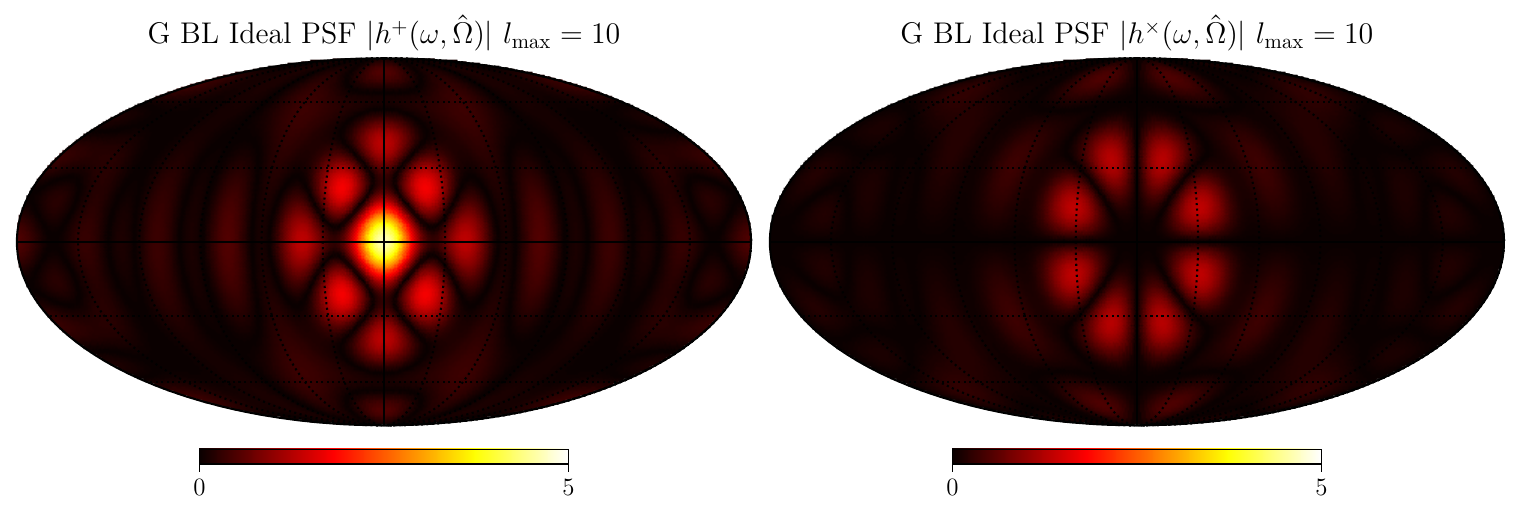}
	\caption{\scriptsize Point-spread function (PSF) of an ideal pulsar timing array for a unit-amplitude, plus-polarized gravitational-wave point source located at $(\theta,\phi)=(90^\circ,180^\circ)$.  The left and right panels show the reconstructed plus- and cross-polarization maps obtained from the gradient-mode coefficients \(a^G_{lm}\) of an ideal delta-function response, Eq.~\eqref{eq:a-h-transform}. The band-limited (BL) reconstruction  truncated at a fiducial $l_{\rm max}=10$. The detector's curl-mode blindness and the finite multipole cutoff produce the characteristic diffraction pattern and side lobes, which define the angular resolution of the PTA sky map. Spurious leakage between polarization channels is visible as signal amplitude appearing in the cross-polarization map despite the purely plus-polarized input source.}
	\label{fig:psf-plus-ideal}

	\vspace{1em}

	\includegraphics[page=3, width=1\textwidth]{figures/psf.pdf}
	\caption{\scriptsize Point-spread function for the MeerKAT PTA. The GW source parameters are identical to those of Figure~\ref{fig:psf-plus-ideal}. White stars mark the positions of PTA pulsars on the sky. The nonuniform pulsar distribution distorts the ideal PSF, broadens the response in preferred directions, and modifies the structure of the side lobes. The spurious leakage effect between polarization channels, visible in the ideal PSF, enhances.}
	\label{fig:psf-plus-real}
\end{figure}

According to Eqs.~\eqref{eq:a-h-transform} and \eqref{eq:RC-analyt}, the point-spread function (PSF) of an ideal PTA is determined solely by the highest recoverable multipole $l_{\rm max}$. This cutoff may originate either from the physical sensitivity limit discussed in Section~\ref{sec:cutoff-regime} or from the finite rank of the PTA response matrix discussed in Section~\ref{sec:conditionning}. The corresponding PSF for an ideal uniformly distributed PTA, constrained at a fiducial $l_{\rm max } = 10$, is shown in Fig.~\ref{fig:psf-plus-ideal}.

In an ideal PTA, all modes with the same multipole number $l$ possess identical sensitivity because the detector response depends only on $l$ through the function $f_l(y)$. A nonuniform pulsar distribution breaks this rotational symmetry and introduces an additional dependence on the azimuthal index $m$. As a result, different modes belonging to the same multipole become unequally constrained by the observations.

To quantify this effect, we define the mode sensitivity
\begin{equation}
	\mathcal{S}
	=
	\left(
	\ds\frac{4\pi}{N_\mathrm{pulsars}}\right)^{1/2}\lVert \mathbf{c}_{lm} \rVert,
\end{equation}
where $\mathbf{c}_{lm}$ is the response-matrix column corresponding to the $(lm)$ mode. In the ideal isotropic limit, $\mathcal{S}$ reduces to a quantity proportional to the analytical sensitivity function, $2\pi |f_l(y)|$, from Eq.~\eqref{eq:RG-analyt}. Deviations of $\mathcal{S}$ from this prediction therefore provide a direct measure of the anisotropy introduced by the finite pulsar distribution.

\begin{figure}
	\centering
	\includegraphics[page=2, width=1\textwidth]{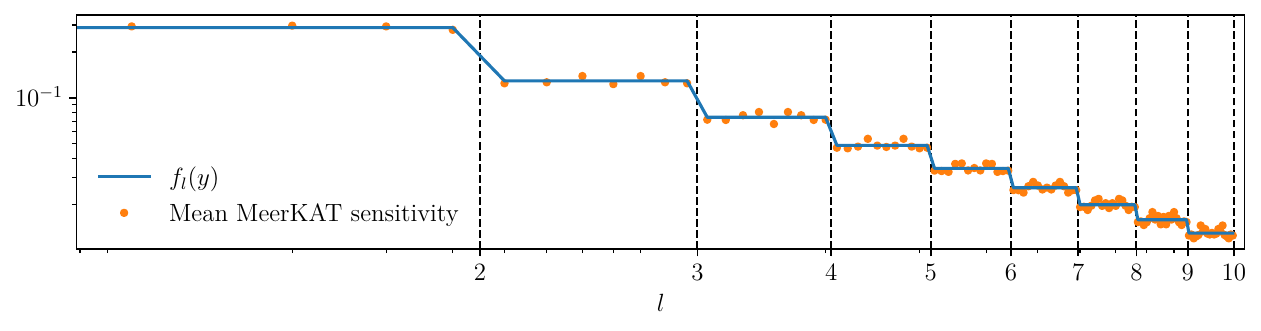}
	\caption{\scriptsize Deviation of individual $m$ modes from the ideal sensitivity function $f_l(y)$ for the MeerKAT PTA pulsar distribution. Vertical lines separate successive multipole orders $l$. In an isotropic PTA all modes with the same $l$ would have identical sensitivity. The observed scatter therefore quantifies the loss of rotational symmetry introduced by the finite and nonuniform pulsar distribution.}
	\label{fig:fl-real-deviation}
\end{figure}

Figure~\ref{fig:fl-real-deviation} compares the ideal sensitivity with that obtained for the MeerKAT PTA (MPTA) pulsar distribution\footnote{\url{https://doi.org/10.57891/j0vh-5g31}} \cite{miles2025}. The scatter between modes with identical $l$ increases with multipole number, demonstrating that nonuniform sky coverage becomes progressively more important as finer angular structure is reconstructed.

The unequal weighting of individual $(lm)$ modes directly affects the PSF. To illustrate this effect, the coefficients $a^G_{lm}$ in Eq.~\eqref{eq:a-h-transform} were artificially weighted by
\[
	\left(
		\ds\frac{\mathcal{S}}
		{2\pi |f_l(y)|}
	\right)^{10},
\]
where the exponent was chosen solely as a visual aid to enhance  spatial distortion of the PSF. The resulting PSF is shown in Fig.~\ref{fig:psf-plus-real}.

Several consequences are immediately apparent. The concentration of pulsars in the Southern Hemisphere introduces a preferred direction into the reconstruction process, causing the PSF to broaden anisotropically. The diffraction side lobes visible in Fig.~\ref{fig:psf-plus-ideal} become partially washed out along one direction while remaining relatively pronounced along the orthogonal direction. Consequently, the effective angular resolution of the PTA becomes direction dependent rather than isotropic.

Another important consequence is reconstruction-induced polarization leakage. As seen in both Fig.~\ref{fig:psf-plus-ideal} and Fig.~\ref{fig:psf-plus-real}, a purely plus-polarized input source produces a spurious measurable response in the reconstructed cross-polarization map. This is not a physical mixing of the gravitational-wave polarizations, but rather a mathematical artifact arising from the incomplete sampling of the tensor-harmonic basis by a finite PTA.  In particular, the exact vanishing of the curl response, Eq.~\eqref{eq:RC-analyt}, makes PTA insensitive to the curl component of the tensor field. Thus, the available observables do not provide a complete basis for separating the two linear polarization components. This incomplete mode information manifests itself as apparent polarization leakage in reconstructed sky maps.

The effects of the finite nonuniform PTA PSF should be substantially addressed while interpreting physical results of sky mapping. Spatial distortion of MPTA PSF was assessed in \cite{grunthal2026}.  Generalized treatment of the PSF is a subject of the subsequent work. 

In summary, the conditioning of the PTA sky-mapping inverse problem is limited by both the geometry of the pulsar distribution and the intrinsic sensitivity cutoff of the elementary detector response. The Fisher-matrix analysis demonstrates that the practical map resolution is bounded by the number of available pulsars through Eq.~\eqref{eq:FIM-max-rank}, while Section~\ref{sec:three-operation-regimes} established a separate physical limitation associated with the maximum accessible multipole. Furthermore, the transition from the Earth-term-dominated regime to the pulsar-term-sensitive regime  corresponds to a heuristic scaling estimate of  $N_{\rm trans}\sim 10^7$ pulsars, Eq.~\eqref{eq:transition-number}, thereby providing grounds for the Earth-term approximation in global full-sky GW sky mapping.

Nonuniform sky coverage introduces an additional source of information loss. It breaks the rotational symmetry of the ideal detector, produces mode-dependent sensitivity variations, distorts the PSF, creates anisotropic side lobes, and 
and exacerbates the apparent inseparability of the two polarization channels.
These effects further degrade the conditioning of the inverse problem and must therefore be taken into account when interpreting PTA sky maps \cite{grunthal2026}.

\emph{Taken together, the results of this section provide a self-consistent mathematical foundation for the principal approximations and sensitivity limits employed throughout the modern PTA literature
} \cite{anholm2009,lentati2013,mingarelli2014,agazie2023,eptacollaboration2023,miles2025,schult2025}. \textit{In particular, they provide a quantitative framework for assessing the applicability of the Earth-term approximation and quantify the geometric factors that limit angular resolution and polarization-mode separability.}

To illustrate the established theoretical conditioning limits, geometric constraints, and regime-switching, we now turn to numerical simulations. The following section employs mock pulsar timing arrays to probe both Earth-term and pulsar-term regimes, directly testing the predicted singular-value spectra, Fisher Information Matrix conditioning, and the practical fidelity of gravitational-wave sky-map recovery.

\section{Numerical Tests of PTA Sky Reconstruction}
\label{sec:numerical-validation}

Having established the analytic properties of the PTA response operator and the conditioning of the corresponding inverse problem in Sections~\ref{sec:three-operation-regimes}--\ref{sec:conditionning}, we now test these results with numerical simulations. The primary objectives are: (i) to provide examples of the predicted transition between the Earth-term dominated and pulsar-term sensitive observational regimes, (ii) to examine how the singular-value spectrum of the response matrix reflects the conditioning of the Fisher Information Matrix (FIM), and (iii) to assess the practical consequences of these properties for reconstructing the angular distribution of gravitational-wave (GW) field coefficients on the sky.

To this end, we construct two mock pulsar timing arrays (PTAs), each containing $500$ pulsars. The first array consists of Galactic-like pulsars uniformly distributed within a spherical shell spanning
\[
	L = 1\text{--}5~\mathrm{kpc}.
\]
At the injected frequency of $10\,\mathrm{nHz}$, the dimensionless parameter
\(
y=\omega L
\)
ranges from $6.5\times10^{3}$ to $3.2\times10^{4}$, placing the array deep within the asymptotic long-arm regime discussed in Section~\ref{sec:three-operation-regimes}. In this regime, pulsar-term contributions are strongly suppressed, and the PTA response is effectively dominated by the Earth term.

The second array contains $500$ pulsars located at the same sky coordinates, but distributed within a much smaller spherical shell,
\[
	L = 0.5\text{--}2~\mathrm{pc}.
\]
We refer to the first and second configurations as the \emph{long-arm} and \emph{short-arm} PTAs, respectively. We emphasize that the short-arm array is not intended to represent an astrophysically realistic PTA; rather, it serves as a controlled numerical experiment designed to isolate the geometric and information-theoretic consequences of decreasing the parameter $y=\omega L$.

To justify this placement, we note that for the nearest pulsars in the short-arm array ($L=0.5$ pc), the dimensionless parameter is $y_{\min} \simeq 3.2$. According to the transition boundaries derived in Section~\ref{sec:transition-regime}, the crossover from Earth-term to pulsar-term dominance occurs at multipoles $l_{1\to2} \simeq \sqrt{y} \approx 1.8$ and $l_{2\to3} \simeq 4\sqrt{y} \approx 7.1$. Since the lowest recoverable tensor multipole is $l=2$, the array operates entirely within the pulsar-term-sensitive regime for all observable modes. Consequently, pulsar-term contributions are no longer strongly suppressed and provide additional directional information that remains observable rather than being projected into the effective null space of the response matrix. To ensure mathematical exactness, the response matrix $\mathbf{R}$ is constructed using the exact closed-form response function $f_l(y)$, Eq.~\eqref{eq:fl-def}, faithfully capturing the pulsar-term sensitivity without relying on asymptotic limits.

The relatively large number of pulsars is chosen to maximize the angular resolution of the reconstructed GW maps and thereby expose the intrinsic limitations imposed by the detector response and the geometry of the inverse problem, rather than those arising from sparse sky sampling.

For each simulation we inject two monochromatic point-like GW sources described by the following parameters:
\begin{itemize}
	\item ~~$\omega = 2\pi\cdot10~\mathrm{nHz}$, $(h_{+},h_{\times})=(5\times10^{-14},0)$, $(\theta,\phi)=(60^\circ,90^\circ)$,
	\item ~~$\omega = 2\pi\cdot10~\mathrm{nHz}$, $(h_{+},h_{\times})=(0,5\times10^{-14})$, $(\theta,\phi)=(120^\circ,270^\circ)$.
\end{itemize}

For the injected frequency the dimensionless parameter $y=\omega L$ spans
$\sim(6.5\div32.3)\times10^3$ for the long-arm array and
$\sim(3.2\div12.9)$ for the short-arm array. The latter therefore lies in the pulsar-term-sensitive region (transition and pulsar-term dominated) identified in Section~\ref{sec:three-operation-regimes}, where the asymptotic suppression of the pulsar term is no longer complete.

The simulated datasets cover $30~\mathrm{yr}$ with an observational cadence of $100$ days. The source amplitudes are chosen to produce timing residuals at the $\mu$s level. Instrumental noise is intentionally reduced to a negligible level; the only stochastic contribution arises from the numerical precision of \texttt{Tempo2} \cite{hobbs2006}, corresponding to white noise with an rms amplitude of approximately $10~\mathrm{ns}$. In contrast to realistic PTA analyses, no red-noise processes, dispersion-measure variations, clock errors, or ephemeris uncertainties are included. 
The pulsar distances are assumed to be known with sufficient precision to
maintain phase coherence of the pulsar term. Since the pulsar-term phase
is proportional to $\omega L$, we require
\(
\omega\,\Delta L\ll1.
\)
The simulations therefore neglect pulsar-term decoherence arising from
distance uncertainties and represent an idealized coherent-response
limit.
The purpose of this idealized setup is to isolate the geometric and information-theoretic aspects of the reconstruction problem and to expose the influence of array configuration on the conditioning of the response operator, 
rather than noise suppression methods and the pulsar term decoherence effects.

We first compare the singular-value decompositions of the response matrices associated with the two PTAs and relate their spectra to the conditioning of the FIM. We then demonstrate GW sky-map reconstruction for both arrays, illustrating how the availability or suppression of pulsar-term information affects the number of recoverable modes and the attainable angular resolution.

The following libraries of \texttt{Python3} are used in this section: 
\texttt{spherical} \cite{boyle2025}, 
\texttt{numpy} \cite{harris2020}, 
\texttt{matplotlib} \cite{hunter2007}, \texttt{libstempo} \cite{vallisneri2020}, 
\texttt{healpy} \cite{zonca2019},
\texttt{scipy} \cite{virtanen2020}, 
\texttt{astropy} \cite{astropycollaboration2022}.

\subsection{SVD Spectrum \& FIM Conditioning}

\begin{figure}
	\centering
	\includegraphics[width=1\textwidth]{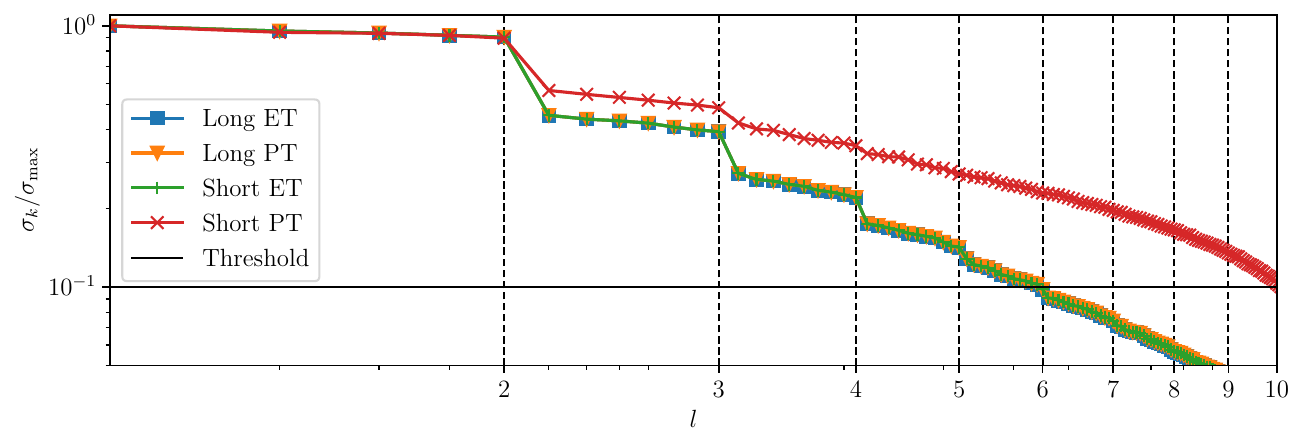}
	\caption{\scriptsize Singular-value spectra of the PTA response matrix for four mock
		arrays. “Long” and “Short” refer to realistic and reduced pulsar-distance
		distributions, respectively, while “ET” and “PT” denote the Earth-term
		approximation and the full Earth--pulsar response. Vertical lines separate
		different multipole orders $l$. The singular values quantify the number
		of independently recoverable sky modes and therefore the conditioning of
		the inverse problem. Retaining pulsar-term information enhances
		sensitivity to higher multipoles and increases the number of
	well-constrained modes.}
	\label{fig:spectrum}
\end{figure}

Figure~\ref{fig:spectrum} shows the singular-value spectra of the PTA response matrix $\mathbf{R}$ for four mock arrays. The plotted quantities are the diagonal elements of the matrix $\mathbf{\Sigma}$ appearing in the singular-value decomposition of the likelihood estimator, Eq.~\eqref{eq:likelihood-svd}. Since the FIM is given by $\mathbf{F}=\tilde{\mathbf{R}}^\dagger\tilde{\mathbf{R}}$, its eigenvalues are equal to the squares of the singular values. The singular-value spectrum therefore directly characterizes the conditioning of the inverse problem and determines how many independent angular modes of the GW sky can be reconstructed with meaningful accuracy.

The four spectra correspond to the long-arm and short-arm PTAs operating either in the Earth-term (ET) approximation or with the full Earth--pulsar response included (PT). In the ET case, the response matrix of Eq.~\eqref{eq:Ra-product} is constructed using the asymptotic large-$y$ response function, Eq.~\eqref{eq:RG-large-y}. In the PT case, the same response matrix is evaluated using the exact Earth--pulsar response, Eq.~\eqref{eq:RG-analyt}.

The long-arm PTA exhibits precisely the behavior predicted by the asymptotic analysis of Section~\ref{sec:three-operation-regimes}. The ET and PT spectra are practically indistinguishable, demonstrating that for
$y\simeq(6.5\div32.3)\times10^{3}$ the asymptotic Earth-term response of Eq.~\eqref{eq:RG-large-y} provides an essentially complete description of the detector. This agrees with the analytic estimates for the multipole number at which the pulsar term sensitivity contribution becomes apparent Eqs.~\eqref{eq:fl-1-2-boundary-number}--\eqref{eq:fl-2-3-boundary-number}, which place the transition at
$l_{1\to2} \sim 80 \div 180$ for the range of pulsar distances considered here \eqref{eq:fl-1-2-boundary-number}. Since all reconstructed multipoles lie well below this transition, the pulsar term contributes negligible additional directional information.

The singular values decrease rapidly with increasing multipole order. For a PTA containing 500 pulsars, only modes up to approximately $l\sim6$ remain comparatively well constrained, corresponding to singular values exceeding the illustrative threshold
$\sigma_k/\sigma_{\max}=0.1$, corresponding to the conditioning number of the regularized matrix. Thus, even an optimistic array with several hundred pulsars provides meaningful constraints on only a limited number of independent angular modes of the GW sky. This behavior is expected because the number of pulsars remains  below the characteristic array size required for the transition from Earth-term-limited to pulsar-term-sensitive sky reconstruction, estimated in Eq.~\eqref{eq:transition-number}. 

The specially constructed short-arm PTA displays a qualitatively different behavior. When the Earth-term approximation is used, its spectrum remains nearly identical to that of the long-arm array. In contrast, inclusion of the full Earth--pulsar response through Eq.~\eqref{eq:RG-analyt} produces a pronounced enhancement of the singular values at higher multipoles. This sensitivity bump is a direct manifestation of the additional directional information carried by the pulsar terms when $y$ is no longer asymptotically large.

Another notable feature of the short-arm PT spectrum is the disappearance of the characteristic ladder-like structure observed in the Earth-term spectra. Beyond approximately $l\gtrsim3$, the singular values form a comparatively smooth distribution. This behavior reflects the nearly multipole-independent sensitivity predicted for the pulsar-term-sensitive regime discussed in Section~\ref{sec:three-operation-regimes}. The onset of this regime occurs near $l\sim \sqrt{y}$ for the nearest pulsars in the array, whose distances correspond to $y_{\min}\simeq3.2$. Consistent with the asymptotic behavior derived from the exact response function, Eq.~\eqref{eq:RG-analyt}, the detector retains appreciable sensitivity to progressively smaller angular scales instead of exhibiting the strong suppression of high multipoles characteristic of the Earth-term response, Eq.~\eqref{eq:RG-large-y}.

The differences among the four spectra translate directly into differences in the effective rank of the FIM and, consequently, in the angular resolution achievable in GW sky reconstruction. In particular, the smoother and more slowly decaying spectrum of the short-arm PTA indicates that a substantially larger number of sky modes can be constrained when pulsar-term information is preserved. The following subsection demonstrates how these conditioning properties manifest themselves in the reconstructed GW maps.

\subsection{Sky-Map Recovery \& Polarization Mode Separability}

\begin{figure}
	\centering
	\includegraphics[width=1\textwidth, page=3]{./figures/pta_image-long-E.pdf}
	\includegraphics[width=1\textwidth]{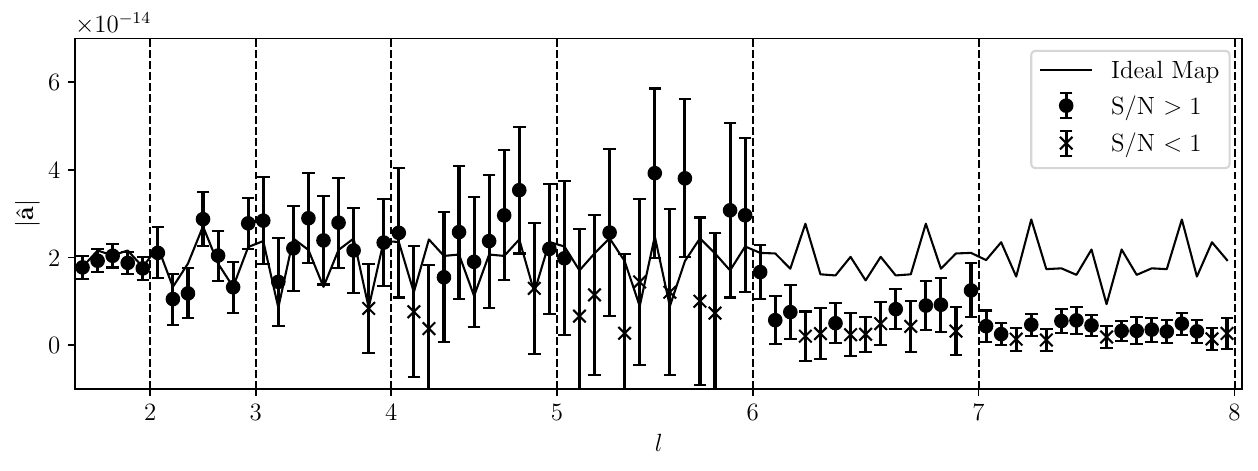}
	\caption{\scriptsize 	
		Sky reconstruction for a realistic PTA using the Earth-term approximation. Upper panels show the reconstructed plus- and cross-polarization maps obtained from the recovered gradient-mode coefficients \(a^G_{l m}\) (G), together with the corresponding band-limited (BL) ideal maps constructed directly from Eq.~\eqref{eq:a-h-transform} with the same truncation multipole. Lower panel compares the magnitudes of the injected and recovered spherical-harmonic coefficients $|a^{G}_{lm}|$. Vertical lines separate multipole orders $l$. Thereconstruction accurately recovers the largest angular scales but loses sensitivity toward progressively higher multipoles.
	}
	\label{fig:map-long-E}

	\vspace{1em}

	\includegraphics[width=1\textwidth, page=3]{figures/pta_image-long-P.pdf}
	\includegraphics[width=1\textwidth]{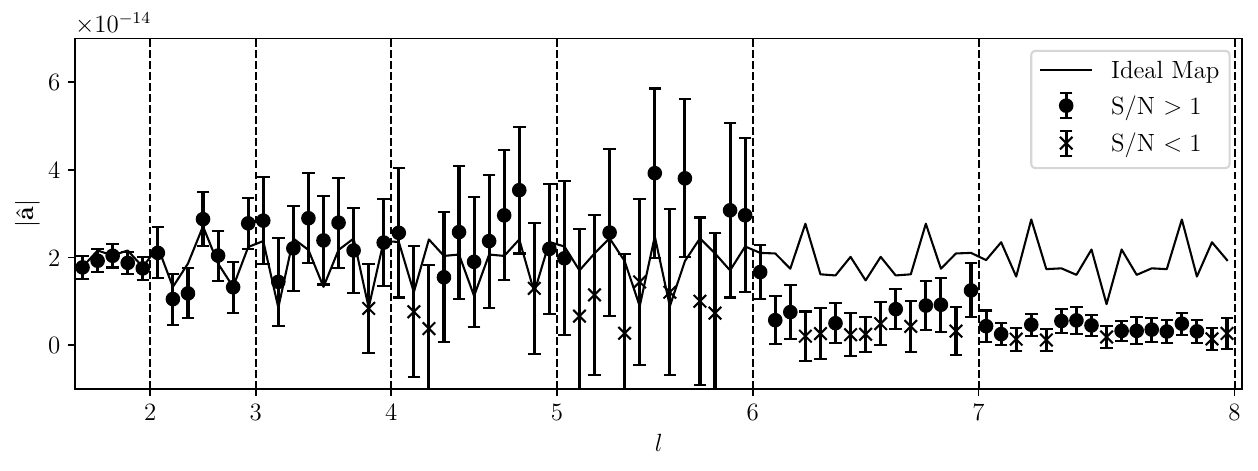}
	\caption{\scriptsize Sky reconstruction for a realistic long-arm PTA using the full
		Earth--pulsar response. The layout repeats the Fig.~\ref{fig:map-long-E}. The close similarity to the Earth-term reconstruction demonstrates that pulsar-term information contributes 	negligibly in the asymptotic long-arm regime.
	}
	\label{fig:map-long-P}
\end{figure}

\begin{figure}
	\centering
	\includegraphics[width=1\textwidth, page=3]{figures/pta_image-short-E.pdf}
	\includegraphics[width=1\textwidth]{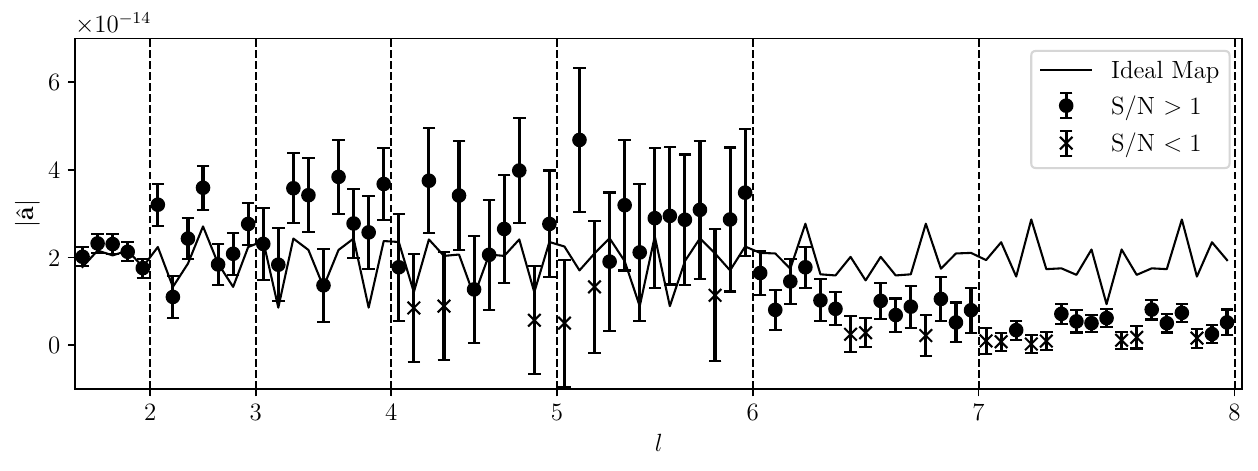}
	\caption{\scriptsize Sky reconstruction for a short-arm PTA in the Earth-term
		approximation. The layout repeats the Fig.~\ref{fig:map-long-E}. Despite the shorter pulsar distances, the Earth-term approximation yields a reconstruction comparable to that of the realistic long-arm PTA.
	}
	\label{fig:map-short-E}

	\includegraphics[width=1\textwidth, page=8]{figures/pta_image-short-P.pdf}
	\includegraphics[width=1\textwidth]{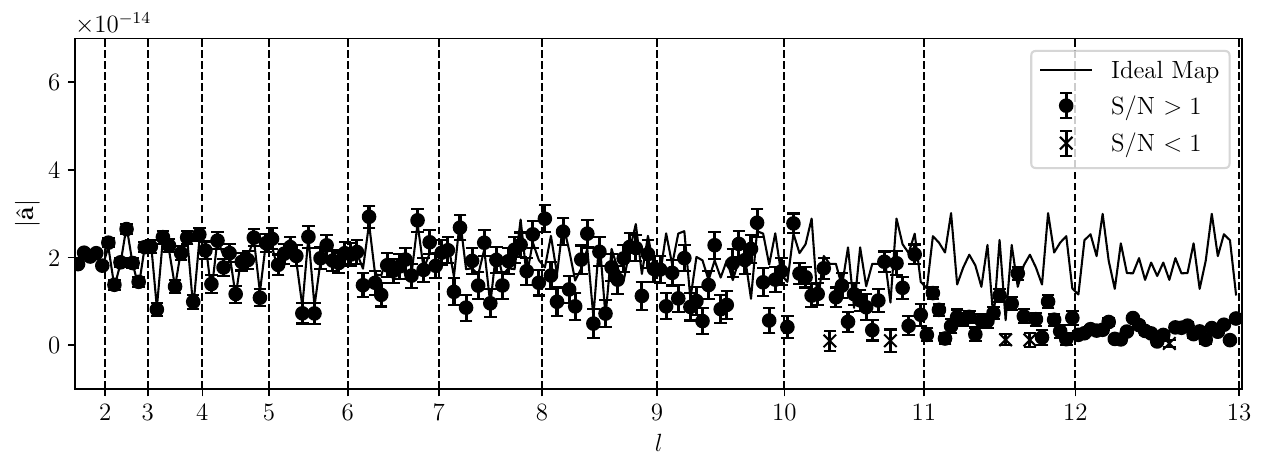}
	\caption{\scriptsize 
		Sky reconstruction in the pulsar-term-sensitive regime for the short-arm PTA in the PT regime. The layout repeats the Fig.~\ref{fig:map-long-E}, except the colorbar limits.	The enhanced high-multipole sensitivity illustrates the theoretical gain in angular resolution obtainable when pulsar-term information is fully exploited.
	}
	\label{fig:map-short-P}
\end{figure}

Figures~\ref{fig:map-long-E}--\ref{fig:map-short-P} show GW sky maps reconstructed from simulated observations with the four PTA configurations considered in this section. The upper panels display the recovered $+$ and $\times$ polarization maps together with the corresponding ideal maps obtained by substituting the injected point-like sources directly into the harmonic expansion of Eq.~\eqref{eq:a-h-transform}. The lower panels compare the magnitudes of the reconstructed spherical-harmonic coefficients $\hat{\mathbf a}$ obtained from the SVD solution of Eq.~\eqref{eq:likelihood-svd} with the coefficients of the ideal maps, restricted to the same number of the recoverable angular modes. The error bars show the formal one-standard-deviation uncertainties of the reconstructed coefficients,
$\sigma_i=\sqrt{\hat C_{ii}}$, derived from the covariance matrix
$\hat{\mathbf C}$ of Eq.~\eqref{eq:likelihood-svd-cov}.

Note, that only gradient-mode coefficients $a_{lm}^G(\omega)$ were used in the reconstruction process as a consequence of detector insensitivity to the curl-modes (see Eqs.~\eqref{eq:RG-analyt}--\eqref{eq:RC-analyt}).

Figures~\ref{fig:map-long-E}, \ref{fig:map-long-P}, and \ref{fig:map-short-E} exhibit remarkably similar angular resolution despite corresponding to different PTA realizations and response models. In all three cases, the injected sources are recovered at the correct sky locations. The reconstructed maps reproduce the broad structure of the ideal point-spread functions, while the deviations from the ideal images arise from the finite number of angular modes that can be constrained by the PTA response. The close similarity between the long-arm and short-arm reconstructions performed in the Earth-term approximation reflects the nearly identical singular-value spectra shown in Figure~\ref{fig:spectrum}.

An important feature visible in all maps is the substantial spurious leakage between the $+$ and $\times$ polarization channels. Even the ideal gradient-mode-only maps, generated directly from Eq.~\eqref{eq:a-h-transform}, exhibit this behavior. The polarization mode separability is therefore not caused by measurement noise, imperfections of the inversion procedure, or deficiencies of the reconstruction algorithm. Rather, it is an intrinsic property of PTA sky imaging resulting from the projection of the tensor GW field onto pulsar timing residuals and from the incomplete separability of the two polarization states on the celestial sphere. The reconstructed maps closely reproduce the leakage pattern present in the ideal maps, demonstrating that it is encoded in the response operator itself.

The coefficient-amplitude comparisons shown in the lower panels provide a quantitative measure of reconstruction fidelity. In Figures~\ref{fig:map-long-E}, \ref{fig:map-long-P}, and \ref{fig:map-short-E}, the reconstructed coefficients closely follow the injected amplitudes at low multipole numbers. The uncertainties increase steadily with increasing $l$ and become comparable to the coefficient amplitudes around $l\simeq6$. Beyond this scale the recovered amplitudes rapidly decrease as a consequence of regularization. These high-$l$ modes correspond to the small singular values identified in Figure~\ref{fig:spectrum}. Their direct inversion would amplify noise and numerical errors, producing unstable solutions with unrealistically large harmonic coefficients. The regularization procedure suppresses such poorly constrained modes and thereby stabilizes the reconstruction.

A qualitatively different behavior appears in Figure~\ref{fig:map-short-P}, which corresponds to the short-arm PTA reconstructed using the full Earth--pulsar response. In this case the recovered harmonic coefficients remain in close agreement with the injected amplitudes up to approximately $l\lesssim7$. Noticeable deviations appear only over the interval $7\lesssim l\lesssim11$, while substantial suppression due to regularization occurs only for higher multipoles. This behavior is fully consistent with the enhanced singular-value spectrum of the short-arm pulsar-term-sensitive PTA shown in Figure~\ref{fig:spectrum}. Because a larger number of singular values remain significant, more angular modes of the GW field can be reconstructed reliably.

The improvement is also visible directly in the sky maps. Compared with the Earth-term reconstructions shown in Figures~\ref{fig:map-long-E}--\ref{fig:map-short-E}, the pulsar-term-sensitive reconstruction of Figure~\ref{fig:map-short-P} reproduces significantly more small-scale structure and exhibits noticeably narrower point-spread functions. The increased high-multipole sensitivity is a direct manifestation of the additional directional information carried by the pulsar terms.

These examples represent an intentionally optimistic observational scenario characterized by negligible measurement noise and a large number of pulsars. Although such PTAs are beyond current observational capabilities, they clearly demonstrate the trends predicted in Sections~\ref{sec:three-operation-regimes}--\ref{sec:conditionning}. The long-arm PTA reproduces the behavior expected for realistic arrays operating deep in the Earth-term-dominated regime. Reducing the number of pulsars or increasing the measurement noise would shift the onset of large coefficient uncertainties toward lower multipole numbers and further degrade the achievable map resolution. Current PTAs, consisting of approximately one hundred pulsars with timing precision at the $\sim1$--$10~\mu\mathrm{s}$ level, are therefore expected to constrain primarily the lowest-order multipole modes of the GW sky.

The short-arm PTA, by contrast, illustrates the transition toward the pulsar-distance-resolved regime. Once the conditions for coherent utilization of pulsar-term information are satisfied, the effective rank of the response operator increases substantially, more harmonic modes become measurable, and the uncertainties of the reconstructed coefficients decrease. An interesting consequence is that lower values of the parameter $y=\omega L$ have two competing effects. On the one hand, decreasing $y$ broadens the  diffraction limit angular response of an individual pulsar. On the other hand, it shifts the transition multipole $l_{1\to 2}$ predicted by Eqs.~\eqref{eq:fl-1-2-boundary-number}--\eqref{eq:fl-2-3-boundary-number} toward lower values, allowing pulsar-term information to contribute at larger angular scales. For PTAs with a finite number of pulsars, the latter effect can dominate, increasing the number of recoverable modes and thereby improving the practical angular resolution of GW sky reconstruction.

Having illustrated the theoretical conditioning limits, sensitivity regimes, and geometric constraints of the deterministic inverse problem through numerical simulations, we now extend the formalism to the primary scientific target of PTA observations: the stochastic gravitational-wave background (SGWB). The tensor spherical-harmonic response operator developed in the preceding sections provides the same geometric advantages in the stochastic case as in the deterministic one. In particular, the conditioning analysis of Section~\ref{sec:conditionning} identifies the subset of sky modes that can be constrained by the PTA and removes those modes that lie effectively in the null space of the detector response. As a result, the formalism retains all statistically recoverable information while preserving a finite-dimensional and computationally efficient representation of the likelihood. This property makes it particularly well suited for anisotropy searches and SGWB inference.

\section{Application to Anisotropic Gravitational-Wave Backgrounds}
\label{sec:anisotropy}

The previous sections considered reconstruction of a deterministic gravitational-wave sky described by the spherical-harmonic coefficients $a^P_{lm}$. The same formalism extends naturally to stochastic gravitational-wave backgrounds (SGWBs), in which the GW field is treated as a random realization drawn from an underlying statistical ensemble. In this case, the observable quantities are no longer the individual spherical-harmonic coefficients but their statistical correlations. The PTA response operator derived in Sections~\ref{sec:three-operation-regimes}--\ref{sec:conditionning} remains unchanged, while the statistical properties of the GW field become the primary target of inference.

Current PTA analyses primarily target the detection of an isotropic, stationary, and unpolarized SGWB, manifesting itself as correlated red timing noise described by the Hellings--Downs correlation curve \cite{hellings1983}. Within the plane-wave expansion formalism, such a field is characterized by the first two statistical moments of the Fourier amplitudes $h_A(\omega,\hat{\Omega})$ \cite{anholm2009, mingarelli2013, lentati2013}:
\begin{equation}
	\langle h_A(\omega,\hat{\Omega}) \rangle = 0,
\end{equation}
\begin{equation}
	\langle h_A(\omega,\hat{\Omega}) h_{A'}^*(\omega',\hat{\Omega}') \rangle
	=
	\ds\frac{1}{2}
	P(\omega,\hat{\Omega})
	\delta^2(\hat{\Omega},\hat{\Omega}')
	\delta(\omega-\omega')
	\delta_{AA'},
\end{equation}
where $P(\omega,\hat{\Omega})$ is the angular power distribution of the background.

Deviations from isotropy are described by expanding the power distribution in the basis of scalar spherical harmonics \cite{mingarelli2013}:
\begin{equation}
	P(\omega,\hat{\Omega})
	=
	\sum_{LM}
	A_{LM}(\omega)
	Y_{LM}(\hat{\Omega}),
\end{equation}
where
\[\sum_{LM}\equiv\sum_{L=0}^{\infty}\sum_{M=-L}^{L}.\]
It is important to distinguish the multipole indices $(L,M)$ of the SGWB power distribution from the indices $(l,m)$ appearing in the tensor-harmonic decomposition of the GW metric perturbation, Eq.~(\ref{eq:h-spherical-decomposition}). Whereas the transverse-traceless tensor field $h_{ab}$ contains only quadrupolar and higher-order modes ($l\ge2$), the power distribution $P(\omega,\hat{\Omega})$ is a scalar field on the sphere and therefore admits a complete spherical-harmonic expansion beginning with the monopole term ($L=0$).

The coefficients $A_{LM}$ characterize the angular structure of the SGWB in direct analogy with the spherical-harmonic multipoles used to describe cosmic microwave background anisotropies and polarization \cite{kamionkowski1997,weinberg2008cosmology}. The isotropic background corresponds to the monopole component alone,
\[
	A_{LM}\propto\delta_{L0}\delta_{M0},
\]
while the dipole ($L=1$) and higher-order multipoles ($L\ge2$) describe progressively finer angular anisotropies in the GW power distribution.

Within the tensor spherical-harmonic decomposition formalism introduced in \ref{app:basis-defs}, the statistical properties of the SGWB are encoded in the covariance matrix of the gradient and curl coefficients $a^P_{lm}$. As shown by \cite{gair2014}, anisotropy generally introduces correlations between different GW sky modes. The indices $(l,m)$ and $(l',m')$ label the tensor spherical-harmonic modes of the GW field itself, whereas $(L,M)$ characterize the angular structure of the SGWB power distribution.  Anisotropy introduces couplings between distinct GW multipoles, with the coupling strength determined by the coefficients $A_{LM}$.

The covariance matrix of the GW field coefficients can therefore be written as
\begin{align}
	\langle a^G_{lm}(\omega)a^{G*}_{l'm'}(\omega') \rangle
	&=
	\langle a^C_{lm}(\omega)a^{C*}_{l'm'}(\omega') \rangle
	=
	\ds\frac{1}{2}
	\sum_{LM}
	A_{LM}(\omega)
	\alpha^{LM}_{lml'm'}
	\delta(\omega-\omega'),
	\\
	\langle a^G_{lm}(\omega)a^{C*}_{l'm'}(\omega') \rangle
	&=
	-\langle a^C_{lm}(\omega)a^{G*}_{l'm'}(\omega') \rangle
	=
	\ds\frac{1}{2}
	\sum_{LM}
	A_{LM}(\omega)
	\beta^{LM}_{lml'm'}
	\delta(\omega-\omega').
\end{align}
The coefficients $\alpha^{LM}_{lml'm'}$ and
$\beta^{LM}_{lml'm'}$ characterize the coupling of the SGWB anisotropy
multipole $(L,M)$ to the GW modes $(l,m)$ and $(l',m')$. They are
obtained from the angular integrals that arise in the spherical-harmonic
decomposition of the anisotropic background and can be written in terms
of Wigner-$3j$ symbols \cite{gair2014,allen2024,varshalovich1988}:

\begin{eqnarray}
	\alpha^{LM}_{lml'm'}
&=&
(-1)^m
\sqrt{\ds\frac{\scriptstyle(2L+1)(2l+1)(2l'+1)}
{\scriptstyle4\pi}}\,
\threej{L}{l}{l'}{M}{-m}{m'}
\left[
	\threej{L}{l}{l'}{0}{2}{-2}
	+
	\threej{L}{l}{l'}{0}{-2}{2}
\right],
\\[1.5ex]
\beta^{LM}_{lml'm'}
&=&
i(-1)^m
\sqrt{\ds\frac{\scriptstyle(2L+1)(2l+1)(2l'+1)}
{\scriptstyle4\pi}}\,
\threej{L}{l}{l'}{M}{-m}{m'}
\left[
	\threej{L}{l}{l'}{0}{2}{-2}
	-
	\threej{L}{l}{l'}{0}{-2}{2}
\right].
\end{eqnarray}

The Wigner-$3j$ symbols encode the angular-momentum selection rules.
Consequently, $\alpha^{LM}_{lml'm'}$ and
$\beta^{LM}_{lml'm'}$ vanish unless the multipoles
$(L,M)$, $(l,m)$, and $(l',m')$ satisfy the usual triangle and
azimuthal-number constraints. In addition, the symmetry properties of
the Wigner-$3j$ symbols imply that $\alpha^{LM}_{lml'm'}$ is nonzero
only when $L+l+l'$ is even, whereas $\beta^{LM}_{lml'm'}$ is nonzero
only when $L+l+l'$ is odd. Thus $\alpha^{LM}_{lml'm'}$ and
$\beta^{LM}_{lml'm'}$ isolate the even- and odd-parity couplings,
respectively. These parity selection rules play a key role in
determining which SGWB anisotropy multipoles contribute to the
gradient--gradient, curl--curl, and gradient--curl covariance blocks.

Assuming the field coefficients $a^P_{lm}$ are Gaussian random variables with covariance matrix
\begin{equation}
	\mathbf K
	= \mathbf K
	\left(A_{LM}\right)
	=
	\langle \mathbf a \mathbf a^\dagger \rangle,
\end{equation}
they can be analytically marginalized, yielding the likelihood function for the anisotropy coefficients (see, e.g., \cite{lentati2013}):
\begin{equation}
	p(A_{LM}|\mathbf s)
	=
	\ds\frac{1}
	{\sqrt{\det(2\pi\mathbf C_K)}}\,
	\exp\!\left[
		-\ds\frac{1}{2}
		\mathbf s^\dagger
		\mathbf C_K^{-1}
		\mathbf s
	\right]
	p(A_{LM}),
	\label{eq:likelihood-anisotropy}
\end{equation}
where
\begin{equation}
	\mathbf C_K
	=
	\mathbf C
	+
	\mathbf R
	\mathbf K
	\mathbf R^\dagger ,
\end{equation}
$\mathbf C$ is the covariance matrix of non-GW timing noise and
$p(A_{LM})$ denotes the prior distribution of the anisotropy coefficients.

The resulting likelihood has the same general structure as that used in conventional Hellings--Downs searches \cite{lentati2013, ellis2020, johnson2024}, but is expressed directly in terms of the PTA response operator and the harmonic covariance of the GW field. This formulation naturally separates the gravitational signal model, encoded in $\mathbf R$ and $\mathbf K$, from other stochastic contributions to the timing residuals.

A particularly important consequence of the analysis presented in Sections~\ref{sec:three-operation-regimes}--\ref{sec:conditionning} is that the likelihood need only include angular modes to which the PTA is sensitive. Truncating the response matrix $\mathbf R$ at a physically motivated value of $l_{\mathrm{max}}$ does not discard measurable information because higher-order modes belong to, or lie very close to, the null space of the response operator and therefore make a negligible contribution to the likelihood. The effective dimensionality of the anisotropy search is thus determined by the rank and conditioning of the PTA response matrix rather than by the formally infinite harmonic expansion of the SGWB.

Model selection between an isotropic SGWB
($A_{LM}\propto\delta_{L0}\delta_{M0}$)
and an anisotropic background with non-vanishing higher-order multipoles can then be carried out using standard Bayesian or frequentist techniques, such as Bayes factors \cite{kass1995, ellis2014} or likelihood-ratio statistics \cite{neyman1933}. Because the anisotropy information is encoded entirely in the covariance matrix $\mathbf K(A_{LM})$, the response-operator framework developed in this work can be incorporated straightforwardly into the existing SGWB analysis pipelines.

\section{DISCUSSION \& OUTLOOK}
\label{sec:conclusion}

In this work we have developed a unified theoretical framework for gravitational-wave sky imaging with pulsar timing arrays. By reformulating the PTA response in the basis of tensor spherical harmonics, we established a linear connection between the spatial structure of the gravitational-wave field and the timing residuals measured by an array of pulsars. Unlike previous treatments based primarily on plane-wave expansions, the present formalism provides an explicit harmonic-space description of the PTA response and enables a systematic analysis of angular resolution, mode coupling, and inverse-problem conditioning.

A central result of this work is the derivation of compact closed-form response functions for an elementary Earth--pulsar detector that retain both the Earth and pulsar terms, which expands the work \cite{gair2015}. The analytic structure of these response functions reveals four distinct operational regimes governed by the dimensionless parameter $y=\omega L$: an Earth-term-dominated regime, a transition regime in which pulsar-term effects become significant, a pulsar-term-dominated regime, and an asymptotic exponential cutoff at sufficiently high multipole numbers. These regimes determine the angular scales to which a PTA is intrinsically sensitive and lead to a quantitative description of the intrinsic diffraction scale, $\theta\sim\lambda/2L$. The analysis also clarifies the physical origin of the PTA point-spread function and demonstrates how the response evolves as the observing frequency or pulsar distance changes.

The tensor spherical-harmonic formalism further provides a natural description of polarization information. In agreement with previous studies, the response of a PTA to curl modes is fully suppressed \cite{gair2014, gair2015}. 
Consequently, only the gradient-mode coefficients $a^G_{lm}$ are observable, while the curl-mode coefficients $a^C_{lm}$ remain unconstrained by PTA timing data.
Our analysis shows that incomplete observability of the tensor-harmonic modes, together with the finite-dimensional reconstruction subspace, can manifest as apparent mixing between the reconstructed plus and cross polarization maps.
The incomplete separability of the polarization modes is therefore not a consequence of noise, imperfect inversion, or finite sampling, but rather follows from the structure of the PTA response operator and its associated null space. Departures from idealized isotropic pulsar distributions further distort the point-spread function and increase the asymmetry of the reconstructed images.

A second major result concerns the conditioning of the PTA inverse problem. Using the Fisher Information Matrix and singular-value decomposition of the response operator, we demonstrated that the practical angular resolution of a PTA is determined not only by the response of individual pulsars but also by the finite dimensionality of the array. In particular, the number of independently recoverable sky modes cannot exceed the number of available pulsars. The singular-value spectrum provides a direct measure of this limitation and naturally identifies the modes that can be reconstructed with statistically meaningful precision.

The analysis also reveals a fundamental distinction between theoretical and practical access to pulsar-term information. Although the pulsar term contains additional directional information and can substantially enhance high-multipole sensitivity, exploiting this information for global sky reconstruction requires an extraordinarily dense PTA. 
Our analysis provides a scaling estimate indicating that within the idealized framework adopted in this paper the transition from the Earth-term-dominated to the pulsar-term-sensitive regime imposes an $l^4$ scaling requirement on the PTA size. This result helps explain why current PTA arrays remain largely insensitive to the additional angular information encoded in pulsar terms and why Earth-term-dominated reconstructions continue to provide an effective description of generic full-sky PTA observations.

The estimates obtained here should be interpreted as  idealized geometric scaling estimates. Finite distance uncertainties, timing noise, ephemeris errors, and realistic measurement covariance are expected to increase the number of pulsars required to reach the same information content or to reduce the accessible multipole range.

The geometric interpretation developed here closely parallels radio interferometric synthesis imaging. The PTA response operator plays a role analogous to the sampling function in the interferometric $uv$ plane, while the singular-value spectrum quantifies the completeness of angular-mode coverage. From this perspective, PTA sky reconstruction is fundamentally an imaging problem whose limitations are controlled by the spatial distribution of pulsars and the conditioning of the measurement operator. The SVD-based reconstruction developed in this work provides a natural and robust regularization strategy by suppressing poorly constrained modes associated with small singular values.

Numerical experiments with mock PTAs illustrate the principal theoretical predictions. Realistic long-arm arrays were found to be accurately described by the Earth-term approximation and to recover only a limited number of low-order multipoles. In contrast, a deliberately constructed short-arm array operating in the pulsar-distance-resolved regime exhibited enhanced sensitivity to high multipoles, a larger effective rank of the response operator, and significantly improved sky-map fidelity. These simulations illustrate how the singular-value spectrum directly controls reconstruction quality and provide an intuitive visualization of the transition between the different operating regimes.

Finally, we showed that the same response-operator formalism extends naturally from deterministic sky reconstruction to searches for anisotropic stochastic gravitational-wave backgrounds. Expressing the SGWB covariance in the tensor spherical-harmonic basis leads to a likelihood formulation in which the anisotropy coefficients enter through the covariance matrix of the GW field, while the geometric PTA response remains encoded in the same response operator derived throughout this work. This separation provides a conceptually transparent framework for anisotropy studies and establishes a direct connection between PTA imaging and stochastic-background analyses.

Several directions for future work naturally emerge from this study. The most immediate extension is the incorporation of realistic timing-noise processes, including red spin noise, dispersion-measure variations, clock errors, and Solar-System ephemeris uncertainties, within a fully Bayesian inference framework. The response formalism developed here can also be generalized to alternative theories of gravity by incorporating scalar and vector polarization states, thereby enabling systematic investigations of non-Einsteinian GW signatures. Some work in this direction was undertaken by \cite{gair2015}. In addition, future PTA facilities, particularly those enabled by the Square Kilometre Array \cite{shannon2025}, may provide sufficiently accurate pulsar-distance measurements to begin probing aspects of the pulsar-distance-resolved regime. The framework developed in this paper offers a natural foundation for such analyses, as well as for future studies of PTA self-calibration, source localization, anisotropic stochastic backgrounds, and high-resolution gravitational-wave sky mapping.

\section*{Acknowledgments}

We are grateful to Prof. Bruce Allen of the Max Planck Institute for Gravitational Physics (Albert Einstein Institute) for suggesting the problem addressed in this paper and for inspiring the present investigation. We thank the anonymous referees of the paper for helping to significantly improve the narrative of the paper.

\appendix

\section{Coordinate Systems and Tensor Spherical Harmonics}
\label{app:basis-defs}

\subsection{Coordinate Conventions}

Throughout this work we assume a flat Minkowski background space-time
perturbed by a weak gravitational-wave field $h_{\mu\nu}$. Two
right-handed coordinate systems are employed: a global (\emph{cosmic})
frame common to all pulsars in the array and a local
(\emph{computational}) frame attached to an individual pulsar. The two
frames share a common time coordinate and differ only by a spatial
rotation.

The global cosmic frame is denoted by $(x,y,z)$ in Cartesian
coordinates and by $(r,\theta,\phi)$ in spherical coordinates. Since
only sky directions are required, all vectors are restricted to the
unit sphere $r=1$ and are parameterized by the angular coordinates
$(\theta,\phi)$. Following the standard PTA convention, the unit vector
$\hat{\Omega}$ 
points toward the GW propagation direction.
The unit vector
$\hat{p}$ points from the Solar-System barycenter toward a pulsar.
Expressed in Cartesian coordinates,

\begin{equation}
	\begin{aligned}
		\hat{\Omega}
	&=
	(\sin\theta\cos\phi,
	\sin\theta\sin\phi,
	\cos\theta)^T,
	\\
	\hat{p}
	&=
	(\sin\zeta\cos\chi,
	\sin\zeta\sin\chi,
	\cos\zeta)^T .
	\end{aligned}
\end{equation}
The polarization basis is constructed from the orthonormal triad
$(\hat l,\hat m,\hat\Omega)$, where

\begin{equation}
	\begin{aligned}
		\hat l
	&=
	(\cos\theta\cos\phi,
	\cos\theta\sin\phi,
	-\sin\theta)^T,
	\\
	\hat m
	&=
	(-\sin\phi,
	\cos\phi,
	0)^T .
	\end{aligned}
	\label{eq:lm-def}
\end{equation}
These vectors satisfy

\begin{equation}
	\hat l\cdot\hat l
	=
	\hat m\cdot\hat m
	=
	\hat\Omega\cdot\hat\Omega
	=
	1,
	\qquad
	\hat l\cdot\hat m
	=
	\hat l\cdot\hat\Omega
	=
	\hat m\cdot\hat\Omega
	=
	0 .
	\label{eq:lm-features}
\end{equation}
The transverse-traceless polarization tensors are

\begin{equation}
	\begin{aligned}
		e^+_{ab}
	&=
	\hat l_a\hat l_b
	-
	\hat m_a\hat m_b,
	\\
	e^\times_{ab}
	&=
	\hat l_a\hat m_b
	+
	\hat m_a\hat l_b ,
	\end{aligned}
	\label{eq:eA-def}
\end{equation}
where $A\in\{+,\times\}$ labels the polarization state.

For analytical calculations it is convenient to introduce, for each
pulsar, a local computational frame whose polar axis is aligned with the
pulsar direction. In this frame,

\begin{equation}
	\overline{\hat p}
	=
	(0,0,1)^T ,
\end{equation}
which considerably simplifies the angular integrations appearing in the
PTA response functions. The corresponding spherical coordinates are
denoted by $(\overline{\theta},\overline{\phi})$, and the GW propagation
direction becomes

\begin{equation}
	\overline{\hat{\Omega}}
	=
	(\sin\overline{\theta}\cos\overline{\phi},
	\sin\overline{\theta}\sin\overline{\phi},
	\cos\overline{\theta})^T .
\end{equation}
The polarization tensors in the computational frame have the same form,
\begin{equation}
	\begin{aligned}
		\overline e^+_{ab}
	&=
	\overline l_a\overline l_b
	-
	\overline m_a\overline m_b,
	\\
	\overline e^\times_{ab}
	&=
	\overline l_a\overline m_b
	+
	\overline m_a\overline l_b,
	\end{aligned}
\end{equation}
where $\overline l$ and $\overline m$ are constructed from
$(\overline\theta,\overline\phi)$ in the same manner as
$\hat l$ and $\hat m$ are constructed from $(\theta,\phi)$.

The computational frame is related to the cosmic frame by a passive
Euler rotation in the ZYZ convention. The Euler angles are chosen as
$(\alpha,\beta,\gamma)=(\chi,\zeta,0)$ so that the pulsar direction
$\hat p(\zeta,\chi)$ is mapped onto the local polar axis
$\overline{\hat p}=(0,0,1)^T$. The transformation of any vector is

\begin{equation}
	\overline{\mathbf v}
	=
	R(\chi,\zeta,0)\,
	\mathbf v
	=
	R_z(0)\,
	R_y(\zeta)\,
	R_z(\chi)\,
	\mathbf v ,
\end{equation}
where, in the adopted convention,

\begin{equation}
	R_z(\chi)
	=
	\left(
		\begin{array}{ccc}
			\cos\chi & \sin\chi & 0\\
			-\sin\chi & \cos\chi & 0\\
			0 & 0 & 1
		\end{array}
	\right),
\end{equation}
and

\begin{equation}
	R_y(\zeta)
	=
	\left(
		\begin{array}{ccc}
			\cos\zeta & 0 & -\sin\zeta\\
			0 & 1 & 0\\
			\sin\zeta & 0 & \cos\zeta
		\end{array}
	\right).
\end{equation}

These coordinate conventions are used throughout the paper. The
computational frame allows the response of an individual Earth--pulsar
baseline to be evaluated analytically, whereas the cosmic frame provides
the common basis in which the GW sky is expanded and reconstructed.

\subsection{Tensor Spherical Harmonics}

The response formalism developed in this paper is based on the
decomposition of the gravitational-wave field into tensor spherical
harmonics on the two-sphere $\mathbb{S}^2$. This construction was introduced into
PTA analyses by \cite{gair2014}, following the analogous
gradient--curl decomposition used in cosmic microwave background
polarization studies \cite{kamionkowski1997}.

Although the tensor harmonics are defined on the sphere, we retain the
same index notation $a,b,\ldots$ used for the spatial metric
perturbation $h_{ab}$. Geometrically, the tensor harmonics are tangent
to the sphere and may be viewed as projections of spatial tensors onto
the plane orthogonal to the propagation direction
$\hat{\Omega}$. The corresponding projection operator is
$P_{ab}=\delta_{ab}-\Omega_a\Omega_b$, which induces the metric on the
unit sphere. In spherical coordinates $(\theta,\phi)$, this induced
metric is
$g_{ab}=\mathrm{diag}(1,\sin^2\theta)$.
With this convention, the tensor harmonics generated below can be used
directly as basis functions in the expansion of the spatial
gravitational-wave field.

Any symmetric trace-free (STF) rank-two tensor field on the sphere can
be decomposed uniquely into a sum of gradient ($G$) and curl ($C$)
components generated by two scalar potentials, conventionally denoted
$A$ and $B$. For an arbitrary scalar field $X$ on $\mathbb{S}^2$, we define the
tensor differential operators
\begin{subequations}\label{eq:GC-operators}
	\begin{align}
		\mathcal{G}_{ab}[X]
	&\equiv
	X_{;ab}
	-
	\ds\frac{1}{2}
	g_{ab}
	X_{;c}^{\ \ c},
	\\
	\mathcal{C}_{ab}[X]
	&\equiv
	\ds\frac{1}{2}
	\left(
		X_{;ac}\epsilon^{c}_{\ b}
		+
		X_{;bc}\epsilon^{c}_{\ a}
	\right),
	\end{align}
\end{subequations}
where $\epsilon_{ab}$ is the Levi--Civita tensor associated with
$g_{ab}$, and the semicolon denotes covariant differentiation with
respect to the spherical metric. The operator
$\mathcal{G}_{ab}$ generates an even-parity (gradient) STF tensor
field, whereas $\mathcal{C}_{ab}$ generates an odd-parity (curl) STF
tensor field.
An arbitrary STF rank-two tensor field on $\mathbb{S}^2$ may therefore be written
in the form

\begin{equation}
	h_{ab}
	=
	\mathcal{G}_{ab}[A]
	+
	\mathcal{C}_{ab}[B].
\end{equation}
Since the scalar potentials $A$ and $B$ can themselves be expanded in
ordinary spherical harmonics $Y_{lm}(\hat{\Omega})$, a complete basis
for STF tensor fields is obtained by applying the operators
$\mathcal{G}_{ab}$ and $\mathcal{C}_{ab}$ to the scalar harmonics. The
resulting tensor spherical harmonics are

\begin{subequations}\label{eq:Ytens-def}
	\begin{align}
		Y^{G}_{(lm)ab}
	&=
	N_l\,\mathcal{G}_{ab}[Y_{lm}]
	=
	N_l
	\left[
		Y_{lm;ab}
		-
		\ds\frac{1}{2}
		g_{ab}
		Y_{lm;c}^{\ \ \ \ c}
	\right],
	\\\nonumber\\
	Y^{C}_{(lm)ab}
	&=
	N_l\,\mathcal{C}_{ab}[Y_{lm}]
	=
	\ds\frac{N_l}{2}
	\Biggl[
		Y_{lm;ac}\epsilon^{c}_{\ b}
		+
		Y_{lm;bc}\epsilon^{c}_{\ a}
	\Biggr],
	\end{align}
\end{subequations}
where the normalization factor

\begin{equation}
	N_l
	=
	\sqrt{
		2\,
		\ds\frac{(l-2)!}{(l+2)!}
	}
	\label{eq:Nl-def}
\end{equation}
is chosen so that the tensor harmonics form an orthonormal basis on
$\mathbb{S}^2$. Because the differential operators
$\mathcal{G}_{ab}$ and $\mathcal{C}_{ab}$ annihilate the monopole
($l=0$) and dipole ($l=1$) scalar harmonics, non-trivial tensor
harmonics exist only for $l\ge2$.

The tensor spherical harmonics satisfy the orthogonality relations

\begin{equation}
	\oint_{\mathbb{S}^2}
	d\hat{\Omega}\,
	Y^{P}_{(lm)ab}(\hat{\Omega})
	Y^{P'\,*}_{(l'm')}{}^{ab}(\hat{\Omega})
	=\delta^{PP'}
	\delta_{ll'}
	\delta_{mm'}
	,
\end{equation}
where $P,P'\in\{G,C\}$. Together, the sets
$\{Y^{G}_{(lm)ab}\}$ and $\{Y^{C}_{(lm)ab}\}$ form a complete
orthonormal basis for symmetric trace-free rank-two tensor fields on
the sphere.

Using this basis, the gravitational-wave field may be expanded as

\begin{equation}
	h_{ab}(\omega,\hat{\Omega})
	=
	\sum_{lm}
	\left[
		a^{G}_{lm}(\omega)\,
		Y^{G}_{(lm)ab}(\hat{\Omega})
		+
		a^{C}_{lm}(\omega)\,
		Y^{C}_{(lm)ab}(\hat{\Omega})
	\right],
\end{equation}
where $a^{G}_{lm}$ and $a^{C}_{lm}$ are the gradient and curl
coefficients of the GW sky. These coefficients constitute the fundamental degrees of freedom reconstructed and analyzed throughout this work.

\subsection{Polarization Basis Transformations}

The gradient and curl tensor spherical harmonics can be expressed in terms of the standard GW polarization tensors
$e^A_{ab}(\hat{\Omega})$ through the scalar kernels
$W_{lm}(\hat{\Omega})$ and $X_{lm}(\hat{\Omega})$:
\begin{subequations}\label{eq:YGC-WX}
	\begin{align}
		Y^G_{(lm)ab}(\hat{\Omega})
	&= \ds\frac{N_l}{2}
	\left[
		W_{lm}(\hat{\Omega}) e_{ab}^+(\hat \Omega)
		+ X_{lm}(\hat{\Omega}) e_{ab}^\times(\hat \Omega)
	\right],
	\\
	Y^C_{(lm)ab}(\hat{\Omega})
	&= \ds\frac{N_l}{2}
	\left[
		W_{lm}(\hat{\Omega}) e_{ab}^\times(\hat \Omega)
		- X_{lm}(\hat{\Omega}) e_{ab}^+(\hat \Omega)
	\right].
	\end{align}
\end{subequations}
The kernels are given by
\begin{subequations}\label{eq:WXlm}
	\begin{align}
		W_{lm}(\hat{\Omega})
	&=
	\left(
		\ds\frac{\partial^2}{\partial \theta^2}
		- \cot\theta\ds\frac{\partial}{\partial\theta}
		+\ds\frac{m^2}{\sin^2\theta}
	\right)
	Y_{lm}(\hat{\Omega})
	=
	\left[
		2\ds\frac{\partial^2}{\partial \theta^2}
		+l(l+1)
	\right]
	Y_{lm}(\hat{\Omega}),
	\\\nonumber\\
	X_{lm}(\hat{\Omega})
	&=
	\ds\frac{2im}{\sin\theta}
	\left(
		\ds\frac{\partial}{\partial\theta}
		-\cot\theta
	\right)
	Y_{lm}(\hat{\Omega}),
	\end{align}
\end{subequations}
where $i=\sqrt{-1}$ denotes the imaginary unit.

Using the standard definition of the scalar spherical harmonics,
\begin{equation}
	Y_{lm}(\hat{\Omega})
	=
	N_{lm} P_l^m(\cos\theta)e^{im\phi},
\end{equation}
with normalization
\begin{equation}
	N_{lm}
	=
	\sqrt{
		\ds\frac{2l+1}{4\pi}
		\ds\frac{(l-m)!}{(l+m)!}
	},
	\label{eq:Nlm-def}
\end{equation}
and where $P_l^m(\cos\theta)$ are the associated Legendre polynomials defined with the Condon--Shortley phase convention, one obtains the explicit expressions
\begin{subequations}\label{eq:WXlm-explicit}
	\begin{align}
		W_{lm}(\hat{\Omega})
	&=
	2N_{lm}
	G^+_{lm}(\cos\theta)
	e^{im\phi},
	\\
	iX_{lm}(\hat{\Omega})
	&=
	-2N_{lm}
	G^-_{lm}(\cos\theta)
	e^{im\phi},
	\end{align}
\end{subequations}\noindent
where the functions $G^{\pm}_{lm}$ are defined by
\begin{subequations}\label{eq:GA-def}
	\begin{align}
		G^+_{lm}(\cos\theta)
	&=
	-\left(
		\ds\frac{l-m^2}{\sin^2\theta}
		+\ds\frac{l(l-1)}{2}
	\right)
	P_l^m(\cos\theta)
	+(l+m)
	\ds\frac{\cos\theta}{\sin^2\theta}
	P_{l-1}^m(\cos\theta),
	\\
	G^-_{lm}(\cos\theta)
	&=
	\ds\frac{m}{\sin^2\theta}
	\left[
		(l-1)\cos\theta\,P_l^m(\cos\theta)
		-(l+m)P_{l-1}^m(\cos\theta)
	\right].
	\end{align}
\end{subequations}

\section{CLOSED-FORM RESPONSE FUNCTIONS}
\label{app:response}

This appendix presents the derivation of the closed-form expressions for the gradient and curl response functions. Starting from the integral representations in Eq.~\eqref{eq:RA-RP-transform},
\begin{subequations}
	\begin{align}
		R^G_{lm}(\omega) &=\ds\frac{N_l}{2}
		\oint_{\mathbb{S}^2} d\hat{\Omega}\,
		\left[
			R^+(\omega, \hat{\Omega}) W_{lm}(\hat{\Omega}) 
			+ R^\times(\omega, \hat{\Omega}) X_{lm}(\hat{\Omega}) 
		\right],\\
		R^C_{lm} (\omega)&=  \ds\frac{N_l}{2}
		\oint_{\mathbb{S}^2} d\hat{\Omega}\,
		\left[
			R^\times(\omega, \hat{\Omega}) W_{lm}(\hat{\Omega}) 
			-  R^+(\omega, \hat{\Omega})X_{lm}(\hat{\Omega})
		\right],
	\end{align}
\end{subequations}\noindent
we employ the coordinate-rotation technique, see, e.g. \cite{anholm2009, gair2014}, to evaluate the angular integrals analytically. The derivation proceeds in three steps. First, we obtain a set of auxiliary integrals that appear repeatedly throughout the calculation. Second, we exploit the rotation between the cosmic and computational frames described in \ref{app:basis-defs} to recast Eq.~\eqref{eq:RA-RP-transform} into a more tractable form. Third, we perform the remaining algebraic reductions, leading to the closed-form results presented in Eqs.~\eqref{eq:RG-analyt}--\eqref{eq:RC-analyt}.

\subsection{A Mathematical Lemma for Evaluating of the Response Functions}
\label{app:calG}

The evaluation of Eq.~\eqref{eq:RA-RP-transform} ultimately reduces to a class of angular integrals containing the product
$R^A(\omega,\hat{\Omega})W_{lm}(\hat{\Omega})$.
After transforming the angular integration to the computational frame, the calculation can be expressed in terms of the auxiliary integral
\begin{equation}
	\mathcal{G}_{l,l'}
	=
	\ds\int_{-1}^{+1} dx\,
	\ds\frac{1-x}{2}\,
	G^+_{l0}(x)\,
	P_{l'}(x),
	\label{eq:calG-def}
\end{equation}
where the function $G^+_{lm}(x)$ is defined in Eq.~\eqref{eq:GA-def}. Substituting its explicit form for $m=0$ yields
\begin{equation}
	\mathcal{G}_{l,l'}
	=
	\ds\frac{1}{2}
	\ds\int_{-1}^{+1} dx
	\left[
		-\ds\frac{l}{1+x}P_l(x)
		-\ds\frac{l(l-1)}{2}(1-x)P_l(x)
		+\ds\frac{lx}{1+x}P_{l-1}(x)
	\right]
	P_{l'}(x).
	\label{eq:calG-expanded}
\end{equation}
The first and third terms in the square brackets contain factors $(1+x)^{-1}$ and therefore appear logarithmically divergent at the lower integration limit $x=-1$. However, this singularity is only apparent. Combining the two terms, Eq.~\eqref{eq:calG-expanded} can be rewritten as
\begin{equation}
	\mathcal{G}_{l,l'}
	=
	-\ds\frac{l}{2}
	\ds\int_{-1}^{+1} dx
	\left[
		\ds\frac{P_l(x)-xP_{l-1}(x)}{1+x}
		+
		\ds\frac{l-1}{2}(1-x)P_l(x)
	\right]P_{l'}(x).
	\label{eq:calG-pre-diff}
\end{equation}
The apparent singularity can be removed by using the recurrence relation \cite[8.832]{gradshtein2007}
\begin{equation}
	P_l(x)-xP_{l-1}(x)=-\ds\frac{1-x^2}{l}\ds\frac{d}{dx}P_{l-1}(x)
	.
\end{equation}
Substituting this identity into Eq.~\eqref{eq:calG-pre-diff}, we obtain
\begin{equation}
	\mathcal{G}_{l,l'}=\mathcal{G}^{(1)}_{l,l'}+\mathcal{G}^{(2)}_{l,l'}\;,
	\label{eq:calG-post-deriv}
\end{equation}
where the two contributions are
\begin{align}
	\mathcal{G}^{(1)}_{l,l'}
	&=
	\ds\frac{1}{2}
	\ds\int_{-1}^{+1} dx\,
	(1-x)
	P_{l'}(x)
	\ds\frac{d}{dx}P_{l-1}(x)
	,
	\\
	\mathcal{G}^{(2)}_{l,l'}
	&=-\ds\frac{l(l-1)}{4}\ds\int_{-1}^{+1} dx\,
	(1-x)P_{l'}(x)P_l(x),
\end{align}

We now evaluate the two contributions in Eq.~\eqref{eq:calG-post-deriv} separately. The term is $\mathcal{G}^{(1)}_{l,l'}$ is calculated by
using the Legendre-polynomial identities \cite[8.832]{gradshtein2007}
\begin{equation}
	\ds\frac{d}{dx}P_n(x)
	=
	\ds\sum_{k=0}^{\left[\frac{n-1}{2}\right]}
	(2n-4k-1)
	P_{n-2k-1}(x),
\end{equation}
\begin{equation}
	xP_n(x)
	=
	\ds\frac{n+1}{2n+1}P_{n+1}(x)
	+
	\ds\frac{n}{2n+1}P_{n-1}(x),
\end{equation}
where $[\dots]$ in the upper limit of the sum is the integer part operator.
We find
\begin{equation}
	\mathcal{G}^{(1)}_{l,l'}
	=
	\ds\frac{1}{2}
	\left[
		\mathcal{I}_{l',\,l-1}
		-
		\ds\frac{l'+1}{2l'+1}\mathcal{I}_{l'+1,\,l-1}
		-
		\ds\frac{l'}{2l'+1}\mathcal{I}_{l'-1,\,l-1}
	\right],
	\label{eq:calG-calI}
\end{equation}
where
\begin{equation}
	\mathcal{I}_{l',l}
	=
	\sum_{k=0}^{\left[\frac{l-1}{2}\right]}
	(2l-4k-1)
	\ds\int_{-1}^{+1} dx\,
	P_{l'}(x) P_{l-2k-1}(x).
\end{equation}
Applying the orthogonality relation of Legendre polynomials,
\begin{equation}
	\ds\int_{-1}^{+1} P_n(x)P_m(x)\,dx
	=
	\ds\frac{2}{2n+1}\delta_{nm},
\end{equation}
the quantity $\mathcal{I}_{l,l'}$ simplifies considerably. A nonzero contribution arises only when one of the indices in the sequence $l-2k-1$ coincides with $l'$. Since this sequence differs from $l$ by odd integers, this occurs only if $(l-l')$ is odd and $l'<l$. In that case, exactly one term survives in the sum, yielding
\begin{equation}
	\mathcal{I}_{l,l'}
	=
	\left\{
		\begin{array}{ll}
			2, & (l-l') \mbox{ is odd and } l'<l,\\
			0, & \mbox{otherwise}.
		\end{array}
	\right.
	\label{eq:calI-sol}
\end{equation}
Substituting Eq.~\eqref{eq:calI-sol} into Eq.~\eqref{eq:calG-calI}, we obtain
\begin{equation}
	\mathcal{G}_{l,l'}^{(1)}
	=
	\left\{
		\begin{array}{ll}
			(-1)^{\,l-l'}, & l'<l-1,\\[1ex]
			-\ds\frac{l-1}{2l-1}, & l'=l-1,\\[2ex]
			0, & l'\ge l.
		\end{array}
	\right.
\end{equation}

The second contribution in Eq.~\eqref{eq:calG-post-deriv} can be evaluated directly using the recurrence relations together with the orthogonality of Legendre polynomials:
\begin{equation}
	\mathcal{G}_{l,l'}^{(2)}
	=
	-\ds\frac{l(l-1)}{2(2l+1)}
	\left[
		\delta_{l,l'}
		-
		\ds\frac{l+1}{2l+3}\delta_{l+1,l'}
		-
		\ds\frac{l}{2l-1}\delta_{l-1,l'}
	\right].
\end{equation}

Combining $\mathcal{G}_{l,l'}^{(1)}$ and $\mathcal{G}_{l,l'}^{(2)}$, we arrive at the final result
\begin{equation}
	\mathcal{G}_{l,l'}
	=
	\left\{
		\begin{array}{ll}
			(-1)^{\,l-l'}, & l'<l-1,
			\\[1ex]
			-\ds\frac{l-1}{2l-1}
			\left[
				1-\ds\frac{l^2}{2(2l+1)}
			\right],
						   & l'=l-1,
						   \\[2ex]
						   -\ds\frac{l(l-1)}{2(2l+1)},
						   & l'=l,
						   \\[2ex]
						   \ds\frac{l(l^2-1)}{2(2l+1)(2l+3)},
						   & l'=l+1,
						   \\[2ex]
						   0,
						   & l'>l+1.
		\end{array}
	\right.
	\label{eq:calG-result}
\end{equation}

\subsection{Evaluation of the Response Integrals}

In the computational reference frame introduced in
\ref{app:basis-defs}, the coordinate axes are chosen such that
the pulsar direction is aligned with the positive $z$-axis,
\[
	\overline{\hat p}=(0,0,1)^T.
\]
The antenna-pattern functions defined in Eqs.~\eqref{eq:R-def}--\eqref{eq:F-def} therefore reduce to
\begin{equation}
	R^+(\omega,\overline{\hat\Omega})
	=
	\left[
		1-e^{-iy(1+x)}
	\right]
	\ds\frac{1-x}{2},
	\qquad
	R^\times(\omega,\overline{\hat\Omega})
	=
	0,
	\label{eq:RA-rotated}
\end{equation}
where $x=\cos\overline{\theta}$.

Using the rotation from the cosmic frame to the computational frame described in \ref{app:basis-defs}, the response integrals in Eq.~\eqref{eq:RA-RP-transform} can be written as \cite{gair2014}
\begin{equation}
	R^G_{lm}(\omega)=
	\sum_{m'=-l}^{l}
	\left[D^l_{mm'}(\chi,\zeta,0)\right]^*
	\ds\frac{N_l}{2}
	\oint_{\mathbb S^2}
	d\overline{\hat\Omega}\,
	R^+(\overline{\hat\Omega})
	W_{lm'}(\overline{\hat\Omega}),
	\label{eq:RG-rotated}
\end{equation}
and
\begin{equation}
	R^C_{lm}(\omega)=
	-\sum_{m'=-l}^{l}
	\left[D^l_{mm'}(\chi,\zeta,0)\right]^*
	\ds\frac{N_l}{2}
	\oint_{\mathbb S^2}
	d\overline{\hat\Omega}\,
	R^+(\overline{\hat\Omega})
	X_{lm'}(\overline{\hat\Omega}),
	\label{eq:RC-rotated}
\end{equation}
where $D^l_{mm'}(\chi,\zeta,0)$ denotes the Wigner $D$ matrix for the Euler
rotation $(\chi,\zeta,0)$ \cite{varshalovich1988}.

The azimuthal dependence of the integrands is entirely contained in the factors
$W_{(lm')}(\overline{\hat\Omega})$ and
$X_{(lm')}(\overline{\hat\Omega})$, both of which are proportional to
$e^{im'\phi}$ [cf. Eq.~\eqref{eq:WXlm-explicit}]. Since
\[
	\ds\int_0^{2\pi} e^{im'\phi}\,d\phi
	=
	2\pi\,\delta_{m'0},
\]
all terms with $m'\neq0$ vanish. Consequently, Eqs.~\eqref{eq:RG-rotated} and \eqref{eq:RC-rotated} reduce to
\begin{equation}
	R^G_{lm}(\omega)
	=
	2\pi
	\left[D^l_{m0}(\chi,\zeta,0)\right]^*
	\ds\frac{N_l}{2}
	\ds\int_{-1}^{+1}
	dx\,
	R^+(\overline{\hat\Omega})
	W_{l0}(\overline{\hat\Omega}),
	\label{eq:RG-azimuth}
\end{equation}
and
\begin{equation}
	R^C_{lm}(\omega)
	=
	-2\pi
	\left[D^l_{m0}(\chi,\zeta,0)\right]^*
	\ds\frac{N_l}{2}
	\ds\int_{-1}^{+1}
	dx\,
	R^+(\overline{\hat\Omega})
	X_{l0}(\overline{\hat\Omega}),
	\label{eq:RC-azimuth}
\end{equation}
where $x=\cos\overline{\theta}$.

The curl response vanishes identically. Indeed, Eq.~\eqref{eq:WXlm-explicit} implies that
\[
	X_{l0}(\overline{\hat\Omega})=0,
\]
and therefore
\begin{equation}
	R^C_{lm}(\omega)=0.
\end{equation}

To evaluate the gradient response, we expand the exponential term in Eq.~\eqref{eq:RA-rotated} using the Rayleigh expansion
\begin{equation}
	e^{-iy(1+x)}
	=
	\sum_{l'=0}^{\infty}
	q_{l'}(y)P_{l'}(x),
	\label{eq:Rayleigh}
\end{equation}
where
\begin{equation}
	q_l(y)
	=
	e^{-iy}(2l+1)(-i)^l j_l(y),
	\label{eq:q-def}
\end{equation}
and $j_l(y)$ are the spherical Bessel functions of the first kind.
Substituting Eqs.~\eqref{eq:Rayleigh} and \eqref{eq:WXlm-explicit} into Eq.~\eqref{eq:RG-azimuth}, we obtain
\begin{equation}
	R^G_{lm}(\omega)
	=
	2\pi
	\left[D^l_{m0}(\chi,\zeta,0)\right]^*
	N_lN_{l0}
	\ds\int_{-1}^{+1}
	dx\,
	\ds\frac{1-x}{2}
	G^+_{(l0)}(x)
	\left[
		1-
		\sum_{l'=0}^{\infty}
		q_{l'}(y)P_{l'}(x)
	\right].
\end{equation}
Introducing the auxiliary integral $\mathcal G_{l,l'}$ defined in \ref{app:calG}, the gradient response can be written as
\begin{equation}
	R^G_{lm}(\omega)
	=
	2\pi
	\left[D^l_{m0}(\chi,\zeta,0)\right]^*
	N_lN_{l0}
	\left[
		\mathcal G_{l,0}
		-
		\sum_{l'=0}^{\infty}
		q_{l'}(y)\,
		\mathcal G_{l,l'}
	\right].
	\label{eq:RG-calG}
\end{equation}
Finally, Eq.~\eqref{eq:calG-result} shows that
$\mathcal G_{l,l'}$ vanishes for $l'>l+1$. Therefore, the formally infinite series in Eq.~\eqref{eq:RG-calG} truncates exactly, reducing to a finite sum over
\[
	0 \le l' \le l+1.
\]
This property enables the response function to be expressed in closed analytical form.

\subsection{Reduction of the Multipole Expansion}

To obtain a closed-form expression for the response function, it remains to evaluate the finite sum appearing in Eq.~\eqref{eq:RG-calG},
\begin{equation}
	S_0^{l+1}
	\equiv
	\sum_{l'=0}^{l+1}
	q_{l'}(y)\,
	\mathcal{G}_{l,l'}.
	\label{eq:S-def}
\end{equation}
We begin with the contribution from the range $0\le l'\le l-2$, for which Eq.~\eqref{eq:calG-result} yields
$\mathcal{G}_{l,l'}=(-1)^{l-l'}$. Substituting Eqs.~\eqref{eq:q-def} and \eqref{eq:calG-result} into Eq.~\eqref{eq:S-def}, we obtain
\begin{equation}
	S_0^{l-2}
	=
	(-1)^l e^{-iy}
	\sum_{l'=0}^{l-2}
	(2l'+1)i^{l'}j_{l'}(y).
	\label{eq:S_0^l-2-def}
\end{equation}
Using the recurrence relation for spherical Bessel functions of the first kind given in Eq.~(8.471) of \cite{gradshtein2007},
\[
	xj_{n-1}(x)+xj_{n+1}(x)
	=
	(2n+1)j_n(x),
\]
Eq.~\eqref{eq:S_0^l-2-def} can be rewritten as
\begin{equation}
	S_0^{l-2}
	=
	(-1)^l e^{-iy}iy
	\left[
		\sum_{p=-1}^{l-3}
		i^p j_p(y)
		-
		\sum_{k=1}^{l-1}
		i^k j_k(y)
	\right],
	\label{eq:S_0^l-2-recurrence}
\end{equation}
where $p=l'-1$ and $k=l'+1$.

For $l=2$, Eq.~\eqref{eq:S_0^l-2-recurrence} reduces immediately to
\begin{equation}
	S_0^{l-2}
	=
	e^{-iy}j_0(y).
	\label{eq:S_0^l-2-special}
\end{equation}

For $l>2$, the two sums in Eq.~\eqref{eq:S_0^l-2-recurrence} cancel term by
term except for the endpoint contributions, yielding
\begin{equation}
	S_0^{l-2}
	=
	(-1)^l e^{-iy}y
	\left[
		j_{-1}(y)
		+
		ij_0(y)
		-
		i^{l-1}j_{l-2}(y)
		-
		i^lj_{l-1}(y)
	\right].
	\label{eq:S_0^l-2-telescopic}
\end{equation}
Using the identity
\[
	j_{-1}(y)+ij_0(y)
	=
	\ds\frac{e^{iy}}{y},
\]
we obtain
\begin{equation}
	S_0^{l-2}
	=
	(-1)^l
	+
	(-i)^l e^{-iy}
	\left[
		iy\,j_{l-2}(y)
		-
		y\,j_{l-1}(y)
	\right].
	\label{eq:S_0^l-2-general}
\end{equation}
The expression in Eq.~\eqref{eq:S_0^l-2-general} reproduces the special case \eqref{eq:S_0^l-2-special} when $l=2$ and is therefore valid for all $l\ge2$.

The remaining contribution to Eq.~\eqref{eq:S-def} is
\begin{equation}
	\begin{aligned}
		e^{iy}S_{l-1}^{l+1}
		=&
		-(l-1)(-i)^{l-1}
		\left[
			1-\ds\frac{l^2}{2(2l+1)}
		\right]
		j_{l-1}(y)
		-(-i)^l
		\ds\frac{l(l-1)}{2}
		j_l(y)
		\\
		 &
		 +(-i)^{l+1}
		 \ds\frac{l(l^2-1)}
		 {2(2l+1)}
		 j_{l+1}(y).
	\end{aligned}
\end{equation}
Eliminating $j_{l+1}(y)$ with the recurrence relation for spherical Bessel functions, we find
\begin{equation}
	S_{l-1}^{l+1}
	=
	e^{-iy}(-i)^l
	\Bigg\{
		i\ds\frac{(l-1)(l-2)}{2}
		j_{l-1}(y)
		-\ds\frac{l(l-1)}{2}
		\left[
			1+i\ds\frac{l+1}{y}
		\right]
		j_l(y)
	\Bigg\}.
	\label{eq:S_l-1^l+1-general}
\end{equation}
Combining Eqs.~\eqref{eq:S_0^l-2-general} and \eqref{eq:S_l-1^l+1-general}, and using the recurrence relation once more to eliminate $j_{l-2}(y)$, we arrive at
\begin{equation}
	\begin{aligned}
		S_0^{l+1}
		=
		(-1)^l
		-
		e^{-iy}(-i)^l
		\Bigg\{
	&
	\left[
		y-i\ds\frac{l(l+1)}{2}
	\right]
	j_{l-1}(y)
	\\
	&
	+
	\left[
		\ds\frac{l(l-1)}{2}
		+i\ds\frac{l(l^2-1)}{2y}
		+iy
	\right]
	j_l(y)
\Bigg\}.
	\end{aligned}
	\label{eq:S-result}
\end{equation}
Substituting Eq.~\eqref{eq:S-result} into Eq.~\eqref{eq:RG-calG}, we note that the constant term $(-1)^l$ cancels exactly against the contribution from $\mathcal{G}_{l,0}$. Indeed, Eq.~\eqref{eq:calG-result} gives
\[
	\mathcal{G}_{l,0}=(-1)^l,
\]
since $0<l-1$ for all $l\ge2$. Therefore,
\begin{equation}
	\begin{aligned}
		\mathcal{G}_{l,0}-S_0^{l+1}
		=
		e^{-iy}(-i)^l
		\Bigg\{
	&
	\left[
		y-i\ds\frac{l(l+1)}{2}
	\right]
	j_{l-1}(y)
	\\
	&
	+
	\left[
		\ds\frac{l(l-1)}{2}
		+i\ds\frac{l(l^2-1)}{2y}
		+iy
	\right]
	j_l(y)
\Bigg\}.
	\end{aligned}
\end{equation}
Equation~\eqref{eq:RG-calG} then reduces to
\begin{equation}
	R^G_{lm}(\omega)
	=
	2\pi
	\left[D^l_{m0}(\chi,\zeta,0)\right]^*
	N_{l0}\,
	e^{-iy}
	(-i)^l
	f_l(y),
	\label{eq:RG-before-D}
\end{equation}
where
\begin{equation}
	f_l(y)
	=N_l
	\left\{
		\left[
			y-i\ds\frac{l(l+1)}{2}
		\right]
		j_{l-1}(y)
		+
		\left[
			\ds\frac{l(l-1)}{2}
			+i\ds\frac{l(l^2-1)}{2y}
			+iy
		\right]
		j_l(y)
	\right\}.
	\label{eq:fl-def}
\end{equation}

Finally, using the Wigner-$D$ matrix identity
\[
	\left[D^{l}_{m0}(\varphi,\theta,\psi)\right]^*
	=
	\ds\frac{1}{N_{l0}}\,Y_{lm}(\theta,\varphi),
\]
we obtain
\begin{equation}
	R^G_{lm}(\omega)
	=
	2\pi
	Y_{lm}(\hat p)
	e^{-iy}
	(-i)^l
	f_l(y),
\end{equation}
which is the desired closed-form expression for the gradient response function.

\section{Asymptotic Analysis of the PTA Sensitivity Function}
\label{app:asymptotics}

The directional response of an elementary Earth--pulsar baseline is
governed by the sensitivity function $f_l(y)$ appearing in the analytical
response function \eqref{eq:RG-analyt}, where
\begin{equation}
	y\equiv \omega L .
\end{equation}
The parameter $y$ measures the Earth--pulsar baseline length in units of
the gravitational-wave wavelength (up to a factor of $2\pi$) and
therefore characterizes the combined effects of the GW frequency and the
pulsar distance.

Since $f_l(y)$ is a linear combination of spherical Bessel functions of
the first kind, its asymptotic behavior is determined by the large-order
and large-argument asymptotics of these functions. The corresponding PTA
response exhibits four physically distinct sensitivity regimes introduced
in Section~\ref{sec:three-operation-regimes}:

\begin{enumerate}

	\item \textit{Earth-term-dominated regime}.

		At sufficiently low multipoles, the rapidly oscillating pulsar term
		averages out, and the detector response is governed predominantly by the
		Earth term. In this regime the sensitivity is well described by the
		Earth-term asymptotic expansion derived below.

	\item \textit{Transition regime}.

		With increasing multipole number, the Earth-term contribution gradually
		decreases while the pulsar-term contribution becomes comparable to it.
		Neither the Earth-term nor the pulsar-term asymptotic approximation alone
		provide an adequate description of the sensitivity function.

	\item \textit{Pulsar-term-dominated regime}.

		At higher multipoles, the pulsar term provides the dominant contribution
		to the detector response. The sensitivity exhibits an oscillatory
		dependence on the multipole number and is described by the large-$l$
		asymptotics of the spherical Bessel functions.

	\item \textit{Effective diffraction cutoff regime}.

		As the multipole number approaches the turning point of the spherical
		Bessel functions, the detector response undergoes a transition from
		oscillatory to exponentially decaying behavior. Beyond this point the
		sensitivity rapidly decreases, rendering progressively finer angular
		structure effectively unobservable.

\end{enumerate}

It is important to distinguish these PTA sensitivity regimes from the
mathematical asymptotic regions of the spherical Bessel functions.
The Earth-term-dominated, transition, and pulsar-term-dominated regimes
all lie within the oscillatory domain of the Bessel functions, whereas
the sensitivity-cutoff regime is associated with the turning-point
(Airy) asymptotics and the subsequent exponential decay of the response.

In this appendix we derive the corresponding asymptotic approximations
for the Earth-term, pulsar-term, and
effective diffraction cutoff
regimes.
Higher-order terms are retained in order to estimate the crossover
multipoles at which the Earth-term and pulsar-term asymptotic expansions
cease to be individually accurate. These crossover multipoles should be
understood as asymptotic estimates of the transition scales rather than
sharp physical boundaries.

The resulting asymptotic approximations and the corresponding transition
multipoles are summarized in Fig.~\ref{fig:fl}.

\subsection{Earth-Term-Dominated Regime}\label{abcde}
We begin with the asymptotic limit
\[
	y\rightarrow\infty ,
\]
while keeping the multipole number $l$ fixed. Physically, this
corresponds to an Earth--pulsar baseline much longer than the
gravitational-wave wavelength. In this limit the phase of the pulsar
term oscillates rapidly and its contribution averages out, leaving a
response dominated by the Earth term.

For fixed $l$ and $y\rightarrow\infty$, the spherical Bessel function
admits the asymptotic expansion in inverse powers of $y$ given in
Eq.~(10.49) of \cite{NIST:DLMF},
\begin{equation}
	j_l(y)
	=
	\sum_{n=0}^{\infty}
	j_l^{[n]}(y),
	\qquad
	y\rightarrow\infty ,
\end{equation}
where $j_l^{[n]}(y)$ denotes the contribution of order
$O(y^{-n})$. More precisely,
\[
	j_l(y)
	-
	\sum_{k=0}^{N}
	j_l^{[k]}(y)
	=
	O\!\left(y^{-N-1}\right),
	\qquad
	y\rightarrow\infty .
\]

The first non-vanishing terms of the expansion are
\begin{subequations}\label{eq:jl-large-y}
	\begin{align}
		j_l^{[0]}(y)
&=
0,
\\\nonumber\\
j_l^{[1]}(y)
&=
\ds\ds\frac{1}{y}
\sin\!\left(
	y-\ds\frac{l\pi}{2}
\right),
\\\nonumber\\
j_l^{[2]}(y)
&=
\ds\ds\frac{l(l+1)}{2y^2}
\cos\!\left(
	y-\ds\frac{l\pi}{2}
\right),
\\\nonumber\\
j_l^{[3]}(y)
&=
-\ds\ds\frac{(l-1)l(l+1)(l+2)}{8y^3}
\sin\!\left(
	y-\ds\frac{l\pi}{2}
\right).
	\end{align}
\end{subequations}

Substituting Eq.~\eqref{eq:jl-large-y} into the definition
\eqref{eq:fl-def} yields a corresponding asymptotic expansion for the
sensitivity function,
\begin{equation}
	f_l(y)
	=
	\sum_{n=0}^{\infty}
	f_l^{[n]}(y),
	\qquad
	y\rightarrow\infty ,
\end{equation}
where $f_l^{[n]}(y)$ denotes the contribution of order $O(y^{-n})$.
The first non-vanishing terms are
\begin{subequations}
	\begin{align}\label{eq:fl-large-y}
		f_l^{[0]}(y)
&=
N_l
\Bigg[
	\cos\!\left(
		y-\ds\frac{l\pi}{2}
	\right)
	+
	i
	\sin\!\left(
		y-\ds\frac{l\pi}{2}
	\right)
\Bigg]
=
N_l e^{iy}(-i)^l,
\\
f_l^{[1]}(y)
&=
0,
\\
f_l^{[2]}(y)
&=
N_l e^{iy}(-i)^{l} \frac{1}{8y^2} 
\frac{(l+2)!}{(l-2)!}
	\end{align}
\end{subequations}
where the normalization factor $N_l$ is defined in Eq. \eqref{eq:Nl-def}.

Retaining only the leading contribution
$f_l^{[0]}(y)$ in Eq.~\eqref{eq:fl-large-y} and substituting it into
Eq.~\eqref{eq:RG-analyt} yields the Earth-term response derived by Gair et al. \cite{gair2014},
\begin{equation}
	R^G_{lm}(\omega)
	=
	2\pi (-1)^l N_l
	Y_{lm}(\hat p).
	\label{eq:RG-large-y}
\end{equation}
Thus, in the Earth-term-dominated regime the sensitivity becomes
independent of the parameter $y$ and is governed entirely by the
normalization factor $N_l$. For large multipole numbers,
\begin{equation}
	N_l
	\sim
	\ds\frac{\sqrt{2}}{l^2},
	\label{eq:N_l-asympt}
\end{equation}
so that the detector response decreases approximately as $l^{-2}$ with
increasing angular complexity of the gravitational-wave sky.
\subsection{Transition Regime}
\label{app:transition-regime}

The transition regime occupies the interval between the
Earth-term-dominated and pulsar-term-dominated regimes.
Unlike the neighboring regimes, it is not associated with a distinct
asymptotic expansion of the sensitivity function. Rather, it arises
because the Earth-term asymptotic approximation gradually loses
validity before the pulsar-term asymptotic approximation becomes
self-consistent.

Consequently, the boundaries of the transition regime cannot be
identified with sharp physical transitions. Instead, they are
estimated from the domains of validity of the asymptotic expansions
describing the neighboring regimes. The resulting transition
multipoles should therefore be interpreted as crossover scales whose
numerical values depend on the level of accuracy required from the
asymptotic approximations.

To quantify this statement, we introduce a dimensionless tolerance
parameter $\epsilon$ that specifies the maximum acceptable fractional
error of an asymptotic expansion. Since the transition boundaries are
used only as order-of-magnitude estimates and are themselves derived
from asymptotic approximations, it is neither necessary nor justified
to adopt a very small value of $\epsilon$. Throughout this work we use
\[
	\epsilon = 
	10\%,
\]
so that the estimated uncertainty remains comparable to the neglected
asymptotic corrections. This choice provides a robust
characterization of the transition region without implying a level of
precision beyond that supported by the asymptotic analysis. The
physical interpretation of the resulting transition boundaries is
discussed in Section~\ref{sec:transition-regime}, whereas the present
appendix derives them from the asymptotic properties of the PTA
sensitivity function.

A natural estimate of the onset of the transition regime is obtained
from the Earth-term asymptotic expansion by comparing the two leading
non-vanishing contributions in Eq.~\eqref{eq:fl-large-y},
\begin{equation}
	\left|
	\ds\frac{f_l^{[2]}(y)}{f_l^{[0]}(y)}
	\right|
	=
	\frac{1}{8y^2} 
	\frac{(l+2)!}{(l-2)!}
	\simeq
	\ds\frac{l^4}{8 y^2},
\end{equation}
where the final expression assumes $l\gg1$.

Requiring the correction term to remain smaller than the prescribed
fractional tolerance yields
\begin{equation}
	\left|
	\ds\frac{f_l^{[2]}(y)}{f_l^{[0]}(y)}
	\right|
	<
	\epsilon
	\qquad\Longrightarrow\qquad
	l_{1\to2}
	<
	2^{3/4}
	y^{1/2}
	\epsilon^{1/4}.
	\label{eq:l-12-boundary}
\end{equation}

Thus the Earth-term approximation gradually loses accuracy once
\(
l=O(y^{1/2}),
\)
and Eq.~\eqref{eq:l-12-boundary} provides an estimate of the onset of
the transition regime. For multipoles below $l_{1\to2}$ the detector
response is accurately described by the Earth-term asymptotic
expansion, whereas for larger multipoles the neglected corrections
become progressively more important.

The upper boundary of the transition regime is determined by the
domain of validity of the pulsar-term asymptotic expansion derived in
the next subsection. Following the notation introduced in
Section~\ref{sec:transition-regime}, we denote this boundary by
$l_{2\to3}$. Between the two boundaries,
\begin{equation}
	l_{1\to2}
	\lesssim
	l
	\lesssim
	l_{2\to3}\;,
\end{equation}
the Earth-term approximation is no longer sufficiently accurate,
whereas the pulsar-term approximation is not yet fully
self-consistent. Consequently, neither asymptotic representation alone
provides an adequate description of the sensitivity function. In this
region the Earth-term and pulsar-term contributions become comparable,
and both must be retained to describe the detector response
accurately.

\subsection{Pulsar-Term-Dominated Regime}
\label{app:pulsar-regime}

We next consider the simultaneous large-$l$, large-$y$ limit in which
the multipole number and the parameter $y$ are of comparable
magnitude. It is convenient to introduce the dimensionless variable
\begin{equation}
	\xi_l\equiv\ds\frac{l+1/2}{y},
	\qquad
	0<\xi_l<1 .
	\label{eq:xi-def}
\end{equation}

The present analysis is restricted to the oscillatory domain of the
spherical Bessel functions, sufficiently far from the turning point,
\begin{equation}
	1-\xi_l^2 \gg y^{-2/3}.
\end{equation}

Physically, this domain corresponds to the pulsar-term-dominated
regime discussed in Section~\ref{sec:three-operation-regimes}. It
extends from the upper boundary of the transition region up to the
vicinity of the turning point of the spherical Bessel functions,
where the oscillatory asymptotics cease to be uniformly valid.

Unlike \ref{abcde}, where the asymptotic expansion was developed for
fixed $l$ and $y\to\infty$, we now consider the Debye limit
\begin{equation}
	l\to\infty,
	\qquad
	y\to\infty,
	\qquad
	\xi_l=\ds\frac{l+1/2}{y}=\mathrm{const}.
\end{equation}
Although in the Debye limit the parameter $\xi_l$ is held constant, in order to derive the expression for $f_l(y)$ \eqref{eq:fl-def}, which contains both spherical Bessel functions of two different multipoles $j_{l}$ and $j_{l-1}$ of the same argument (i.e. two different parameters $\xi_l$), we consider $\xi_l$ to be a function of $l$ for $y=\mathrm{const}$.

Since this is a different asymptotic regime, a new asymptotic
expansion must be introduced. We therefore write
\begin{equation}
	j_l(y)
	=
	\sum_{n=0}^{\infty}
	j_l^{\langle n\rangle}(y),
	\label{eq:debye-jl-series}
\end{equation}
where $j_l^{\langle n\rangle}(y)$ denotes the contribution whose
magnitude is smaller than the leading term by a factor
$O(l^{-n})$ at fixed $\xi_l$. Equivalently,
\begin{equation}
	\ds\frac{j_l^{\langle n\rangle}(y)}
	{j_l^{\langle0\rangle}(y)}
	=
	O(l^{-n}) .
\end{equation}

In this limit the spherical Bessel functions admit the Debye
asymptotic expansion \eqref{eq:debye-jl-series}. Retaining the first
two terms and using Eq.~(8.453) of \cite{gradshtein2007} together with
Eq.~(10.19) of \cite{NIST:DLMF}, one obtains
\begin{subequations}\label{eq:jl-xi-asympt}
	\begin{align}
		j_l^{\langle0\rangle}(y)
&=
\ds\frac{1}
{y(1-\xi_l^2)^{1/4}}\,\cos\alpha_l\;,
\\
j_l^{\langle1\rangle}(y)
&=
\ds\frac{1}
{y(1-\xi_l^2)^{1/4}}\ds\frac{\Xi(\xi_l)}{l}\,\sin\alpha_l\;,
	\end{align}
\end{subequations}
where
\begin{equation}
	\Xi(\xi_l)
	\equiv
	\ds\frac{\xi_l(3+2\xi_l^2)}
	{24(1-\xi_l^2)^{3/2}}\;,
\end{equation}
is the first-order Debye correction coefficient, and the oscillation phase 
\begin{equation}
	\alpha_l
	=
	y
	\sqrt{1-\xi_l^2}
	-
	\left(
		l+\ds\frac12
	\right)
	\arccos\xi_l
	-
	\ds\frac{\pi}{4}.
	\label{eq:alpha-def}
\end{equation}

Since both $j_l(y)$ and $j_{l-1}(y)$ enter the definition
\eqref{eq:fl-def}, the relative phase shift between the two functions
must be retained. To leading orders,
\begin{equation}
	\alpha_{l-1}
	\simeq
	\alpha_l+\arccos\xi _l
	+
	\frac{1}{2y\sqrt{1-\xi_l^2}}
	+ O(y^{-2})
	,
\end{equation}
which yields
\begin{align}
	\cos\alpha_{l-1}
&\simeq
\left[\xi_l - \frac{1}{2y}\right]\cos(\alpha_{l})
-
\left[
\sqrt{1-\xi_l^2} + \frac{\xi_l}{2y\sqrt{1-\xi_l^2}}\right]
\sin(\alpha_{l}) + O(y^{-2}),
\\\nonumber
\sin\alpha_{l-1}
&\simeq
\left[\xi_l - \frac{1}{2y}\right]\sin(\alpha_{l})
+
\left[
\sqrt{1-\xi_l^2} + \frac{\xi_l}{2y\sqrt{1-\xi_l^2}}\right]
\cos(\alpha_{l}) + O(y^{-2}).
\end{align}

Corresponding decopmposition of the multiplication factor gives
\begin{subequations}
	\begin{align}
		\label{eq:Al-def}
		A_l(y) &\equiv \frac{1}
		{y(1-\xi_l^2)^{1/4}}\\
		A_{l-1}(y) &\simeq A_l(y) 
		\left[
			1 - \frac{\xi_l}{2y(1-\xi_l^2)}
		\right]  + O(y^{-2}) 
		\\
		\label{eq:jm1-def}
		j_{l-1}(y) &\simeq \frac{1}{y(1-\xi_l^2)^{1/4}} \left[ \left( \xi_l - \frac{1}{2y(1-\xi_l^2)} + \frac{\Xi(\xi_l)\sqrt{1-\xi_l^2}}{y\xi_l} \right) \cos\alpha_l + \left( -\sqrt{1-\xi_l^2} + \frac{\Xi(\xi_l)}{y} \right) \sin\alpha_l \right]
	\end{align}
\end{subequations}

Substituting Eqs.~\eqref{eq:jl-xi-asympt}--\eqref{eq:alpha-def}
and \eqref{eq:Al-def}--\eqref{eq:jm1-def}
into Eq.~\eqref{eq:fl-def}, we obtain a
corresponding Debye expansion for the sensitivity function,
\begin{equation}
	f_l(y)
	\sim
	\sum_{n=0}^{\infty}
	f_l^{\langle n\rangle}(y),
	\label{eq:debye-fl-series}
\end{equation}
with
\begin{subequations}
	\label{eq:fl-ly-asympt}
	\begin{align}
		f_l^{\langle0\rangle}(y)
			&\simeq
			\frac{1}{2^{1/2}y(1-\xi_l^2)^{1/4}}
			\Big[
				\cos\alpha_l
				+
				i\sqrt{1-\xi_l^2}\sin\alpha_l
			\Big],
			\\
			f_l^{\langle1\rangle}(y)
			&\simeq
			\frac{1}{2^{1/2}y(1-\xi_l^2)^{1/4}}
			\frac{1}{l} \times
			\\
			&
			\left\{
				i\left[
					\frac{3(1-\xi_l^2)}{2\xi_l}
					+
					\frac{1}{2\xi_l(1-\xi_l^2)}
					-
					\Xi(\xi_l)\sqrt{1-\xi_l^2}
				\right]
				\cos\alpha_l
				+
				\left[
					\Xi(\xi_l) - \frac{2}{\xi_l}\sqrt{1-\xi_l^2}
				\right]
				\sin\alpha_l
			\right\}.
			\nonumber
	\end{align}
	\label{eq:fl-ly-asympt2}
\end{subequations}

The upper envelope of the leading Debye contribution is
\begin{equation}
	|f_l^{\langle0\rangle}(y)|_{\rm env}
	=
	2^{-1/2}
	y^{-1}
	(1-\xi_l^2)^{-1/4}.
	\label{eq:fl-assimp-jy-assymp}
\end{equation}

Equation~\eqref{eq:fl-assimp-jy-assymp} shows that, away from the
turning point $\xi_l=1$, the sensitivity depends only weakly on the
multipole number and is controlled primarily by the parameter $y$.
In contrast to the Earth-term-dominated regime, where the response
decreases approximately as $l^{-2}$, the sensitivity remains nearly
constant over a broad interval of multipoles and exhibits an
oscillatory modulation governed by the phase
\eqref{eq:alpha-def}. This behavior is responsible for the extension
of PTA sensitivity toward significantly finer angular scales discussed
in Section~\ref{sec:three-operation-regimes}.

The asymptotic expansion derived above also permits an estimate of the
onset of the pulsar-term-dominated regime. Because the leading-order
term oscillates and possesses zeros, the pointwise ratio
$f_l^{\langle1\rangle}/f_l^{\langle0\rangle}$ is not a reliable measure
of asymptotic accuracy throughout the entire oscillatory domain.
Instead, we examine the lower boundary of the Debye regime where
\begin{equation}
	l\ll y,
	\qquad
	\xi_l=\ds\frac{l+1/2}{y}\ll1.
\end{equation}
In this limit
\begin{equation}
	\Xi(\xi_l)
	=
	\ds\frac{\xi_l}{8}
	+
	O(\xi_l^3),
\end{equation}
and the terms in the square brackets of Eq.~\eqref{eq:fl-ly-asympt2} are approximated as follows: 
\begin{equation}
	f_l^{\langle1\rangle}(y) 
	=
	\frac{1}{2^{1/2}y(1-\xi_l^2)^{1/4}}
	\frac{\sqrt{2}}{\xi_l l}
	ie^{i\alpha_{l}}+
	O\!\left(\frac{\xi_l^2}{y^2}\right).
	\label{eq:f1-small-xi}
\end{equation}
Similarly, the leading Debye contribution becomes
\begin{equation}
	f_l^{\langle0\rangle}(y)
	=
	\frac{1}{2^{1/2}y(1-\xi_l^2)^{1/4}}e^{i\alpha_l}
	+
	O\!\left(\frac{\xi_l^2}{y}\right).
	\label{eq:f0-small-xi}
\end{equation}

To evaluate the fractional error, from Eqs.~\eqref{eq:f1-small-xi}and~\eqref{eq:f0-small-xi}, we obtain
\begin{equation}
	\frac{f_l^{\langle1\rangle}(y)}
	{f_l^{\langle0\rangle}(y)}
	\simeq
	\frac{i\sqrt{2}}{\xi_l l}.
\end{equation}
Using
\[
	\xi_l=\ds\frac{l+1/2}{y}
	\simeq
	\ds\frac{l}{y},
\]
we finally obtain
\begin{equation}
	\left|
	\ds\frac{f_l^{\langle1\rangle}(y)}
	{f_l^{\langle0\rangle}(y)}
	\right|_{\rm env}
	\simeq
	\sqrt{2}\frac{y}{l^2}
	\label{eq:debye-small-parameter}
\end{equation}

Equation~\eqref{eq:debye-small-parameter} shows that, near the lower
boundary of the Debye domain, the effective expansion parameter is
$\sqrt{2}y/l^2$. The leading Debye approximation therefore becomes
self-consistent when
\[
	\sqrt{2}\frac{y}{l^2}<\epsilon ,
\]
where $\epsilon$ is the prescribed fractional tolerance. This yields
\begin{equation}
	l
	>
	2^{1/4}
	y^{1/2}
	\epsilon^{-1/2}.
	\label{eq:l-23-boundary}
\end{equation}

Equation~\eqref{eq:l-23-boundary} therefore defines the characteristic
multipole $l_{2\to3}$ at which the leading pulsar-term asymptotic
contribution begins to dominate its first correction. For
\(
l \gtrsim l_{2\to3},
\)
the pulsar-term asymptotic expansion becomes self-consistent and the
detector enters the pulsar-term-dominated regime.

Combining Eqs.~\eqref{eq:l-12-boundary}
and~\eqref{eq:l-23-boundary}, we identify the transition interval
\begin{equation}
	l_{1\to2}
	\lesssim
	l
	\lesssim
	l_{2\to3},
\end{equation}
through which the detector response evolves continuously from
Earth-term dominance to pulsar-term dominance. Within this interval
neither asymptotic representation is individually adequate, and both
contributions remain important.

For the adopted value $\epsilon=10\%$, the transition boundaries are
approximately
\begin{equation}
	l_{1\to2}\simeq 
	y^{1/2},
	\qquad
	l_{2\to3}\simeq 
	4\,y^{1/2},
\end{equation}
demonstrating that the entire transition region is confined to
multipoles of order $y^{1/2}$. Although the exact numerical values of
the boundaries depend on the chosen tolerance, their scaling with
$y^{1/2}$ is a robust consequence of the asymptotic analysis.

Substituting $y=\omega L$ yields the numerical estimates quoted in
Section~\ref{sec:transition-regime}. The boundary $l_{1\to2}$ marks
the practical limit of applicability of the Earth-term approximation,
whereas $l_{2\to3}$ marks the onset of the pulsar-term-dominated
regime. Thus, Section~\ref{sec:transition-regime} provides the
physical interpretation of these scales, while the present appendix
establishes their asymptotic origin.
\subsection{Effective Diffraction-Cutoff Regime}

A second qualitative change in the detector response occurs in the
vicinity of the turning point of the spherical Bessel functions,
where the oscillatory Debye expansion derived in the previous subsection
ceases to be uniformly valid. The turning point is defined by the condition
\begin{equation}
	l+\ds\frac12=y ,
\end{equation}
for which the radial wavenumber appearing in the Debye phase
vanishes. It separates the oscillatory ($l<y$) and exponentially
decaying ($l>y$) domains of the spherical Bessel functions. This region marks the transition from the
pulsar-term-dominated regime to the high-multipole sensitivity-cutoff
regime.

We therefore consider the limit
\begin{equation}
	y\gg1,
	\qquad
	\xi_l=\ds\frac{l+1/2}{y}\simeq1 ,
\end{equation}
so that the turning point lies within the asymptotic domain.
In this region the assumptions underlying the Debye expansion are no
longer satisfied and a different asymptotic description is required.

Substituting $l\simeq y$ into Eq.~\eqref{eq:fl-def} shows that all
contributions become comparable in magnitude. Retaining only the
leading-order terms yields
\begin{equation}
	f_l(y)
	\sim
	\ds\frac{j_l(y)}{\sqrt{2}}.
	\label{eq:fl-jl-approx}
\end{equation}

The transition from oscillatory to exponentially decaying behavior does
not occur over a scale $\Delta l=O(1)$. Instead, the natural width of
the turning-point region grows as
\(
\Delta l=O(y^{1/3}).
\)
Accordingly, we introduce the scaled turning-point coordinate
\begin{equation}
	a
	\equiv
	\ds\frac{y-l}{l^{1/3}},
	\label{eq:airy-scaling}
\end{equation}
which remains finite,
\(
a=O(1),
\)
throughout the transition region. Thus, $a$ measures the distance from
the turning point in units of the natural turning-point width
$l^{1/3}$.

In terms of $a$, the turning point corresponds to
\(
a=0,
\)
while
\(
a>0
\; (y>l)
\)
describes the oscillatory side of the transition and
\(
a<0
\; (y<l)
\)
describes the exponentially decaying side.

Near the turning point,
\begin{equation}
	l-y=O(y^{1/3}),
\end{equation}
the spherical Bessel functions admit the uniform Airy-function
approximation given in Eq.~(10.20.4) of \cite{NIST:DLMF},
\begin{equation}
	j_l(y)=j_l(l+a\,l^{1/3})
	\sim
	\sqrt{\pi}\,
	2^{-1/6}
	y^{-5/6}
	\,
	\mathrm{Ai}
	\!\left(
		-2^{1/3}a
	\right),
	\label{eq:jl-airy}
\end{equation}
which remains valid on both sides of the turning point and therefore
provides a uniform description of the transition region.

Substituting Eq.~\eqref{eq:jl-airy} into
Eq.~\eqref{eq:fl-jl-approx} gives the corresponding asymptotic form of
the sensitivity function,
\begin{equation}
	f_l(y)
	\sim
	\sqrt{\pi}
	\,2^{-2/3}
	\,y^{-5/6}
	\,
	\mathrm{Ai}
	\!\left(
		-2^{1/3}a
	\right).
	\label{eq:fl-Ai}
\end{equation}
Equation~\eqref{eq:fl-Ai} provides a uniform approximation across the
transition from oscillatory to exponentially decaying behavior and
therefore describes the onset of the PTA apparent sensitivity cutoff.

The Airy function approximation, Eq.~\eqref{eq:fl-Ai}, shows that in this limit function $f_l(y)$, Eq.~\eqref{eq:fl-def}, exhibits a maxima. At the turning point,
\(
a=0,
\)
the Airy function takes the value
\(
\mathrm{Ai}(0)=0.355028\ldots .
\)
However, the global maximum of the Airy function is attained at a negative value of its argument (see Sec.~9.9 of \cite{NIST:DLMF}):
\begin{equation}
	\max_x \mathrm{Ai}(x)
	=
	0.53565666\ldots
	\simeq
	\ds\frac12 \,>\,{\rm Ai}(0)\;.
\end{equation}
This maximum is attained at
\(
x_{\rm max}\simeq -1.01879,
\)
which corresponds to
\begin{equation}
	a_{\rm max}
	=
	-\ds\frac{x_{\rm max}}{2^{1/3}}
	\simeq
	0.809 .
\end{equation}
Therefore the maximum response occurs not exactly at the turning point,
but slightly on its oscillatory side,
\[
	y-l
	=
	a_{\rm max}\,l^{1/3}
	=
	O(l^{1/3}).
\]
Consequently, the largest PTA sensitivity attained within the turning-point
region scales as
\begin{equation}
	\max_a f_l(y)=\sqrt{\pi}
	\,2^{-2/3}
	\,y^{-5/6}
	\,
	\mathrm{Ai}
	\!\left(
		-2^{1/3}a_{\rm max}
	\right)
	\sim
	\sqrt{\pi}
	\,2^{-5/3}
	\,y^{-5/6}\;.
	\label{eq:fl-secondary-peak}
\end{equation}

For
\(
a>0
\; (l<y),
\)
the argument of the Airy function ${\rm Ai}\left(-2^{1/3}a\right)$ is negative, and $\mathrm{Ai}$ is oscillatory,
\[
	\mathrm{Ai}(x)
	\sim
	\ds\frac{1}{\sqrt{\pi}}
	(-x)^{-1/4}
	\sin\!\left[
		\ds\frac{2}{3}(-x)^{3/2}+\ds\frac{\pi}{4}
	\right],
	\qquad
	x\to-\infty .
\] 
In this limit
Eq.~\eqref{eq:fl-Ai} matches smoothly onto the Debye asymptotic
expansion derived in the previous subsection.

For
\(
a<0
\; (l>y),
\)
the Airy-function argument becomes positive and
\[
	\mathrm{Ai}(x)
	\sim
	\ds\frac{1}{\sqrt{\pi}}
	x^{-1/4}
	\exp\!\left(
		-\ds\frac{2}{3}x^{3/2}
	\right),
	\qquad
	x\to+\infty .
\]
The detector response is therefore exponentially suppressed beyond the
turning point.This characteristic behavior of the Airy function is mirrored in the corresponding sensitivity function $f_l(y)$, as seen in Region IV of Fig.~\ref{fig:fl}.

The transition from oscillatory to exponentially decaying behavior
occurs at the turning point of the spherical Bessel functions,
\(
l+1/2=y,
\)
which, to leading order, defines the characteristic multipole
\begin{equation}
	l_{3\to4}
	\simeq
	y .
	\label{eq:l-34-boundary}
\end{equation}
For a PTA composed of Galactic pulsars, an estimate of this scale is
given in Eq.~\eqref{eq:fl-3-4-boundary-number}.

This multipole marks the onset of the 
effective diffraction cutoff
regime and
therefore defines the  angular-resolution limit,
\begin{equation}
	l_{\rm cut}\simeq\omega L,
\end{equation}
of an individual Earth--pulsar baseline detector. Multipoles
significantly exceeding
\(l_{\rm cut}\)
are exponentially suppressed and are therefore effectively
unobservable.
Combining this result with the multipoles
\(l_{1\to2}\) and \(l_{2\to3}\) derived in the preceding subsection
yields the asymptotic hierarchy
\begin{equation}
	l_{1\to2}\lesssim l_{2\to3}
	\ll
	l_{3\to4}\equiv l_{\rm cut}.
\end{equation}

This hierarchy shows that the Earth-term-dominated, transition, and
pulsar-term-dominated regimes all lie within the oscillatory domain of
the underlying spherical Bessel functions. By contrast, the
sensitivity-cutoff regime is governed by the turning-point and
exponential asymptotics. The corresponding regime boundaries,
defined by the characteristic multipoles
$l_{1\to2}$, $l_{2\to3}$ and $l_{3\to4}$
are illustrated in Fig.~\ref{fig:fl}.

\section{Relation to TAM\&LOS-integral formalisms}
\label{app:TAM-LOS-adaptation}
The total-angular-momentum (TAM) representation was introduced into PTA analyses by \cite{qin2019} and subsequently employed in studies of finite-distance, anisotropic, and polarized gravitational-wave backgrounds \cite{ng2022, liu2022, bernardo2023, domenech2024}. Since these approaches are widely used in the PTA literature, it is useful to establish an explicit correspondence between the tensor-spherical-harmonic formalism developed in this paper and both the TAM and line-of-sight (LOS) integral representations. 

In this appendix we show a way to unify the presented frameworks. The main results of the appendix are Eqs.~\eqref{eq:tam-spin2-translation} and \eqref{eq:jl-integral-analyt}. The former establishes the relation of the spin-0 and spin-2 decomposition coefficients, while the letter allows to analytically compute the LOS-integral.

The TAM formalism is based on the decomposition of the timing residuals in the (spin-0) spherical harmonics basis:
\begin{equation}
	\delta t(\hat{p})
	=
	\sum_{l=0}^{\infty}\sum_{m=-l}^{l}
	a^{\rm TAM}_{lm} Y_{lm}(\hat{p}),
	\label{eq:tam-decomposition}
\end{equation}
where all coefficients $a^{\rm TAM}_{lm}$ are referred to a single frequency bin $\omega$ as it was described in Sec.~\ref{sec:formalism}.
This decomposition seamlessly translates to the spin-2 spherical harmonic formalism used in this paper by comparing Eqs.~\eqref{eq:h-spherical-decomposition}--\eqref{eq:RC-analyt} to Eq.~\eqref{eq:tam-decomposition}:
\begin{equation}
	a^{\rm TAM}_{lm}(\omega) =
	a^G_{lm}(\omega) \cdot
	2\pi\,
	(-i)^l\,
	e^{-iy}\,
	f_l(y).
	\label{eq:tam-spin2-translation}
\end{equation}
Physically, Eq.~\eqref{eq:tam-spin2-translation} provides the mapping between the GW field (spin-2) and the observed timing residual (spin-0) spaces. From here, the sense of the response function $f_l(y)$ becomes apparent.

The line-of-sight (LOS) integral formalism is based on the Sachs--Wolfe representation of the timing residuals' redshift \cite{sachs1967}. Starting from the LOS integral for the redshift fluctuation in a flat static background:
\begin{equation}
	z(t_r, \hat{p}) = -\frac{1}{2} \int_{t_e}^{t_r} dt' \, \hat{p}^a \hat{p}^b \frac{\partial}{\partial t'} h_{ab}(t', \vec{x}(t')),
	\label{eq:sachs-wolfe}
\end{equation}
we expand the metric perturbation in plane waves using Eq.~\eqref{eq:h-fourier}:
\begin{equation}
	h_{ab}(t, \vec{x}) = \sum_A \int d\omega \oint d\hat\Omega \, h_A(\omega, \hat\Omega) e^A_{ab}(\hat\Omega) e^{i\omega(t - \hat\Omega \cdot \vec{x})}.
\end{equation}
The time derivative yields $\partial_{t'} h_{ab} = i\omega h_{ab}$. The photon trajectory is $\vec{x}(t') = \vec{x}_r - \hat{p}(t_r - t')$. Substituting this and changing the integration variable to $x = \omega(t_r - t')$ (with $dx = -\omega dt'$ and limits from $0$ to $y = \omega L$), we obtain:
\begin{equation}
	\begin{aligned}
		z(t_r, \hat{p}) &= \frac{i}{2} 
		\hat{p}^a \hat{p}^b 
		\int_0^y dx \, e^{-ix} 
		\int d\omega \oint d\hat\Omega \, h_A(\omega, \hat\Omega) e^A_{ab}(\hat\Omega) e^{i\omega \hat\Omega \cdot \vec{x}_r + i\omega t_r} e^{-ix \hat\Omega \cdot \hat{p}} 
		\\
						&= \frac{i}{2} 	
						\int_0^y dx \, e^{-ix} 
						\int d\omega \oint d\hat\Omega \, 
						e^{i\omega \hat\Omega \cdot \vec{x}_r + i\omega t_r}
						\sum_A
						h_A(\omega, \hat\Omega)
						e^A_{ab}(\hat\Omega)
						\hat{p}^a \hat{p}^b e^{-ix \hat\Omega \cdot \hat{p}},
	\end{aligned}
\end{equation}
where $\hat\Omega$ is the GW propagation direction. For a single GW frequency, choosing the origin of the coordinates to be $\vec{x_r} = 0$, this gives
\begin{equation}
	z(\hat{p}) =
	\frac{i}{2} 	
	\int_0^y dx \, e^{-ix} 
	\sum_A
	h_A(\omega, \hat\Omega)
	e^A_{ab}(\hat\Omega)
	\hat{p}^a \hat{p}^b e^{-ix \hat\Omega \cdot \hat{p}},
\end{equation}

Projecting $ z$ onto the spherical harmonic basis $Y_{lm}(\hat{p})$, we need to evaluate the angular integral:
\begin{equation}
	\mathcal{I}^A =
	\int d\hat{p}\,
	\hat{p}^a \hat{p}^b  	
	e^A_{ab}(\hat\Omega)
	Y^*_{lm} (\hat{p})
	e^{-ix \hat\Omega \cdot \hat{p}}.
\end{equation}
Note that
\begin{equation}
	\hat{p}^a \hat{p}^b 
	e^{-ix \hat\Omega \cdot \hat{p}}
	= 
	-\frac{1}{x^2} 
	\frac{\partial^2}{\partial \hat{\Omega}^a \partial \hat{\Omega}^b}
	e^{-ix \hat\Omega \cdot \hat{p}}.
\end{equation}
Thus, the integral becomes
\begin{equation}
	\begin{aligned}
		\mathcal{I}^A &=
		-\frac{1}{x^2} 
		\int d\hat{p}\,
		Y^*_{lm} (\hat{p})	
		e^A_{ab}(\hat\Omega)
		\frac{\partial^2}{\partial \hat{\Omega}^a \partial \hat{\Omega}^b}
		e^{-ix \hat\Omega \cdot \hat{p}}
		\\
					  &=
					  -\frac{1}{x^2}
					  e^A_{ab}(\hat\Omega) 
					  \frac{\partial^2}{\partial \hat{\Omega}^a \partial \hat{\Omega}^b}
					  \int d\hat{p}\,
					  Y^*_{lm} (\hat{p})	
					  e^{-ix \hat\Omega \cdot \hat{p}}.
	\end{aligned}
\end{equation}
Using the Rayleigh plane-wave expansion
\begin{equation}
	e^{-ix \hat\Omega \cdot \hat{p}}=
	4\pi \sum_{LM}
	(-i)^{L} j_L(x) 
	Y^*_{LM}(\hat\Omega)
	Y_{LM} (\hat{p}),
\end{equation}
the integration over $\hat{p}$ yields $4\pi (-i)^l j_l(x) Y^*_{lm}(\hat\Omega)$, so that
\begin{equation}
	\mathcal{I}^A =
	-\frac{4\pi (-i)^l j_l(x) }{x^2} 
	e^A_{ab}(\hat\Omega)
	\frac{\partial^2}{\partial \hat{\Omega}^a \partial \hat{\Omega}^b}
	Y^*_{lm}(\hat\Omega).
\end{equation}

To evaluate the derivatives, we decompose the gradient operator as
\begin{equation}
	\frac{\partial}{\partial \hat{\Omega}^a} = 
	\hat\Omega_a \frac{\partial}{\partial r}
	+
	\nabla_a,
\end{equation}
where $\nabla_a$ is the angular part of the derivative, corresponding to the covariant derivative on the sphere. The second derivative is
\begin{equation}
	\frac{\partial^2}{\partial \hat{\Omega}^a \partial \hat{\Omega}^b} =
	\hat\Omega_a \hat\Omega_b 
	\frac{\partial^2}{\partial r^2}
	+
	\nabla_a
	\hat\Omega_b \frac{\partial}{\partial r}
	+
	\hat\Omega_a \frac{\partial}{\partial r}
	\nabla_b
	+
	\nabla_a \nabla_b.
\end{equation}
Since the polarization tensor is transverse, 
\begin{equation}
	\hat\Omega^a e^A_{ab}(\hat\Omega) = 0,
\end{equation}
the contraction simplifies to
\begin{equation}
	e^A_{ab}(\hat\Omega)
	\frac{\partial^2}{\partial \Omega^a \partial \Omega^b} =
	e^A_{ab}(\hat\Omega)
	\nabla^a \nabla^b.
\end{equation}
Therefore,
\begin{equation}
	\mathcal{I}^A =
	-\frac{4\pi (-i)^l j_l(x) }{x^2} 
	e^A_{ab}(\hat\Omega)
	\nabla^a \nabla^b
	Y^*_{lm}(\hat\Omega).
\end{equation}

Using the standard relation between the tensor-harmonic operators and spin-weighted spherical harmonics (see \ref{app:basis-defs}), contraction with the polarization tensors is equivalent to the action of the spin-raising and spin-lowering operators. Therfore,
\begin{subequations}
	\begin{align}
		e^+_{ab}(\hat\Omega)
		\nabla^a \nabla^b
		Y^*_{lm}(\hat\Omega) &= 
		\frac{1}{\sqrt{2} N_l}
		\left(
			{}_{2}Y^*_{lm} (\hat\Omega) +
			{}_{-2}Y^*_{lm} (\hat\Omega)
		\right) = 
		W^*_{lm}(\hat\Omega),
		\\
		e^\times_{ab}(\hat\Omega)
		\nabla^a \nabla^b
		Y^*_{lm}(\hat\Omega) &= 
		\frac{i}{\sqrt{2} N_l }
		\left(
			{}_{2}Y^*_{lm} (\hat\Omega) -
			{}_{-2}Y^*_{lm} (\hat\Omega)
		\right)
		=
		X^*_{lm}(\hat\Omega).
	\end{align}
\end{subequations}
This gives
\begin{subequations}
	\begin{align}
		\mathcal{I}^+ &=
		-\frac{4\pi}{N_l}  (-i)^l
		\frac{ j_l(x) }{x^2} 
		W^*_{lm}(\hat\Omega),
		\\
		\mathcal{I}^\times &=
		-\frac{4\pi}{N_l}  (-i)^l
		\frac{ j_l(x) }{x^2} 
		X^*_{lm}(\hat\Omega).
	\end{align}
\end{subequations}
Defining $c_l(x) = -4\pi (-i)^l j_l(x)/x^2$, and using the definition of the spherical harmonic coefficients $a^G_{lm}(\omega)$ from Eq.~\eqref{eq:h-a-transform}, the angular integration yields
\begin{equation}
	\oint d\hat\Omega\,
	\sum_A h_A(\omega, \hat\Omega)
	\mathcal{I}^A
	= \frac{c_l(x)}{N_l}
	a^G_{lm}(\omega).
\end{equation}

In our paper, the origin is chosen to be $\vec{x}_r = 0$. Substituting the above result into the expression for $z$, the single-frequency Fourier harmonic of the $a_{lm}$ coefficient is
\begin{equation}
	\begin{aligned}
		a_{lm}^{\rm TAM} (\omega) &= 
		\frac{i}{2} 	
		\int_0^y dx \, e^{-ix} 
		\sum_A h_A(\omega', \hat\Omega)
		\mathcal{I}^A
		\\
								  &=
								  \frac{i}{2} 	
								  \int_0^y dx \, e^{-ix}  
								  \frac{c_l(x)}{N_l}
								  a^G_{lm}(\omega')
								  \\
								  &=
								  a^G_{lm} (\omega) 
								  i
								  \frac{-2\pi(-i)^l}{N_l}
								  \int_0^y dx \, e^{-ix} 
								  \frac{ j_l(x) }{x^2}. 
	\end{aligned}
\end{equation}
Comparing this result to the connection between TAM and spin-2 spherical harmonic formalisms, Eq.~\eqref{eq:tam-spin2-translation}, we obtain
\begin{equation}
	i e^{-iy} N_l f_l(y) = 
	\int_0^y dx \, e^{-ix} 
	\frac{ j_l(x) }{x^2},
	\label{eq:jl-integral-analyt}
\end{equation}
where the plane-wave relation $\delta t(\hat{p}) = z(\hat{p})/i\omega$ is used.

Equation \eqref{eq:jl-integral-analyt} provides an exact analytical representation of the LOS kernel in terms of the closed-form response function $f_l(y)$. The result therefore establishes the equivalence between the LOS-integral approach and the tensor-harmonic formalism developed in this paper, while replacing the LOS integral by a compact analytic expression involving spherical Bessel functions. Furthermore, analogous analytical derivations for the results from \cite{gair2015} can be adapted to calculate the integrals for non-Einsteinian GW modes.

\section*{References}
\bibliographystyle{iopart-num}
\bibliography{literature}{}

\providecommand{\newblock}{}
\begin{thebibliography}{100}
\expandafter\ifx\csname url\endcsname\relax
  \def\url#1{{\tt #1}}\fi
\expandafter\ifx\csname urlprefix\endcsname\relax\def\urlprefix{URL }\fi
\providecommand{\eprint}[2][]{\url{#2}}

\bibitem{einstein1918}
Einstein A 1918 {\em Sitzungsberichte der Königlich Preussischen Akademie der
  Wissenschaften\/}  154--167
  \urlprefix\url{https://ui.adsabs.harvard.edu/abs/1918SPAW.......154E}

\bibitem{MTW1973}
Misner C~W, Thorne K~S and Wheeler J~A 1973 {\em Gravitation\/} (San Francisco:
  W. H. Freeman)

\bibitem{Maggiore2008}
Maggiore M 2008 {\em Gravitational Waves. Volume 1: Theory and Experiments\/}
  (Oxford: Oxford University Press)

\bibitem{Abbott2017GW170817}
Abbott B~P {\em et~al.\/} (LIGO Scientific Collaboration and Virgo
  Collaboration) 2017 {\em Physical Review Letters\/} {\bf 119} 161101

\bibitem{Monitor2017}
Abbott B~P {\em et~al.\/} (LIGO Scientific Collaboration, Virgo Collaboration,
  Fermi GBM, INTEGRAL, and many others) 2017 {\em Astrophysical Journal
  Letters\/} {\bf 848} L12

\bibitem{taylor1982}
Taylor J~H and Weisberg J~M 1982 {\em The Astrophysical Journal\/} {\bf 253}
  908--920 ISSN 0004-637X
  \urlprefix\url{https://ui.adsabs.harvard.edu/abs/1982ApJ...253..908T}

\bibitem{ligo2016}
Abbott B~P {\em et~al.\/} (LIGO Scientific Collaboration and Virgo
  Collaboration) 2016 {\em Physical Review Letters\/} {\bf 116} 061102

\bibitem{abbott2021gwtc3}
Abbott R {\em et~al.\/} (LIGO Scientific Collaboration, Virgo Collaboration,
  and KAGRA Collaboration) 2023 {\em Physical Review X\/} {\bf 13} 041039

\bibitem{KamionkowskiKosowsky1999}
Kamionkowski M and Kosowsky A 1999 {\em Annual Review of Nuclear and Particle
  Science\/} {\bf 49} 77--123

\bibitem{CapriniFigueroa2018}
Caprini C and Figueroa D~G 2018 {\em Classical and Quantum Gravity\/} {\bf 35}
  163001

\bibitem{Andresen2022}
Andresen H, O'Connor E~P, Janka H~T, Müller E and Summa A 2022 {\em Living
  Reviews in Computational Astrophysics\/} {\bf 8} 1

\bibitem{weber1960}
Weber J 1960 {\em Physical Review\/} {\bf 117} 306--313

\bibitem{weber1969}
Weber J 1969 {\em Physical Review Letters\/} {\bf 22} 1320--1324

\bibitem{garwin1974}
Garwin R~L 1974 {\em Nature\/} {\bf 250} 316--317

\bibitem{schutz2009}
Schutz B~F 2009 {\em Classical and Quantum Gravity\/} {\bf 26} 094020

\bibitem{hulse1975}
Hulse R~A and Taylor J~H 1975 {\em The Astrophysical Journal\/} {\bf 195}
  L51--L53 ISSN 0004-637X
  \urlprefix\url{https://ui.adsabs.harvard.edu/abs/1975ApJ...195L..51H}

\bibitem{peters1963}
Peters P~C and Mathews J 1963 {\em Physical Review\/} {\bf 131} 435--440

\bibitem{peters1964}
Peters P~C 1964 {\em Physical Review\/} {\bf 136} B1224--B1232

\bibitem{landaulifshitz1975}
Landau L~D and Lifshitz E~M 1975 {\em The Classical Theory of Fields\/} 4th ed
  (Oxford: Pergamon Press)

\bibitem{weisberg2016}
Weisberg J~M and Huang Y 2016 {\em Astrophysical Journal\/} {\bf 829} 55

\bibitem{wolszczan1991}
Wolszczan A 1991 {\em Nature\/} {\bf 350} 688--690

\bibitem{kaspi2000}
Kaspi V~M, Lyne A~G, Manchester R~N {\em et~al.\/} 2000 {\em Astrophysical
  Journal\/} {\bf 543} 321--327

\bibitem{burgay2003}
Burgay M {\em et~al.\/} 2003 {\em Nature\/} {\bf 426} 531--533

\bibitem{lyne2004}
Lyne A~G {\em et~al.\/} 2004 {\em Science\/} {\bf 303} 1153--1157

\bibitem{stairs2003}
Stairs I~H 2003 {\em Living Reviews in Relativity\/} {\bf 6} 5

\bibitem{kramer2006}
Kramer M {\em et~al.\/} 2006 {\em Science\/} {\bf 314} 97--102

\bibitem{wex2014}
Wex N 2014 Testing relativistic gravity with radio pulsars {\em Frontiers in
  Relativistic Celestial Mechanics: Applications and Experiments\/} ed Kopeikin
  S~M (Berlin and Boston: De Gruyter) pp 39--102

\bibitem{kramer2021}
Kramer M, Stairs I~H, Manchester R~N, Wex N, Deller A~T, Coles W~A, Ali M,
  Burgay M, Camilo F, Cognard I, Damour T, Desvignes G, Ferdman R~D, Freire
  P~C~C, Grondin S, Guillemot L, Hobbs G~B, Janssen G, Karuppusamy R, Lorimer
  D~R, Lyne A~G, McKee J~W, McLaughlin M, Münch L~E, Perera B~B~P, Pol N,
  Possenti A, Sarkissian J, Stappers B~W and Theureau G 2021 {\em Physical
  Review X\/} {\bf 11} 041050
  \urlprefix\url{https://ui.adsabs.harvard.edu/abs/2021PhRvX..11d1050K}

\bibitem{sazhin1978}
Sazhin M~V 1978 {\em Soviet Astronomy\/} {\bf 22} 36--38 ISSN 0038-5301
  \urlprefix\url{https://ui.adsabs.harvard.edu/abs/1978SvA....22...36S}

\bibitem{detweiler1979}
Detweiler S 1979 {\em The Astrophysical Journal\/} {\bf 234} 1100--1104 ISSN
  0004-637X
  \urlprefix\url{https://ui.adsabs.harvard.edu/abs/1979ApJ...234.1100D}

\bibitem{kaspi1994}
Kaspi V~M, Taylor J~H and Ryba M~F 1994 {\em The Astrophysical Journal\/} {\bf
  428} 713 ISSN 0004-637X
  \urlprefix\url{https://ui.adsabs.harvard.edu/abs/1994ApJ...428..713K}

\bibitem{jenet2006}
Jenet F~A, Hobbs G~B, van Straten W, Manchester R~N, Bailes M, Verbiest J~P~W,
  Edwards R~T, Hotan A~W, Sarkissian J~M and Ord S~M 2006 {\em Astrophysical
  Journal\/} {\bf 653} 1571--1576

\bibitem{kopeikin1997}
Kopeikin S~M 1997 {\em Physical Review D\/} {\bf 56} 4455--4469

\bibitem{kopeikin1999}
Kopeikin S~M 1999 {\em Monthly Notices of the Royal Astronomical Society\/}
  {\bf 305} 563--568

\bibitem{shannon2015}
Shannon R~M, Ravi V, Lentati L~T, Lasky P~D, Hobbs G, Kerr M, Manchester R~N,
  Coles W~A, Levin Y, Bailes M {\em et~al.\/} 2015 {\em Science\/} {\bf 349}
  1522--1525

\bibitem{hellings1983}
Hellings R~W and Downs G~S 1983 {\em The Astrophysical Journal\/} {\bf 265}
  L39--L42 ISSN 0004-637X
  \urlprefix\url{https://ui.adsabs.harvard.edu/abs/1983ApJ...265L..39H}

\bibitem{foster1990}
Foster R~S and Backer D~C 1990 {\em The Astrophysical Journal\/} {\bf 361} 300
  ISSN 0004-637X
  \urlprefix\url{https://ui.adsabs.harvard.edu/abs/1990ApJ...361..300F}

\bibitem{allen2023PhRvDa}
{Allen} B 2023 {\em \prd\/} {\bf 107} 043018 (\textit{Preprint}
  \eprint{2205.05637})

\bibitem{allen2023PhRvDb}
{Allen} B and {Romano} J~D 2023 {\em \prd\/} {\bf 108} 043026
  (\textit{Preprint} \eprint{2208.07230})

\bibitem{allen2024CQGra}
{Romano} J~D and {Allen} B 2024 {\em Classical and Quantum Gravity\/} {\bf 41}
  175008 (\textit{Preprint} \eprint{2308.05847})

\bibitem{allen2025PhRvL}
{Allen} B and {Romano} J~D 2025 {\em \prl\/} {\bf 134} 031401
  (\textit{Preprint} \eprint{2407.10968})

\bibitem{agazie2023}
Agazie G, Anumarlapudi A, Archibald A~M, Arzoumanian Z, Baker P~T, Bécsy B,
  Blecha L, Brazier A, Brook P~R, {Burke-Spolaor} S, Burnette R, Case R,
  Charisi M, Chatterjee S, Chatziioannou K, Cheeseboro B~D, Chen S, Cohen T,
  Cordes J~M, Cornish N~J, Crawford F, Cromartie H~T, Crowter K, Cutler C~J,
  Decesar M~E, Degan D, Demorest P~B, Deng H, Dolch T, Drachler B, Ellis J~A,
  Ferrara E~C, Fiore W, Fonseca E, Freedman G~E, {Garver-Daniels} N, Gentile
  P~A, Gersbach K~A, Glaser J, Good D~C, Gültekin K, Hazboun J~S, Hourihane S,
  Islo K, Jennings R~J, Johnson A~D, Jones M~L, Kaiser A~R, Kaplan D~L, Kelley
  L~Z, Kerr M, Key J~S, Klein T~C, Laal N, Lam M~T, Lamb W~G, Lazio T~J~W,
  Lewandowska N, Littenberg T~B, Liu T, Lommen A, Lorimer D~R, Luo J, Lynch
  R~S, Ma C~P, Madison D~R, Mattson M~A, McEwen A, McKee J~W, McLaughlin M~A,
  McMann N, Meyers B~W, Meyers P~M, Mingarelli C~M~F, Mitridate A, Natarajan P,
  Ng C, Nice D~J, Ocker S~K, Olum K~D, Pennucci T~T, Perera B~B~P, Petrov P,
  Pol N~S, Radovan H~A, Ransom S~M, Ray P~S, Romano J~D, Sardesai S~C,
  Schmiedekamp A, Schmiedekamp C, Schmitz K, Schult L, {Shapiro-Albert} B~J,
  Siemens X, Simon J, Siwek M~S, Stairs I~H, Stinebring D~R, Stovall K, Sun
  J~P, Susobhanan A, Swiggum J~K, Taylor J, Taylor S~R, Turner J~E, Unal C,
  Vallisneri M, {van Haasteren} R, Vigeland S~J, Wahl H~M, Wang Q, Witt C~A,
  Young O and {Nanograv Collaboration} 2023 {\em The Astrophysical Journal\/}
  {\bf 951} L8 ISSN 0004-637X
  \urlprefix\url{https://ui.adsabs.harvard.edu/abs/2023ApJ...951L...8A}

\bibitem{eptacollaboration2023}
{EPTA Collaboration}, {InPTA Collaboration}, Antoniadis J, Arumugam P, Arumugam
  S, Babak S, Bagchi M, Bak~Nielsen A~S, Bassa C~G, Bathula A, Berthereau A,
  Bonetti M, Bortolas E, Brook P~R, Burgay M, Caballero R~N, Chalumeau A,
  Champion D~J, Chanlaridis S, Chen S, Cognard I, Dandapat S, Deb D, Desai S,
  Desvignes G, {Dhanda-Batra} N, Dwivedi C, Falxa M, Ferdman R~D, Franchini A,
  Gair J~R, Goncharov B, Gopakumar A, Graikou E, Grießmeier J~M, Guillemot L,
  Guo Y~J, Gupta Y, Hisano S, Hu H, Iraci F, {Izquierdo-Villalba} D, Jang J,
  Jawor J, Janssen G~H, Jessner A, Joshi B~C, Kareem F, Karuppusamy R, Keane
  E~F, Keith M~J, Kharbanda D, Kikunaga T, Kolhe N, Kramer M, Krishnakumar M~A,
  Lackeos K, Lee K~J, Liu K, Liu Y, Lyne A~G, McKee J~W, Maan Y, Main R~A,
  Mickaliger M~B, Niţu I~C, Nobleson K, Paladi A~K, Parthasarathy A, Perera
  B~B~P, Perrodin D, Petiteau A, Porayko N~K, Possenti A, Prabu T,
  Quelquejay~Leclere H, Rana P, Samajdar A, Sanidas S~A, Sesana A, Shaifullah
  G, Singha J, Speri L, Spiewak R, Srivastava A, Stappers B~W, Surnis M,
  Susarla S~C, Susobhanan A, Takahashi K, Tarafdar P, Theureau G, Tiburzi C,
  {van der Wateren} E, Vecchio A, Venkatraman~Krishnan V, Verbiest J~P~W, Wang
  J, Wang L and Wu Z 2023 {\em Astronomy and Astrophysics\/} {\bf 678} A50 ISSN
  0004-6361
  \urlprefix\url{https://ui.adsabs.harvard.edu/abs/2023A&A...678A..50E}

\bibitem{reardon2023}
Reardon D~J, Zic A, Shannon R~M, Hobbs G~B, Bailes M, Di~Marco V, Kapur A,
  Rogers A~F, Thrane E, Askew J, Bhat N~D~R, Cameron A, Curyło M, Coles W~A,
  Dai S, Goncharov B, Kerr M, Kulkarni A, Levin Y, Lower M~E, Manchester R~N,
  Mandow R, Miles M~T, Nathan R~S, Osłowski S, Russell C~J, Spiewak R, Zhang S
  and Zhu X~J 2023 {\em The Astrophysical Journal\/} {\bf 951} L6 ISSN
  0004-637X
  \urlprefix\url{https://ui.adsabs.harvard.edu/abs/2023ApJ...951L...6R}

\bibitem{xu2023}
Xu H, Chen S, Guo Y, Jiang J, Wang B, Xu J, Xue Z, Caballero R~N, Yuan J, Xu Y,
  Wang J, Hao L, Luo J, Lee K, Han J, Jiang P, Shen Z, Wang M, Wang N, Xu R, Wu
  X, Manchester R, Qian L, Guan X, Huang M, Sun C and Zhu Y 2023 {\em Research
  in Astronomy and Astrophysics\/} {\bf 23} 075024 ISSN 1674-4527
  \urlprefix\url{https://ui.adsabs.harvard.edu/abs/2023RAA....23g5024X}

\bibitem{miles2025}
Miles M~T, Shannon R~M, Reardon D~J, Bailes M, Champion D~J, Geyer M, Gitika P,
  Grunthal K, Keith M~J, Kramer M, Kulkarni A~D, Nathan R~S, Parthasarathy A,
  Singha J, Theureau G, Thrane E, Abbate F, Buchner S, Cameron A~D, Camilo F,
  Moreschi B~E, Shaifullah G, Shamohammadi M, Possenti A and Krishnan V~V 2025
  {\em Monthly Notices of the Royal Astronomical Society\/} {\bf 536}
  1489--1500 ISSN 0035-8711
  \urlprefix\url{https://ui.adsabs.harvard.edu/abs/2025MNRAS.536.1489M}

\bibitem{jaffe2003}
Jaffe A~H and Backer D~C 2003 {\em Astrophysical Journal\/} {\bf 583} 616--631

\bibitem{sesana2008}
Sesana A, Vecchio A and Colacino C~N 2008 {\em Monthly Notices of the Royal
  Astronomical Society\/} {\bf 390} 192--209 ISSN 0035-8711
  \urlprefix\url{https://ui.adsabs.harvard.edu/abs/2008MNRAS.390..192S}

\bibitem{sesana2010}
Sesana A and Vecchio A 2010 {\em Classical and Quantum Gravity\/} {\bf 27}
  084016 ISSN 0264-9381
  \urlprefix\url{https://doi.org/10.1088/0264-9381/27/8/084016}

\bibitem{sesana2013}
Sesana A 2013 {\em MNRAS\/} {\bf 433} L1--L5

\bibitem{grishchuk1976}
Grishchuk L~P 1976 {\em Soviet Journal of Experimental and Theoretical Physics
  Letters\/} {\bf 23} 293 ISSN 0021-3640
  \urlprefix\url{https://ui.adsabs.harvard.edu/abs/1976JETPL..23..293G}

\bibitem{vilenkin1981}
Vilenkin A 1981 {\em Physics Letters B\/} {\bf 107} 47--50 ISSN 0370-2693
  \urlprefix\url{https://ui.adsabs.harvard.edu/abs/1981PhLB..107...47V}

\bibitem{damour2005}
Damour T and Vilenkin A 2005 {\em Physical Review D\/} {\bf 71} 063510

\bibitem{caprini2018}
Caprini C and Figueroa D~G 2018 {\em Classical and Quantum Gravity\/} {\bf 35}
  163001

\bibitem{mingarelli2013}
Mingarelli C~M~F, Sidery T, Mandel I and Vecchio A 2013 {\em Physical Review
  D\/} {\bf 88} 062005 ISSN 1550-79980556-2821
  \urlprefix\url{https://ui.adsabs.harvard.edu/abs/2013PhRvD..88f2005M}

\bibitem{grunthal2024}
Grunthal K, Nathan R~S, Thrane E, Champion D~J, Miles M~T, Shannon R~M,
  Kulkarni A~D, Abbate F, Buchner S, Cameron A~D, Geyer M, Gitika P, Keith M~J,
  Kramer M, Lasky P~D, Parthasarathy A, Reardon D~J, Singha J and Krishnan V~V
  2024 {\em Monthly Notices of the Royal Astronomical Society\/} {\bf 536}
  1501--1517 ISSN 0035-8711, 1365-2966 (\textit{Preprint} \eprint{2412.01214})
  \urlprefix\url{http://arxiv.org/abs/2412.01214}

\bibitem{curylo2026}
Curyło M, Thrane E, Lasky P~D and Gaynor D~S 2026 A comprehensive framework
  for phase-coherent mapping of the gravitational-wave sky with pulsar timing
  arrays \urlprefix\url{https://ui.adsabs.harvard.edu/abs/2026arXiv260419073C}

\bibitem{anholm2009}
Anholm M, Ballmer S, Creighton J~D~E, Price L~R and Siemens X 2009 {\em
  Physical Review D\/} {\bf 79} 084030 ISSN 1550-79980556-2821
  \urlprefix\url{https://ui.adsabs.harvard.edu/abs/2009PhRvD..79h4030A}

\bibitem{lee2011}
Lee K~J, Wex N, Kramer M, Stappers B~W {\em et~al.\/} 2011 {\em MNRAS\/} {\bf
  414} 3251--3264

\bibitem{ellis2012}
Ellis J~A, Jenet F~A and McLaughlin M~A 2012 {\em ApJ\/} {\bf 753} 96

\bibitem{mingarelli2014}
Mingarelli C~M~F and Sidery T 2014 {\em Physical Review D\/} {\bf 90} 062011
  ISSN 1550-7998, 1550-2368
  \urlprefix\url{https://link.aps.org/doi/10.1103/PhysRevD.90.062011}

\bibitem{tsai2026}
Tsai A~C, Jow D~L and Pen U~L  {\bf 114} 022003 ISSN 1550-7998,
  1550-7998,0556-2821
  \urlprefix\url{https://ui.adsabs.harvard.edu/abs/2026PhRvD.114b2003T/abstract}

\bibitem{janssen2015}
Janssen G~H {\em et~al.\/} 2015 {\em Proceedings of Science\/} {\bf AASKA14}
  037

\bibitem{shannon2025}
Shannon R~M, Bhat N~D~R, Chalumeau A, Chen S, Cromartie H~T, Gopukumar A,
  Grunthal K, Hazboun J~S, Iraci F, Joshi B~C, Kato R, Keith M~J, Lee K, Liu K,
  Middleton H, Miles M~T, Mingarelli C~M~F, Parthasarathy A, Reardon D~J,
  Shaifullah G~M, Takahashi K, Tiburzi C, Truant R~J, Xue X, Zic A and {The
  SKAO Pulsar Science Working Group} 2025 The {{SKAO Pulsar Timing Array}}
  \urlprefix\url{https://ui.adsabs.harvard.edu/abs/2025arXiv251216163S}

\bibitem{kramer2022}
Kramer M {\em et~al.\/} 2022 {\em Experimental Astronomy\/} {\bf 53} 217--252

\bibitem{deller2011}
Deller A~T, Brisken W~F, Chatterjee S, Cordes J~M, Goss W~M, Janssen G~H,
  Kovalev Y~Y, Lazio T~J~W, Petrov L and Stappers B~W 2011 {{PSR$\pi$}}: {{A}}
  large {{VLBA}} pulsar astrometry program
  \urlprefix\url{https://ui.adsabs.harvard.edu/abs/2011evga.conf..178D}

\bibitem{ng2022}
Ng K~W 2022 {\em Physical Review D\/} {\bf 106} 043505 ISSN 1550-79980556-2821
  \urlprefix\url{https://ui.adsabs.harvard.edu/abs/2022PhRvD.106d3505N}

\bibitem{liu2022}
Liu G~C and Ng K~W 2022 {\em Physical Review D\/} {\bf 106} 064004 ISSN
  1550-79980556-2821
  \urlprefix\url{https://ui.adsabs.harvard.edu/abs/2022PhRvD.106f4004L}

\bibitem{bernardo2023}
Bernardo R~C and Ng K~W 2023 {\em Physical Review D\/} {\bf 107} 044007 ISSN
  1550-79980556-2821
  \urlprefix\url{https://ui.adsabs.harvard.edu/abs/2023PhRvD.107d4007B}

\bibitem{sachs1967}
Sachs R~K and Wolfe A~M 1967 {\em The Astrophysical Journal\/} {\bf 147} 73
  ISSN 0004-637X
  \urlprefix\url{https://ui.adsabs.harvard.edu/abs/1967ApJ...147...73S}

\bibitem{domenech2024}
Domènech G and Tsabodimos A 2024 {\em European Physical Journal C\/} {\bf 84}
  1005 ISSN 1434-6044
  \urlprefix\url{https://ui.adsabs.harvard.edu/abs/2024EPJC...84.1005D}

\bibitem{qin2019}
Qin W, Boddy K~K, Kamionkowski M and Dai L 2019 {\em Physical Review D\/} {\bf
  99} 063002 ISSN 1550-79980556-2821
  \urlprefix\url{https://ui.adsabs.harvard.edu/abs/2019PhRvD..99f3002Q}

\bibitem{jow2026}
Jow D~L, Tsai A and Pen U~L  {\bf 113} 043034 ISSN 1550-79980556-2821
  \urlprefix\url{https://ui.adsabs.harvard.edu/abs/2026PhRvD.113d3034J}

\bibitem{grunthal2026}
Grunthal K, Champion D~J, Thrane E, Nathan R~S, Kramer M and Miles M~T 2026
  {\em Astronomy and Astrophysics\/} {\bf 706} A352 ISSN 0004-6361
  \urlprefix\url{https://ui.adsabs.harvard.edu/abs/2026A&A...706A.352G}

\bibitem{agarwal2026a}
Agarwal D, Romano J~D, {Ali-Haïmoud} Y and Smith T~L 2026 Addressing leakage
  and mode suppression in angular power spectrum estimation for
  gravitational-wave backgrounds using pulsar timing arrays
  \urlprefix\url{https://ui.adsabs.harvard.edu/abs/2026arXiv260220075A}

\bibitem{semenzato2026}
Semenzato F, Bellomo N, Raccanelli A and Mingarelli C~M~F  {\bf 2026} 021 ISSN
  1475-7516
  \urlprefix\url{https://ui.adsabs.harvard.edu/abs/2026JCAP...09..021S}

\bibitem{gair2014}
Gair J, Romano J~D, Taylor S and Mingarelli C~M~F 2014 {\em Physical Review
  D\/} {\bf 90} 082001 ISSN 1550-79980556-2821
  \urlprefix\url{https://ui.adsabs.harvard.edu/abs/2014PhRvD..90h2001G}

\bibitem{cornish2014}
Cornish N~J and {van Haasteren} R 2014 Mapping the nano-{{Hertz}} gravitational
  wave sky
  \urlprefix\url{https://ui.adsabs.harvard.edu/abs/2014arXiv1406.4511C}

\bibitem{gair2015}
Gair J~R, Romano J~D and Taylor S~R 2015 {\em Physical Review D\/} {\bf 92}
  102003 ISSN 1550-7998, 1550-2368 (\textit{Preprint} \eprint{1506.08668})
  \urlprefix\url{http://arxiv.org/abs/1506.08668}

\bibitem{lentati2013}
Lentati L, Alexander P, Hobson M~P, Taylor S, Gair J, Balan S~T and {van
  Haasteren} R 2013 {\em Physical Review D\/} {\bf 87} 104021 ISSN
  1550-79980556-2821
  \urlprefix\url{https://ui.adsabs.harvard.edu/abs/2013PhRvD..87j4021L}

\bibitem{kamionkowski1997}
Kamionkowski M, Kosowsky A and Stebbins A 1997 {\em Physical Review D\/} {\bf
  55} 7368--7388 ISSN 1550-79980556-2821
  \urlprefix\url{https://ui.adsabs.harvard.edu/abs/1997PhRvD..55.7368K}

\bibitem{allen1999}
Allen B and Romano J~D 1999 {\em Physical Review D\/} {\bf 59} 102001
  \urlprefix\url{https://link.aps.org/doi/10.1103/PhysRevD.59.102001}

\bibitem{cornish2015}
{Cornish} N~J and {van Haasteren} R 2014 {\em arXiv e-prints\/} arXiv:1406.4511
  (\textit{Preprint} \eprint{1406.4511})

\bibitem{estabrook1975}
Estabrook F~B and Wahlquist H~D 1975 {\em General Relativity and Gravitation\/}
  {\bf 6} 439--447 ISSN 0001-7701
  \urlprefix\url{https://ui.adsabs.harvard.edu/abs/1975GReGr...6..439E}

\bibitem{thorne1980}
Thorne K~S 1980 {\em Rev. Mod. Phys.\/} {\bf 52} 299--339

\bibitem{ellis2014}
Ellis J~A 2014 {\em Searching for Gravitational Waves Using Pulsar Timing
  Arrays\/} Ph.D. thesis University of Wisconsin--Milwaukee Milwaukee,
  Wisconsin \urlprefix\url{https://minds.wisconsin.edu/handle/1793/93937}

\bibitem{agazie2023a}
Agazie G, Anumarlapudi A, Archibald A~M, Arzoumanian Z, Baker P~T, Bécsy B,
  Blecha L, Brazier A, Brook P~R, {Burke-Spolaor} S, Charisi M, Chatterjee S,
  Cohen T, Cordes J~M, Cornish N~J, Crawford F, Cromartie H~T, Crowter K,
  DeCesar M~E, Demorest P~B, Dolch T, Drachler B, Ferrara E~C, Fiore W, Fonseca
  E, Freedman G~E, {Garver-Daniels} N, Gentile P~A, Glaser J, Good D~C, Guertin
  L, Gültekin K, Hazboun J~S, Jennings R~J, Johnson A~D, Jones M~L, Kaiser
  A~R, Kaplan D~L, Kelley L~Z, Kerr M, Key J~S, Laal N, Lam M~T, Lamb W~G,
  W~Lazio T~J, Lewandowska N, Liu T, Lorimer D~R, Luo J, Lynch R~S, Ma C~P,
  Madison D~R, McEwen A, McKee J~W, McLaughlin M~A, McMann N, Meyers B~W,
  Mingarelli C~M~F, Mitridate A, Ng C, Nice D~J, Ocker S~K, Olum K~D, Pennucci
  T~T, Perera B~B~P, Pol N~S, Radovan H~A, Ransom S~M, Ray P~S, Romano J~D,
  Sardesai S~C, Schmiedekamp A, Schmiedekamp C, Schmitz K, {Shapiro-Albert}
  B~J, Siemens X, Simon J, Siwek M~S, Stairs I~H, Stinebring D~R, Stovall K,
  Susobhanan A, Swiggum J~K, Taylor S~R, Turner J~E, Unal C, Vallisneri M,
  Vigeland S~J, Wahl H~M, Witt C~A, Young O and {The NANOGrav Collaboration}
  2023 {\em The Astrophysical Journal Letters\/} {\bf 951} L10 ISSN 2041-8205,
  2041-8213
  \urlprefix\url{https://iopscience.iop.org/article/10.3847/2041-8213/acda88}

\bibitem{eptacollaboration2023a}
{EPTA Collaboration}, {InPTA Collaboration}, Antoniadis J, Arumugam P, Arumugam
  S, Babak S, Bagchi M, Nielsen A~S~B, Bassa C~G, Bathula A, Berthereau A,
  Bonetti M, Bortolas E, Brook P~R, Burgay M, Caballero R~N, Chalumeau A,
  Champion D~J, Chanlaridis S, Chen S, Cognard I, Dandapat S, Deb D, Desai S,
  Desvignes G, {Dhanda-Batra} N, Dwivedi C, Falxa M, Ferdman R~D, Franchini A,
  Gair J~R, Goncharov B, Gopakumar A, Graikou E, Grießmeier J~M, Guillemot L,
  Guo Y~J, Gupta Y, Hisano S, Hu H, Iraci F, {Izquierdo-Villalba} D, Jang J,
  Jawor J, Janssen G~H, Jessner A, Joshi B~C, Kareem F, Karuppusamy R, Keane
  E~F, Keith M~J, Kharbanda D, Kikunaga T, Kolhe N, Kramer M, Krishnakumar M~A,
  Lackeos K, Lee K~J, Liu K, Liu Y, Lyne A~G, McKee J~W, Maan Y, Main R~A,
  Mickaliger M~B, Niţu I~C, Nobleson K, Paladi A~K, Parthasarathy A, Perera
  B~B~P, Perrodin D, Petiteau A, Porayko N~K, Possenti A, Prabu T, Leclere H~Q,
  Rana P, Samajdar A, Sanidas S~A, Sesana A, Shaifullah G, Singha J, Speri L,
  Spiewak R, Srivastava A, Stappers B~W, Surnis M, Susarla S~C, Susobhanan A,
  Takahashi K, Tarafdar P, Theureau G, Tiburzi C, {van der Wateren} E, Vecchio
  A, Krishnan V~V, Verbiest J~P~W, Wang J, Wang L and Wu Z 2023 {\em Astronomy
  and Astrophysics\/} {\bf 678} A49 ISSN 0004-6361
  \urlprefix\url{https://ui.adsabs.harvard.edu/abs/2023A&A...678A..49E}

\bibitem{vanhaasteren2009}
van Haasteren R~e~a 2009 {\em MNRAS\/} {\bf 395} 1005--1014

\bibitem{ellis2013apj}
Ellis J~A, Siemens X and Creighton J~D~E 2012 {\em ApJ\/} {\bf 756} 175

\bibitem{talbot2021}
Talbot C, Thrane E, Biscoveanu S and Smith R 2021 {\em Physical Review
  Research\/} {\bf 3} 043049
  \urlprefix\url{https://link.aps.org/doi/10.1103/PhysRevResearch.3.043049}

\bibitem{golub2013}
Golub G~H and Van~Loan C~F 2013 {\em Matrix Computations\/} 4th ed (Baltimore:
  Johns Hopkins University Press)

\bibitem{hansen1998}
Hansen P~C 1998 {\em Rank-Deficient and Discrete Ill-Posed Problems\/} (SIAM)

\bibitem{moore1920}
Moore E~H 1920 {\em Bull. Amer. Math. Soc.\/} {\bf 26} 394--395
  \urlprefix\url{https://cir.nii.ac.jp/crid/1583387450082342272}

\bibitem{penrose1955}
Penrose R 1955 {\em Math. Proc. Cambridge Philos. Soc.\/} {\bf 51}(3) 406--413

\bibitem{taylor2021}
Taylor S~R 2021 The {{Nanohertz Gravitational Wave Astronomer}}
  \urlprefix\url{https://ui.adsabs.harvard.edu/abs/2021arXiv210513270T}

\bibitem{foucart2013}
Foucart S and Rauhut H {\em A {{Mathematical Introduction}} to {{Compressive
  Sensing}}\/} Applied and {{Numerical Harmonic Analysis}} (Springer New York)
  ISBN 978-0-8176-4947-0 978-0-8176-4948-7
  \urlprefix\url{https://link.springer.com/10.1007/978-0-8176-4948-7}

\bibitem{hogbom1974}
Högbom J~A 1974 {\em Astronomy and Astrophysics Supplement Series\/} {\bf 15}
  417 ISSN 0365-01380004-6361
  \urlprefix\url{https://ui.adsabs.harvard.edu/abs/1974A&AS...15..417H}

\bibitem{lorimer2013}
Lorimer D~R 2013 The {{Galactic Millisecond Pulsar Population}} {\em Neutron
  {{Stars}} and {{Pulsars}}: {{Challenges}} and {{Opportunities}} after 80
  Years\/} vol 291 (eprint: arXiv:1210.2746) pp 237--242
  \urlprefix\url{https://ui.adsabs.harvard.edu/abs/2013IAUS..291..237L}

\bibitem{ellis2013}
Ellis J~A 2013 {\em Classical and Quantum Gravity\/} {\bf 30} 224004 ISSN
  0264-9381
  \urlprefix\url{https://ui.adsabs.harvard.edu/abs/2013CQGra..30v4004E}

\bibitem{taylor2026}
Taylor S~R 2026 On the angular localization of gravitational-wave signals by
  pulsar timing arrays (\textit{Preprint} \eprint{2603.10120})
  \urlprefix\url{http://arxiv.org/abs/2603.10120}

\bibitem{schult2025}
Schult L, Petrov P, Taylor S~R, Pol N, Laal N, Charisi M and Ma C~P 2025
  Expectations for the first supermassive black-hole binary resolved by {{PTAs
  I}}: {{Model}} efficacy
  \urlprefix\url{https://ui.adsabs.harvard.edu/abs/2025arXiv251001317S}

\bibitem{hobbs2006}
Hobbs G~B, Edwards R~T and Manchester R~N 2006 {\em Monthly Notices of the
  Royal Astronomical Society\/} {\bf 369} 655--672 ISSN 0035-8711
  \urlprefix\url{https://ui.adsabs.harvard.edu/abs/2006MNRAS.369..655H}

\bibitem{boyle2025}
Boyle M 2025 The spherical package Zenodo
  \urlprefix\url{https://zenodo.org/records/17081336}

\bibitem{harris2020}
Harris C~R, Millman K~J, {van der Walt} S~J, Gommers R, Virtanen P, Cournapeau
  D, Wieser E, Taylor J, Berg S, Smith N~J, Kern R, Picus M, Hoyer S, {van
  Kerkwijk} M~H, Brett M, Haldane A, {del Río} J~F, Wiebe M, Peterson P,
  {Gérard-Marchant} P, Sheppard K, Reddy T, Weckesser W, Abbasi H, Gohlke C
  and Oliphant T~E 2020 {\em Nature\/} {\bf 585} 357--362 ISSN 0028-0836
  \urlprefix\url{https://ui.adsabs.harvard.edu/abs/2020Natur.585..357H}

\bibitem{hunter2007}
Hunter J~D 2007 {\em Computing in Science and Engineering\/} {\bf 9} 90--95
  \urlprefix\url{https://ui.adsabs.harvard.edu/abs/2007CSE.....9...90H}

\bibitem{vallisneri2020}
Vallisneri M 2020 {\em Astrophysics Source Code Library\/}  ascl:2002.017
  \urlprefix\url{https://ui.adsabs.harvard.edu/abs/2020ascl.soft02017V}

\bibitem{zonca2019}
Zonca A, Singer L, Lenz D, Reinecke M, Rosset C, Hivon E and Gorski K 2019 {\em
  The Journal of Open Source Software\/} {\bf 4} 1298
  \urlprefix\url{https://ui.adsabs.harvard.edu/abs/2019JOSS....4.1298Z}

\bibitem{virtanen2020}
Virtanen P, Gommers R, Oliphant T~E, Haberland M, Reddy T, Cournapeau D,
  Burovski E, Peterson P, Weckesser W, Bright J, {van der Walt} S~J, Brett M,
  Wilson J, Millman K~J, Mayorov N, Nelson A~R~J, Jones E, Kern R, Larson E,
  Carey C~J, Polat İ, Feng Y, Moore E~W, VanderPlas J, Laxalde D, Perktold J,
  Cimrman R, Henriksen I, Quintero E~A, Harris C~R, Archibald A~M, Ribeiro A~H,
  Pedregosa F, {van Mulbregt} P and {SciPy 1 0 Contributors} 2020 {\em Nature
  Medicine\/} {\bf 17} 261--272
  \urlprefix\url{https://ui.adsabs.harvard.edu/abs/2020NatMe..17..261V}

\bibitem{astropycollaboration2022}
{Astropy Collaboration}, {Price-Whelan} A~M, Lim P~L, Earl N, Starkman N,
  Bradley L, Shupe D~L, Patil A~A, Corrales L, Brasseur C~E, Nöthe M, Donath
  A, Tollerud E, Morris B~M, Ginsburg A, Vaher E, Weaver B~A, Tocknell J,
  Jamieson W, {van Kerkwijk} M~H, Robitaille T~P, Merry B, Bachetti M, Günther
  H~M, Aldcroft T~L, {Alvarado-Montes} J~A, Archibald A~M, Bódi A, Bapat S,
  Barentsen G, Bazán J, Biswas M, Boquien M, Burke D~J, Cara D, Cara M, Conroy
  K~E, Conseil S, Craig M~W, Cross R~M, Cruz K~L, D'Eugenio F, Dencheva N,
  Devillepoix H~A~R, Dietrich J~P, Eigenbrot A~D, Erben T, Ferreira L,
  {Foreman-Mackey} D, Fox R, Freij N, Garg S, Geda R, Glattly L, Gondhalekar Y,
  Gordon K~D, Grant D, Greenfield P, Groener A~M, Guest S, Gurovich S, Handberg
  R, Hart A, {Hatfield-Dodds} Z, Homeier D, Hosseinzadeh G, Jenness T, Jones
  C~K, Joseph P, Kalmbach J~B, Karamehmetoglu E, Kałuszyński M, Kelley M~S~P,
  Kern N, Kerzendorf W~E, Koch E~W, Kulumani S, Lee A, Ly C, Ma Z, MacBride C,
  Maljaars J~M, Muna D, Murphy N~A, Norman H, O'Steen R, Oman K~A, Pacifici C,
  Pascual S, {Pascual-Granado} J, Patil R~R, Perren G~I, Pickering T~E, Rastogi
  T, Roulston B~R, Ryan D~F, Rykoff E~S, Sabater J, Sakurikar P, Salgado J,
  Sanghi A, Saunders N, Savchenko V, Schwardt L, {Seifert-Eckert} M, Shih A~Y,
  Jain A~S, Shukla G, Sick J, Simpson C, Singanamalla S, Singer L~P, Singhal J,
  Sinha M, Sipőcz B~M, Spitler L~R, Stansby D, Streicher O, Šumak J, Swinbank
  J~D, Taranu D~S, Tewary N, Tremblay G~R, {de Val-Borro} M, Van~Kooten S~J,
  Vasović Z, Verma S, {de Miranda Cardoso} J~V, Williams P~K~G, Wilson T~J,
  Winkel B, {Wood-Vasey} W~M, Xue R, Yoachim P, Zhang C, Zonca A and {Astropy
  Project Contributors} 2022 {\em The Astrophysical Journal\/} {\bf 935} 167
  ISSN 0004-637X
  \urlprefix\url{https://ui.adsabs.harvard.edu/abs/2022ApJ...935..167A}

\bibitem{weinberg2008cosmology}
Weinberg S 2008 {\em Cosmology\/} (Oxford: Oxford University Press) ISBN
  9780198526827

\bibitem{allen2024}
Allen B 2024 {\em Physical Review D\/} {\bf 110} 043043 ISSN 1550-79980556-2821
  \urlprefix\url{https://ui.adsabs.harvard.edu/abs/2024PhRvD.110d3043A}

\bibitem{varshalovich1988}
Varshalovich D~A, Moskalev A~N and Khersonskii V~K 1988 {\em Quantum Theory of
  Angular Momentum\/} (World Scientific)

\bibitem{ellis2020}
Ellis J~A, Vallisneri M, Taylor S~R and Baker P~T 2020 {ENTERPRISE}: Enhanced
  numerical toolbox enabling a robust pulsar inference suite
  \urlprefix\url{https://zenodo.org/records/4059815}

\bibitem{johnson2024}
Johnson A~D, Meyers P~M, Baker P~T, Cornish N~J, Hazboun J~S, Littenberg T~B,
  Romano J~D, Taylor S~R, Vallisneri M, Vigeland S~J, Olum K~D, Siemens X,
  Ellis J~A, {van Haasteren} R, Hourihane S, Agazie G, Anumarlapudi A,
  Archibald A~M, Arzoumanian Z, Blecha L, Brazier A, Brook P~R, {Burke-Spolaor}
  S, Bécsy B, {Casey-Clyde} J~A, Charisi M, Chatterjee S, Chatziioannou K,
  Cohen T, Cordes J~M, Crawford F, Cromartie H~T, Crowter K, Decesar M~E,
  Demorest P~B, Dolch T, Drachler B, Ferrara E~C, Fiore W, Fonseca E, Freedman
  G~E, {Garver-Daniels} N, Gentile P~A, Glaser J, Good D~C, Gültekin K,
  Jennings R~J, Jones M~L, Kaiser A~R, Kaplan D~L, Kelley L~Z, Kerr M, Key J~S,
  Laal N, Lam M~T, Lamb W~G, Lazio T~J~W, Lewandowska N, Liu T, Lorimer D~R,
  Luo J, Lynch R~S, Ma C~P, Madison D~R, McEwen A, McKee J~W, McLaughlin M~A,
  McMann N, Meyers B~W, Mingarelli C~M~F, Mitridate A, Ng C, Nice D~J, Ocker
  S~K, Pennucci T~T, Perera B~B~P, Pol N~S, Radovan H~A, Ransom S~M, Ray P~S,
  Sardesai S~C, Schmiedekamp C, Schmiedekamp A, Schmitz K, {Shapiro-Albert}
  B~J, Simon J, Siwek M~S, Stairs I~H, Stinebring D~R, Stovall K, Susobhanan A,
  Swiggum J~K, Turner J~E, Unal C, Wahl H~M, Witt C~A, Young O and {Nanograv
  Collaboration} 2024 {\em Physical Review D\/} {\bf 109} 103012 ISSN
  1550-79980556-2821
  \urlprefix\url{https://ui.adsabs.harvard.edu/abs/2024PhRvD.109j3012J}

\bibitem{kass1995}
Kass R~E and Raftery A~E 1995 {\em Journal of the American Statistical
  Association\/} {\bf 90} 773--795 (\textit{Preprint}
  \eprint{https://doi.org/10.1080/01621459.1995.10476572})
  \urlprefix\url{https://doi.org/10.1080/01621459.1995.10476572}

\bibitem{neyman1933}
Neyman J and Pearson E~S 1933 {\em Philosophical Transactions of the Royal
  Society of London. Series A, Containing Papers of a Mathematical or Physical
  Character\/} {\bf 231} 289--337 ISSN 0264-3952 (\textit{Preprint}
  \eprint{91247}) \urlprefix\url{https://www.jstor.org/stable/91247}

\bibitem{gradshtein2007}
Gradshteĭn I~S, Ryzhik I~M and Jeffrey A 2007 {\em Table of Integrals, Series,
  and Products\/} 7th ed (Amsterdam ; Boston: Academic Press) ISBN
  978-0-12-373637-6

\bibitem{NIST:DLMF}
 2026 {{{\emph{NIST}}}}{\emph{ digital library of mathematical functions}}
  \urlprefix\url{https://dlmf.nist.gov/}

\end{thebibliography}

\end{document}